\documentclass[11pt]{article}

\usepackage{jcapmod}
\usepackage{booktabs}
\usepackage[english]{babel}
\usepackage{amsmath,amssymb,amsbsy,amstext,amsthm,simplewick}
\usepackage{multirow}
\usepackage{epsfig,multicol,bbm,simplewick}
\usepackage{graphicx}
\usepackage{amsfonts}
\usepackage{amssymb}
\usepackage{upgreek}
\usepackage{framed}
\usepackage{soul}
\usepackage{enumitem}
\usepackage{bm}
\usepackage{braket}
\usepackage{wrapfig}
\usepackage{graphicx}
\usepackage{amsfonts}
\usepackage{upgreek}
\usepackage{exscale,relsize}
\usepackage[makeroom]{cancel}
\usepackage{soul}
\usepackage{lipsum}
\usepackage{mdframed}
\usepackage{mathtools}
\usepackage[export]{adjustbox}
\usepackage{colortbl}
\usepackage{fontawesome}
\usepackage{xcolor}
\usepackage{floatrow}
\newfloatcommand{capbtabbox}{table}[][\FBwidth]

\newcommand{\circled}[2][]{%
\tikz[baseline=(char.base)]{%
\node[shape = circle, draw, inner sep = 1pt]
    (char) {\phantom{\ifblank{#1}{#2}{#1}}};%
\node at (char.center) {\makebox[0pt][c]{#2}};}}
\robustify{\circled}
\newcommand*\diff{\mathop{}\!\mathrm{d}}

\definecolor{lightgreen}{cmyk}{0.2, 0, 0.2, 0.2}
\definecolor{lightgray}{cmyk}{0.1,0.2,0,0.1}
\definecolor{lightgray2}{cmyk}{0.1,0.1,0,0.1}
\makeatletter
\newlength{\apb@width}
\newcommand{\autoparbox}[2][c]{\settowidth{\apb@width}{#2}\parbox[#1]{\apb@width}{#2}}

\newcommand{\Cen}[2]{%
  \ifmeasuring@
    #2%
  \else
    \makebox[\ifcase\expandafter #1\maxcolumn@widths\fi]{$\displaystyle#2$}%
  \fi
}
\makeatother
\newcommand{\beq}{\begin{equation}\begin{aligned}}
\newcommand{\eeq}{\end{aligned}\end{equation}}

\numberwithin{equation}{section}
\def\beq{\begin{equation}}
\def\eeq{\end{equation}}
\def\Beq{\begin{equation}\begin{aligned}}
\def\Eeq{\end{aligned}\end{equation}}
\def\bea{\begin{eqnarray}}
\def\eea{\end{eqnarray}}

\def\d{{\rm d}}

\def\beq{\begin{equation}}
\def\eeq{\end{equation}}
\def\bea{\begin{eqnarray}}
\def\eea{\end{eqnarray}}

\def\d{{\rm d}}

\def\d{{\rm d}}

\definecolor{darkviolet}{rgb}{0.58, 0.0, 0.83}

\def\eq#1{{(\ref{#1})}}

\newenvironment{bottompar}{\par\vspace*{\fill}}{\clearpage}

\newcommand*{\xdash}[1][3em]{\rule[0.5ex]{#1}{0.55pt}}

\DeclareRobustCommand{\SkipTocEntry}[4]{}
\DeclareSymbolFont{extraup}{U}{zavm}{m}{n}
\DeclareMathSymbol{\varheart}{\mathalpha}{extraup}{86}
\DeclareMathSymbol{\vardiamond}{\mathalpha}{extraup}{87}
\DeclareMathSymbol{\test}{\mathalpha}{extraup}{88}
\definecolor{darkblue}{rgb}{0.15,0.35,0.75}
\definecolor{redd}{rgb}{0.1, 0.1, 0.1}
\definecolor{vdev}{cmyk}{0.9,0,0.8,0.1}
\usepackage{hyperref}
\hypersetup{pageanchor=false,
linktocpage=true,
colorlinks=true,
citecolor=darkblue,
linkcolor=redd,
urlcolor=darkblue,
pdfauthor={},
pdftitle={},
pdfsubject={}
}

\newcommand{\chic}{\chi_{\bk}}

\newcommand{\opchic}{\hat{\chi}_{\bk}}

\newcommand{\delp}{\delta\phi}
\newcommand{\ddp}{{\delta\phi_{\bm{k}}}}
\newcommand{\ddpp}{{\delta\phi_{\bm{k}'}}}

\newcommand{\ddr}{{\delta\rho_{r,\bm{k}}}}
\newcommand{\ddrp}{{\delta\rho_{r,\bm{k}'}}}

\newcommand{\rr}{\rho_r}
\newcommand{\rp}{\rho_\phi}
\newcommand{\mm}{M_P}
\newcommand{\dv}{V_{\phi}}

\newcommand{\RR}{\mathcal{R}}
\newcommand{\PP}{\mathcal{P}}
\newcommand{\ev}{\epsilon_V}
\newcommand{\etv}{\eta_V}
\newcommand{\ddnc}{\Delta N_{\text{cut}}}
\newcommand{\bk}{{\bm{k}}}

\newcommand{\bx}{\bm{x}}

\DeclareMathOperator{\sinc}{sinc}

\newcommand{\idd}{\mathbb{I}}
\newcommand{\ann}{\hat{a}_{\bm{k}}}
\newcommand{\cre}{\hat{a}^\dagger_{\bm{k}}}
\newcommand{\annp}{\hat{a}_{\bm{k}'}}
\newcommand{\crep}{\hat{a}^\dagger_{\bm{k}'}}
\newcommand{\phih}{{\phi^{(h)}_k}}
\newcommand{\phii}{{\phi^{(i)}_{\bm{k}}}}

\newcommand{\qcl}{Q_{\text{cl}}}
\newcommand{\ecl}{\epsilon_{\text{cl}}}

\newcommand{\vcl}{\left.V\right|_{\phi_{\text{cl}}}}
\newcommand{\dvcl}{\left.V_{\phi}\right|_{\phi_{\text{cl}}}}
\newcommand{\cl}{\text{cl}}
\newcommand{\mn}{M_P}
\newcommand{\drphi}{\mathcal{D}_{\phi}}
\newcommand{\drpi}{\mathcal{D}_{\pi}}
\newcommand{\drphicl}{\mathcal{D}_{\phi}^{\text{(cl)}}}
\newcommand{\drpicl}{\mathcal{D}_{\pi}^{\text{(cl)}}}
\newcommand{\drdphi}{\mathcal{D}_{\phi}^\text{(st)}}
\newcommand{\drdpi}{\mathcal{D}_{\pi}^\text{(st)}}
\newcommand{\dphis}{\delta\phi_{\text{st}}}
\newcommand{\dpis}{\delta\pi_{\text{st}}}
\newcommand{\phcl}{\phi_{\text{cl}}}
\newcommand{\pcl}{\pi_{\text{cl}}}
\newcommand{\bphi}{\bar{\phi}}
\newcommand{\bpi}{\bar{\pi}}

\newcommand{\noise}{\tilde  \xi_\phi}
\newcommand{\pnoise}{\tilde \xi_\pi}
\newcommand{\noisef}{\tilde  \xi_f}
\newcommand{\noiseg}{\tilde  \xi_g}
\newcommand{\sigmak}{{k_\sigma}}

\newcommand{\ddv}{V_{\phi\phi}}
\newcommand{\bareval}{\right|_{\bar{\phi}}}
\newcommand{\cleval}{\right|_{\phi_{\text{cl}}}}

\newcommand{\prefactor}{\sqrt{\frac{2\Gamma T}{(aH)^3}}}

\newcommand{\be}{\text{BE}}

\newcommand{\en}{\text{end}}
\newcommand{\bbn}{\text{BBN}}

\newcommand{\infla}{\text{inf}}
\newcommand{\rh}{\text{rh}}
\newcommand{\rad}{\text{rad}}
\newcommand{\mat}{\text{mat}}

\definecolor{gbcolor}{rgb}{.6,.2,.5}

\begin{document}

\begin{titlepage}

\setcounter{page}{1} \baselineskip=15.5pt \thispagestyle{empty}

\bigskip\

IFT UAM-CSIC 23-37  \hfill DESY-23-046 

\vspace{2cm}
\begin{center}

{\fontsize{20.74}{24}\selectfont  
\bfseries  
Monomial warm inflation revisited}

\end{center}

\vspace{0.2cm}

\begin{center}
{\fontsize{12}{30} \bf   Guillermo Ballesteros$^{1,2}$, Alejandro P\'erez Rodr\'iguez$^{1,2}$ and Mathias Pierre$^{3}$}
\setcounter{footnote}{0}
\end{center}

\begin{center}

\vskip 7pt 
\textsl{$^1$ Departamento de F\'{\i}sica Te\'{o}rica, Universidad Aut\'{o}noma de Madrid (UAM), \\Campus de Cantoblanco, 28049 Madrid, Spain}\\
\textsl{$^2$ Instituto de F\'{\i}sica Te\'{o}rica UAM-CSIC,  Campus de Cantoblanco, 28049 Madrid, Spain}\\
\textsl{$^3$ Deutsches Elektronen-Synchrotron DESY, Notkestr. 85, 22607 Hamburg, Germany}
\vskip 7pt

\end{center}

\vspace{0.3cm}
\centerline{\bf Abstract}
\vspace{0.3cm}

\noindent
We revisit the idea that the inflaton may have dissipated part of its energy into a thermal bath during inflation, considering monomial inflationary potentials and three different forms of dissipation rate. Using a numerical Fokker-Planck approach to describe the stochastic dynamics of inflationary fluctuations, we confront this scenario with current bounds on the spectrum of curvature fluctuations and primordial gravitational waves. We also obtain purely analytical approximations that improve over previously used ones in the small dissipation regime for the amplitude of the spectrum and its tilt.
We show that only our numerical Fokker-Planck method is accurate, fast and precise enough to test these models against current data. We advocate its use in future studies of warm inflation. We also apply the stochastic inflation formalism to this scenario, finding that the resulting spectrum is the same as the one obtained with standard perturbation theory. We discuss the origin and convenience of using a commonly implemented large thermal correction to the primordial spectrum and 
the implications of such a term for a specific scenario. Improved bounds on the scalar spectral index will further constrain warm inflation in the near future.

\begin{bottompar}
\noindent\xdash[15em]\\
\small{
guillermo.ballesteros@uam.es\\
alejandro.perezrodriguez@uam.es\\
mathias.pierre@desy.de}
\end{bottompar}

\end{titlepage}

\hypersetup{pageanchor=true}

\tableofcontents

\section{Introduction}

Inflation is the leading framework to solve the flatness and horizon problems of the standard hot big bang scenario and to account for the temperature anisotropies measured in the cosmic microwave background (CMB). It assumes a phase of accelerated expansion during which the Universe grows its volume by a factor of $\sim e^{3 N}$, where the number of e-folds of inflation, $N$, satisfies $N\gtrsim 50$. The most common way of implementing inflation makes use a single scalar field, $\phi$, the inflaton, with a very flat potential $V(\phi)$. The energy density stored in the potential makes the expansion accelerate as long as the potential is flat enough and its curvature is sufficiently small. This is the standard slow-roll inflation scenario. In this picture, inflation ends when the energy stored in the inflaton is released into other species (and eventually the Standard Model) while the inflaton oscillates around the minimum of its potential. This occurs by means of a process called reheating, during which the expansion of the Universe slows down becoming radiation-dominated. 

The improvement of the CMB measurements has allowed to rule out an important subset of single-field slow-roll inflationary models (concrete choices of $V(\phi)$) that two decades ago were still viable \cite{Planck:2018jri}. This progress has been possible thanks to significantly stronger bounds on the amount of primordial gravitational waves generated during inflation at CMB scales. Much of the experimental effort that will take place in the near future in this area is going to be dedicated to set even more constraining bounds on this quantity, specifically through measurements of the CMB polarization. This may lead to constraining even further the space of allowed single-field slow-roll models, or to a momentous discovery. 

The previously discussed standard description of inflation assumes that the Universe is populated by the inflaton and devoid of any other species which the inflaton may produce. This is justified by small couplings and by the accelerated expansion itself, which quickly dilutes any of the inflaton's offspring. In this paper we will entertain the possibility that particle production during inflation is strong enough that it becomes necessary to consider a thermal bath during inflation. The thermal bath is thought to originate from the inflation gradually releasing part of its energy into other species, which then thermalize (or, more commonly in concrete implementations, lead to yet other particles which finalize thermalize, see \cite{Kamali:2023lzq} for a review). This scenario is known as {\it warm} inflation \cite{Berera:1995ie}, as opposed to the standard picture of {\it (cold)} inflation. The presence of the thermal bath changes the course of inflation as well as the spectrum of primordial fluctuations. Moreover, the reheating of the Universe may occur when the radiation energy density surpasses that of the inflaton, as the latter diminishes continuously. 

Warm inflation has been invoked to salvage notable inflationary potentials from being excluded from the bound on primordial gravitational waves that we mentioned above. The relevant quantity is the tensor-to-scalar ratio, $r$, which compares the amplitudes of the tensor and scalar primordial spectra at CMB scales. The current upper bound on $r$ is $r < 0.032$ at $k=$ 0.05 Mpc$^{-1}$ and 95$\%$ confidence level \cite{Tristram:2021tvh}. This bound strongly excludes in the standard slow-roll picture all monomial potentials $V \propto \phi^n$ (with $n\geq 0$). For sufficiently small $n<1$ with $n \in \mathbb{Q}$, the exclusion comes instead from the value of the tilt of the scalar spectrum. In this paper we explore the consistency of the cases $n=2,4,6$ with current CMB bounds in the context of warm inflation. 

In warm inflation, the transference of energy from the inflaton to the thermal bath is determined by a function of the inflaton and the temperature of the bath: $\Gamma(\phi, T)$. In this paper we consider three different choices for $\Gamma(\phi, T)$, which have been proposed in previous works in concrete implementations of warm inflation. Specifically, we consider the cases $\Gamma\propto T$ (with $\partial\Gamma/\partial \phi =0$)~\cite{Berera:1998px,Bastero-Gil:2016qru,Bastero-Gil:2018yen,Kitazawa:2022gzk}, $\Gamma\propto T^3$ (with $\partial\Gamma/\partial \phi =0$)~\cite{Berera:2008ar,Berghaus:2019whh,Berghaus:2020ekh} and, finally, $\Gamma\propto T^3/\phi^2$~\cite{Bastero-Gil:2012akf,Bastero-Gil:2011zxb}. In total, we have $3\times 3$ combinations of choices of $V(\phi)$ and $\Gamma(\phi,T)$, as shown in Tab.\,\ref{tab:my_table}. In this table, the symbols \faCheck \, and \faTimes \, indicate, respectively, the consistency and inconsistency of each choice with current cosmological data. The most notable information from this table is that the case $n=2$ is incompatible with the data for all three choices of $\Gamma$. This occurs not because of the tensor-to-scalar ratio, $r$, but due to the scalar spectral index, $n_s$, which is too large. This result can be understood analytically, which we detail in~Sec.\,\ref{sec:analyticalestimates}. The cases $n=4,6$ are compatible with the data for specific choices of $\Gamma$. Again, the most relevant cosmological parameter turns out to be $n_s$, whereas $r$ can be smaller than current bounds (and smaller than the precision of next-generation CMB experiments).

\begin{table}[h!] \vspace{0.3cm}
    \centering
    \begin{tabular}{c||c|c|c}
   &  $\phi^6$  &  $\phi^4$  &    $\phi^2$  
\\
        \hline
        \hline
   $T$    & \faCheck  & \faCheck & \faTimes  \\ \hline
      $T^3$  & \faTimes & \faCheck & \faTimes  \\  \hline
       $T^3/\phi^2$ & \faTimes & \faTimes & \faTimes  \\ 
    \end{tabular}
    \caption{\small \it Summary of the compatibility with CMB data of the dissipation rates (rows) and inflaton potentials (columns) considered in this work. \faCheck$\,$ means compatible. \faTimes$\,$   means incompatible. {These results assume no thermal correction due to the occupation number of inflaton fluctuations (see point 3 below). }}
    \label{tab:my_table}
\end{table}

The models we study have also been considered in earlier papers, see \cite{Bartrum:2013fia, Bastero-Gil:2016qru} and
\cite{Benetti:2016jhf,Bastero-Gil:2017wwl,Arya:2017zlb,Arya:2018sgw,Bastero-Gil:2018uep}.  Our work goes beyond them in three key aspects:

\begin{enumerate}

\item {\bf Computation of the primordial power spectrum.} 

 In order to obtain the spectrum of primordial scalar perturbations in warm inflation it is necessary to solve a system of stochastic differential equations. The stochasticity of the system arises from a statistical treatment of the many degrees of freedom contained in the thermal bath and becomes manifest as a source of random noise. In principle, the system of equations needs to be solved over many realizations of the noise and then an average should be taken from the ensemble of solutions. This is numerically costly. For this reason, previous works have used a semi-analytic fitting formula for the averaged power spectrum~\cite{Ramos:2013nsa,Bastero-Gil:2014jsa, Bastero-Gil:2016qru}. Unfortunately, this fitting formula can be inaccurate and is model-dependent. Instead, we apply a Fokker-Planck method which is highly accurate and valid for any model of warm inflation. We refer to this method as the ``matrix formalism''. This method was introduced in~\cite{Ballesteros:2022hjk} and we advocate its use in future studies of warm inflation.

\item {\bf Analytic exploration of the models.}

We provide a careful analytic study of the primordial spectrum which allows to understand its properties for any combination of $V(\phi)\propto \phi^n$ and $\Gamma(\phi,T) \propto \phi^\alpha\, T^\beta$. Considering specifically the cases listed in Table 3 and focusing in particular in the quartic potential with dissipation linear in $T$, we find that previous purely analytical estimates fail to give a sufficiently good approximation to the properties of the power spectrum in the region of parameter space that is relevant for CMB data, which requires a precision $ \lesssim 1\%$.\footnote{This led to the introduction of the aforementioned numerically-fitted semi-analytic formula, see the previous point.} Whereas our purely analytical approximations are still not as good as it would be desirable (although better than previous attempts) for the amplitude of the primordial spectrum, they reach the desired precision for the scalar spectral index for most of the relevant parameter space, see Fig.~\ref{fig:comparewithprevious}.

  \item {\bf Stochastic inflation formalism.}

Calculations of the primordial spectrum of scalar fluctuations in warm inflation have often been done including a term that aims to take into account the occupation number of thermalized inflaton fluctuations. The origin of this term was discussed using stochastic inflation in~\cite{Ramos:2013nsa}, where it was already mentioned that such a term is model-dependent. We reconsider this issue, emphasizing that such a term is not strictly required by stochastic inflation. Using standard perturbation theory, we also show how this term may appear.

We show that the application of the stochastic inflation formalism in the context of warm inflation leads to the same result for the primordial spectrum, at lowest order in fluctuations,  as the one obtained from standard perturbation theory. The correction is model-dependent and may become relevant only for Lagrangians which generate a significant occupation number of inflaton particles, which needs to be checked on a case by case basis. This point and the previous ones imply that our results differ from earlier ones for the same combinations of potential and dissipation rate.  We {discuss} the impact of such non-vanishing occupation number {on}the scalar power spectrum and compare our results with earlier works for a specific scenario.
\end{enumerate}

In the next section we review the basics of warm inflation at background level, paying particular attention to estimating the number of e-folds of inflation required to solve the horizon and flatness problems, depending on the equation of state of the universe after inflation ends. In Sec.\,\ref{sec:perturbations} we review the dynamics of metric, inflaton and radiation perturbations in warm inflation. We present the two methods (Langevin and Fokker-Planck) that we use to compute the primordial spectrum of curvature fluctuations. In Sec.\,\ref{ss:pheno} we discuss our methodology to compare our numerical predictions with current cosmological bounds on the spectrum of primordial fluctuations and we proceed to study the nine cases listed in Tab.\,\ref{tab:my_table}. We perform an analysis to validate the application of the Fokker-Planck method by comparing it to the more standard Langevin one. And we compare our results to previous works, explaining in further detail the main differences. In Sec.\,\ref{sec:analyticalestimates} we present our analytical estimates for the amplitude and the spectral index of the spectrum of curvature fluctuations. We also discuss the quantization of the fluctuations of the inflaton in presence of a classical noise {source,} { considering {as well a} potentially significant occupation number for the inflaton fluctuations}. In Sec.\,\ref{sec:stoch}, we apply the formalism of stochastic inflation, showing that the result from standard perturbation theory is recovered at lowest order in fluctuations. We present our conclusions in Sec.\,\ref{summconc}.

This work also contains three appendices. In App.\,\ref{matrices} we present a more efficient implementation of the Fokker-Planck formalism. It gives the same results as the one discussed in Sec.\ \ref{sec:perturbations}, but saves some computation time. In App.\,\ref{app:statistics} we discuss the statistical limitations of the Langevin approach. This illustrates the need of solving the system of stochastic differential equations for the fluctuations a large number of times in this approach to obtain a sufficient precision. In App.\,\ref{app:reheating} we perform an analytical description of the dynamics of the Universe after warm inflation. This discussion informs our method to relate the fiducial scale of the CMB to the number of e-folds of inflation. Finally, in App.\,\ref{app:stochasticv2}, we follow the stochastic inflation approach introduced in Ref.\,\cite{Ramos:2013nsa} and show that a careful analysis yields results consistent with linear perturbation theory, and also with our stochastic inflation approach of Sec.\,\ref{sec:stoch} as well as our analytical approach of Sec.\,\ref{sec:analyticalestimates}. We identify a source of discrepancy between our results and those presented initially in Ref.\,\cite{Ramos:2013nsa}.

Throughout the paper we set $c=\hbar=k_{\rm B}=1$.

\section{Dissipation during inflation} \label{dissipsec}

We assume that the dynamics of inflation is such that the Universe can be described as a homogeneous expanding background in which small fluctuations live. In this paper we will not be concerned about the initial conditions (prior to inflation) leading to this state in the context of warm inflation. See however Sec.\,\ref{sec:computationspectrum} for a brief discussion about the consistency of the initial conditions for the fluctuations during warm inflation. 

\subsection{Background equations}\label{s:bck}

The background dynamics of the inflaton and the thermal bath obeys a system of ordinary differential equations which determine the expansion rate of the Universe, $H$, via 
\begin{align} \label{fried}
3\mn^2H^2 = \rho_r+V(\phi)+\frac{1}{2}\dot{\phi}^2\,.
\end{align}
In this equation, $\mn = (8\pi G)^{-1/2} \simeq 2.45 \times 10^{18}~\text{GeV}$ is the reduced Planck mass (and $G$ is Newton's gravitational constant). The function $\dot\phi$ is the (cosmic) time derivative of the background inflaton and $\rho_r$ is the energy density of the radiation bath, which is also a function of time. The time evolution of the latter is given by
\begin{align}
\dot{\rho}_r+4H\rho_r&=\Gamma\dot{\phi}^2\,,
\end{align}
where $\Gamma$ is a function of $\phi$ and the temperature of the thermal bath $T$, related to $\rho_r$ by
\begin{align}
\rho_r=\frac{\pi^2}{30}g_\star T^4\,, 
\label{eq:rhor}
\end{align}
where $g_\star$ is the effective number of relativistic degrees of freedom in the bath. The inflaton dynamics follows
\begin{align}
\ddot{\phi}+(3H+\Gamma)\dot{\phi}+V_\phi&=0\,,
\label{eq:KGequation}
\end{align}
where $V_\phi$ denotes the derivative of the potential with respect to the inflaton field and we define
\begin{align}
Q = \frac{\Gamma}{3H}\,.
\end{align}
At the level of the background, the dissipation rate, $\Gamma$, acts as an extra source of friction for the inflaton, which transfers energy into the radiation bath.

\subsection{The duration of inflation}  
\label{sec:durationinflation}

The amount of expansion that the Universe undergoes is quantified by the number of e-folds
\begin{align}
N = \int H\, \diff t\,,
\end{align} 
with $N$ growing as time progresses, by convention. We define the slow-roll parameter: 
\begin{align}
\epsilon = - \frac{\diff \log H}{\diff N} = -\frac{\dot H}{H^2}\,.
\end{align}
Inflation happens as long as $\epsilon  < 1$. The number of e-folds between the time a scale $k$ becomes superhorizon during inflation and the end of the latter can be estimated as 
\begin{align} \label{implicitNk}
N_k  = -\log\frac{k}{H_k} -\frac{1}{3(1+w)}\log\frac{\rho_\en}{\rho_{\bbn}} - \frac{1}{4}\log\frac{\rho_\bbn}{\rho_{\rm eq}}-\log(1+z_{\rm eq})\,, 
\end{align} 
where $H_k$ is the Hubble function at the crossing time and $w$ is the average equation of state of the Universe from the end of inflation to the time of the big bang nucleosynthesis (BBN). The quantities $\rho_\bbn$, $\rho_\en$ and $\rho_{\rm eq}$ represent the energy density of the Universe at BBN, the end of inflation and the time of matter-radiation equality (corresponding to redshift $z_{\rm eq}$), respectively. This approximation assumes sudden transitions between each of these phases. Notice that \eq{implicitNk} is actually an equation for $N_k$ since $H_k$ is nothing but $H(N_k)$.   

\begin{figure}[t!]
    \centering
  \includegraphics[width=0.52\textwidth]{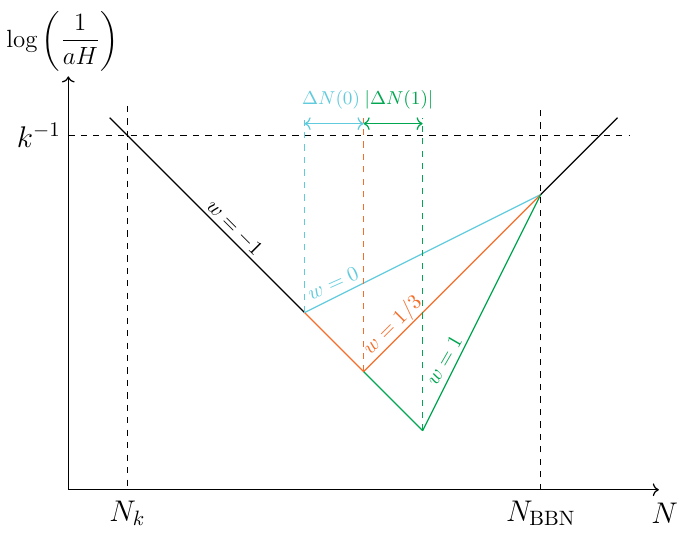}\hspace{0.3cm}
  \includegraphics[width=0.45\textwidth]{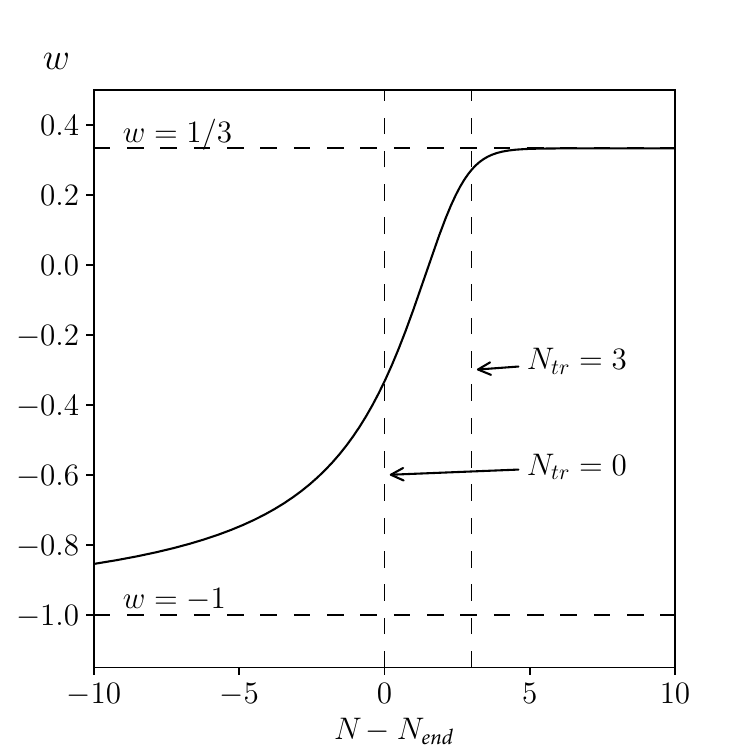}
\caption{\em \label{fig:efolds} {{\bf Left:} Schematic evolution of the comoving Hubble radius with the number of e-folds between crossing and re-entry for a given scale $k$. This illustrates the effect of a matter domination ($w=0$, in blue), radiation domination ($w=1/3$, in orange) or kination ($w=1$, in green) phases on the duration between Hubble radius crossing $N_k$ and the end of inflation (corresponding to the minimum Hubble radius in each case)
 {\bf Right:} Equation-of-state evolution with the number of e-folds before and after the end of inflation $N_\text{end}$.}}
\end{figure}

A particular case of interest consists in taking $w=1/3$. This corresponds to radiation domination from the end of inflation to BBN. The inflationary e-fold difference between this case and a Universe expanding with equation of state $w$ right after inflation (with all the other parameters appearing in \eq{implicitNk} being equal) is\footnote{Here we have assumed that $H_k$ is the same in both universes, which is a reasonable approximation as long as $H$ evolves slowly during inflation.}
\begin{align} 
\Delta N(w) =\frac{1-3w}{12(1+w)}\log\frac{\rho_\en}{\rho_\bbn}\,.
\end{align}
This is illustrated in Fig.\,\ref{fig:efolds} (left). For $\rho_\bbn^{1/4}\sim 10^{-3}$~GeV and  values of $\rho_\en^{1/4}$ like the ones we will encounter in the models we consider in this work (e.g. $10^{13}$~GeV -- $10^{14}$~GeV), we have that $\Delta N(0)\sim 10$ (matter domination after inflation) and $\Delta N(1)+\Delta N(0)\sim 0$. Therefore, a universe that enters into kination domination right after inflation and maintains this equation of state until BBN, expands $\sim  10$ e-folds less than one that covers that period expanding as radiation. We will use this estimates later to obtain conservative limits on the validity of specific combinations of $V(\phi)$ and $\Gamma(\phi, T)$. 

Let us now return to the case in which $w=1/3$ after inflation. In this case, it is possible to describe the transition towards the end of inflation from $w=-1$ to $w=1/3$ more accurately than assuming a sudden transition at the end of inflation, while at the same time still having an expression for the number of e-folds that is useful for analytical estimates. This can be done by simply assuming that the sudden transition happens at a later time, when the actual equation of state is closer to that of radiation than to $-1$.	Then, the equation for $N_k$ (as defined above) is now
\begin{align}
N_k + N_{\text{tr}} + \log\frac{k}{H(N_k)} = -\frac{1}{4}\log\frac{\rho(N_\en+N_{\text{tr}})}{\rho_{\rm eq}} -\log(1+z_{\rm eq})\,,
\label{eq:forNk}
\end{align}
where $N_{\rm tr}$ accounts for that extra time needed to describe the transition more accurately. In practice, we will solve numerically the equations for the background and perturbations until a $N_{\rm tr}$ that is chosen ad-hoc as a proxy for the intermediate phase between $w=-1$ and $w=1/3$. This is useful in models of warm inflation in which the Universe is automatically reheated when the radiation bath overcomes the energy density of the inflaton.  Concretely, we will be working later with $N_{\rm tr}=5$ in models in which there is a gradual transition from inflation to radiation. Fig.\,\ref{fig:efolds} (right) illustrates that integrating the background equations numerically until $N_{\rm tr}=3$ or further provides a better characterization of the system than stopping the integration at the end of inflation (which needs to be done in the absence of a model for reheating). 

Let us now recall the amount of inflation that is needed to solve the horizon and flatness problems. In order to solve the horizon problem, inflation needs to last at least long enough for scales reentering the horizon today to be in causal contact at the beginning of it. This is equivalent to saying that the lapse of conformal time from the moment such scale exited the horizon during inflation must be, at least, the same as the interval of conformal time from the end of inflation until today. Formally, this condition reads
\begin{equation}\label{ec:0}
\int_\infla \diff \eta= \int_\rh \diff \eta + \int_\rad \diff \eta + \int_\mat \diff \eta + \int_{\Lambda} \diff \eta\,,
\end{equation}
where $ \diff\eta = (aH)^{-1}\diff N$ is  conformal time. The integration domains refer to inflation, reheating (with an equation of state $w$), radiation domination, matter domination and a latter epoch dominated by a cosmological constant $\Lambda$. Expressing the duration of these phases in e-folds we have: 
\begin{align}
\delta N_\rh &= \frac{1}{3(1+w)}\log\frac{\rho_\en}{\rho_\bbn}\simeq \frac{48}{(1+w)}\\
\delta N_\rad &= \log\frac{a_{\rm eq}}{a_\bbn} = \log\frac{T_\bbn}{T_{\rm eq}}\simeq 11.5\\
\delta N_\mat &= \log\frac{a_{\Lambda}}{a_{\rm eq}} = 2\log\frac{T_{\rm eq}}{T_{\Lambda}}\approx 16\\
\delta N_\Lambda &= \log\frac{1}{a_\Lambda} = \log(1+z_\Lambda)\simeq 0.3\,.
\end{align}
Since $\delta N_{\Lambda}$ is much smaller than the other three, Eq.\,\eq{ec:0} becomes
\begin{align}
\delta N_\infla \simeq \frac{1+3w}{2}\delta N_\rh + \delta N_\rad + \frac{1}{2}\delta N_\mat + \log\left(3 - e^{-\delta N_\Lambda}\right)\simeq \frac{24(1+3w)}{1+w} + 20\,.
\end{align}
We conclude that the minimum number of e-folds of inflation necessary to solve the horizon problem is approximately 44 assuming that reheating lasts until BBN with $w=0$. Similarly, 56 e-folds are sufficient for $w=1/3$ and 68 for $w=1$. 

In order to solve the flatness problem, inflation has to last long enough to dilute $\Omega_k = |K| (aH)^{-2} $ (where $K$ is the spatial curvature parameter of the FLRW metric) to values compatible with constraints from CMB and other data. Today, $|\Omega_k^{(0)}|\lesssim 10^{-3}$. This quantity is given by 
\begin{equation}
|\Omega_k^{(0)}| = |\Omega_k^{(i)}|\exp\left[{-2\delta N_\infla + (1+3w)\delta N_\rh + 2\delta N_\rad + \delta N_\mat - 2\delta N_\Lambda}\right]\,,
\end{equation}
where we take $|\Omega_k^{(i)}|$ to be the value of $|\Omega_k|$ at the time at which scales that are currently reentering the horizon exited it during inflation. Assuming $|\Omega_k^{(i)}|\sim \mathcal{O}(1)$, the minimum number of e-folds of inflation needed to solve the flatness problem is 
\begin{equation}
\delta N_\infla \simeq \frac{24(1+3w)}{1+w} + 23\,.
\end{equation}
This amounts to 47, 59 and 71 e-folds, depending on whether the equation of state $w$  between the end of inflation and BBN is 0 (matter), 1/3 (radiation) or 1 (kination). Therefore, having enough inflation to solve the flatness problem guarantees solving the horizon problem as well. In Sec.\,\ref{ss:pheno} we will compare these numbers with the amount of inflation happening in models of warm inflation that are compatible with the measurements of the primordial scalar and tensor spectra.

\section{Perturbations}\label{s:pert}
 
\label{sec:perturbations}

\subsection{System of equations}

We work in the Newtonian gauge: $\d s^2=-(1+2\psi)\d t^2+a^2(1-2\psi)\delta_{ij}\d x^i\d x^j $, where we have a single metric perturbation, $\psi$, due to the absence of anisotropic stress. The conservation of the energy-momentum tensor of the system composed by the inflaton and the thermal bath, $T^{\mu\nu}_{(\phi)}+T^{\mu\nu}_{(r)}$, is guaranteed by the pair of equations~\cite{Gleiser:1993ea,Bastero-Gil:2014jsa}:
\beq
\nabla_{\mu}T^{\mu\nu}_{(\phi)} = -\Gamma\, u^{\mu}\nabla_{\mu}\phi\,\nabla^{\nu}\phi + \sqrt{\frac{2\Gamma T}{a^3}}\xi_t \nabla^{\nu}\phi= - \nabla_{\mu}T^{\mu\nu}_{(r)}\,.
 \eeq 
The 4-velocity of the radiation component is $u^{\nu}$ and $\xi_t\equiv \d W_t/\d t$, where $\d W_t$ is a Wiener increment which satisfies $\langle \xi_t({\bm x})\xi_{t'}({\bm x}')\rangle=\delta^{(3)}({\bm x}-{\bm x}')\delta(t-t')$, being $\langle\cdots\rangle$ a stochastic average over different realizations. For a brief practical summary of the definition of a Wiener process in stochastic differential equations see App.\,C of Ref.\,\cite{Ballesteros:2022hjk}. In Fourier space, and denoting with primes the derivatives with respect to $N$, and partial derivatives with subscripts (e.g. $\Gamma_T=\partial\Gamma/\partial T$), the full system of equations for linear perturbations reads\footnote{It has been argued in \cite{Bastero-Gil:2014jsa} that there is an ambiguity in the definition of the stochastic source in Eq.\ \eq{ec:b02}. It stems from the identification of the source of stochasticity in the interaction between the inflaton field and the radiation bath: it can be associated to the energy flux between the field and the bath, the momentum flux or some intermediate possibility. The differences are expected to be relevant for weak dissipation and become negligible in the strong dissipative regime. }
\begin{align} \nonumber
\delta\phi'' & +\left(3+\frac{\Gamma}{H}+\frac{H'}{H}\right)\delta\phi'+\left(\frac{k^2}{a^2H^2}+\frac{V_{\phi\phi}}{H^2}+\frac{\phi'\Gamma_{\phi}}{H}\right)\delta\phi+\Gamma_{T}\frac{\phi'T}{4H\rho_r}\delta\rho_r  -4\psi'\phi'
+\left(\frac{2V_{\phi}}{H^2} +  \frac{\Gamma\phi'}{H}\right)\psi \\  & = \sqrt{\frac{2\Gamma T}{a^3H^3}}\,\xi, \label{ec:b01}
\end{align}
\begin{align} \nonumber
\delta\rho_r' & +\left(4-\Gamma_T\frac{H\phi'^2T}{4\rho_r}\right)\delta\rho_r-H \frac{k^2}{a^2\,H^2}\delta q_r+\Gamma H\phi'^2\psi-4\rho_r\psi'-\Gamma_{\phi}H\phi'^2\delta\phi-2\Gamma H\phi'\delta\phi' \\ & = -\sqrt{\frac{2\Gamma TH}{a^3}}	\phi'\xi,\label{ec:b02}
\end{align}
\begin{align}
\psi'+\psi+\frac{1}{2\mn^2}\left(\frac{\delta q_r}{H}-\phi'\delta\phi\right)&=0,\label{ec:b03}
\end{align}
where $\delta q_r=\frac{4}{3}\rho_r\delta v_r,$ and $\delta v_r$ is the velocity perturbation of the radiation. $\delta q_r$ obeys the following differential equation
\begin{equation}\label{ec:b05}
\delta q_r^\prime  + \frac{4}{3H}\rr\psi + 3 \delta q_r + \frac{1}{3H} \delta \rho_r + \Gamma\phi'\delta\phi = 0\,.
\end{equation}
However, combining Eqs.\,(\ref{ec:b01}), (\ref{ec:b02}) and (\ref{ec:b03}), it
can be written as follows
\begin{equation}
\delta q_r=\frac{1}{3H}\left(2\mn^2\frac{k^2}{a^2}-H^2\phi'^2\right)\psi+\frac{\delta\rho_r}{3H}+\frac{1}{3}H\phi'\delta\phi'+\frac{1}{3H}(V_\phi+3H^2\phi')\delta\phi\,.\label{ec:b04}
\end{equation}
In total, there are four independent differential equations that need to be solved for the scalar perturbations. The initial conditions are 
\begin{equation}
\label{eq:inicondsperts}
\delta\rho_r=0,\qquad\psi=0,\qquad\delta\phi=-\frac{\phi'}{2\mn\,a\sqrt{k\epsilon}}\exp\bigg(-ik\int\frac{\d N}{aH}\bigg),
\end{equation}
 where we are assuming that the inflaton field fluctuations are in the Bunch-Davies vacuum. In Sec.\,\ref{sec:analyticalestimates} we justify this choice. 

\subsection{Computation of the primordial spectra}
\label{sec:computationspectrum}

In this section we present the computational methods we use for the primordial scalar and tensor spectra. 

\subsubsection{Scalar modes}
The primordial scalar spectrum can be computed either a) solving multiple times the system of stochastic differential equations discussed above and averaging over realizations or b) by means of a Fokker-Planck approach in which a system of (non-stochastic) differential equations is solved once. The second method is faster\footnote{In our implementation of the Fokker-Planck approach, computing the power spectrum for one Fourier mode takes around one minute with a four-core laptop. Alternatively, we computed $\mathcal{O}(10^4)$ realizations of the Langevin equation required to achieve an expected precision at the percent level on the averaged power spectrum (details are provided in the following and in App.~\ref{app:limitationLangevin}). We used a Runge Kutta method with a fixed time-step of $10^{-5}$ $e$-folds implemented in the Wolfram Mathematica method “StochasticRungeKutta”. Computing $10^4$ stochastic realizations on a cluster with $45$ cores and 256 GB of RAM takes about $\sim 24$ hours. The matrix formalism is much faster than the Langevin approach.}. We explain both methods below. In Sec.\,\ref{ss:pheno} we compare the results of both procedures for a specific choice of $V$ and $\Gamma$. We use the first method to validate the use of the second one. We recall that the comoving curvature fluctuation in warm inflation is
\begin{equation}
\label{eq:curvatureperturbation}
\mathcal{R}=\frac{H}{\rho+p}\big(\delta q_r-\dot{\phi}\delta\phi\big)-\psi \,,
\end{equation}
and we define its power spectrum as
\begin{equation}
\mathcal{P}_\mathcal{R}=\frac{k^3}{2\pi^2}|\mathcal{R}|^2\bigg|_{k\ll aH}\,.
\label{eq:powerspectrum_curvatureperturbation}
\end{equation}
We stress that $p$ and $\rho$ in the definition for $\mathcal{R}$ are the total pressure and energy density of the inflaton {\it and} radiation.\vspace{0.1cm}

\subsubsection{The system of Langevin equations and the matrix formalism}\label{seclav}

\noindent The variables we need to solve in order to compute the spectrum of curvature fluctuations \eq{eq:curvatureperturbation} are $\delta\phi$, $\delta\rho_r$ and $\psi$. As discussed above, the variable $\delta q_r$ can be obtained knowing the previous three. This allows to recast the system of stochastic differential equations, using the number of e-folds as time variable, in the following form: 
\begin{align}
\d\bm \Phi	+\boldsymbol{A}\bm \Phi	 \d N&=\frac{1}{\sqrt{2}} \boldsymbol{B}\left(\d W_N^{r}+i \d W_N^{i}\right) = \boldsymbol{B}\, \xi_N\, dN\,,
\label{eq:Langevin}
\end{align}
where $\d W_N^{r}\equiv\sqrt{2}\,{\rm Re}(\xi_N)\d N$ and $\d W_N^{i}\equiv\sqrt{2}\,{\rm Im}(\xi_N)\d N$ denote real-valued and independent Wiener increments and
\begin{align}
\langle\xi_N({\bm k})\xi^*_{\tilde N}({\bm k}')\rangle = \delta(N-\tilde N)\delta^3({\bm k}-{\bm k}')\,. 
\end{align}
In Eq.\ \eq{eq:Langevin} we use the four-component ``vector"
\begin{equation}
\bm \Phi	 \equiv \bigg(\psi,\delta\rho_r,\frac{\d\delta\phi}{\d N},\delta\phi\bigg)^{\rm T}\,,
\end{equation}
where $^{T}$ indicates matrix transposition. With this notation, $\boldsymbol{A}$ is a $4\times 4$ matrix, $\boldsymbol{B}$ is a column vector (both given in App.\,\ref{matrices}) and \eq{eq:Langevin} is a system of Langevin equations. In order to solve it we use the initial conditions \eq{eq:inicondsperts} evaluated at some initial time $N_i$, which can be expressed as 
\begin{equation}
\bm \Phi	(N_i)=\Big( 0, 0 , \dfrac{i\sqrt{k}}{\sqrt{2}ak_i}+\dfrac{1}{a}\dfrac{1}{\sqrt{2k}} ,  -\dfrac{1}{\sqrt{2k}a} \Big)^{\rm T}\,,
\end{equation}
where $k_i$ is the scale that crosses the horizon at the time at which the initial conditions are set $k_i=H(N_i)a(N_i)$. For the results in this paper, we take $k/k_i \simeq 100$ unless stated otherwise (which is only the case in Sec.\,\ref{quantumn}). The system of Langevin equations can be solved using, e.g.\ a fixed time-step Runge Kutta method. Averaging over many solutions (for different realizations of the noise) we can obtain the (averaged) power spectrum of curvature fluctuations\,\cite{Ballesteros:2022hjk}. In App.\,\ref{app:statistics} we discuss the statistics of the primordial fluctuations obtained from the system of Langevin equations.

Defining the two-point statistical moments 
\begin{equation}
\label{moments}
{\bm Q}\equiv\langle \bm \Phi	\bm \Phi^\dagger\rangle(N)\equiv \int \prod_i \diff \bm \Phi_i\, \int  \prod_j	\diff \bm \Phi^\star_j  \; P(\bm \Phi,\bm \Phi^\star,N)\;\bm \Phi	\bm \Phi^\dagger\,,
\end{equation}
where $P(\bm \Phi	,\bm \Phi^\star,N)$ is the probability density  for the system to be in state $\left\{\bm \Phi	,\bm \Phi^\star\right\}$ at time $N$,
the system of Langevin equations yields a deterministic differential equation for $\bm{Q}$:
\begin{equation}
\label{matrix_eq}
\frac{\d{\bm Q}}{\d N}=-{\bm A}{\bm Q}-{\bm Q}{\bm A}^{\rm T}+{\bm B}{\bm B}^{\rm T}.
\end{equation}
This equation allows to circumvent the problem of solving the full system of stochastic differential equations for the perturbations multiple times, provided that we are interested solely in the stochastic average of the power spectrum. The latter is given by 
\begin{equation}
\label{eq:pspec_projection}
\langle \mathcal{P}_\mathcal{R} \rangle =\frac{k^3}{2\pi^2}{\bm C}^{\rm T}\langle \bm \Phi	\bm \Phi^\dagger\rangle{\bm C}\bigg|_{k\ll aH}\,,
\end{equation}
where the column matrix ${\bm C}$ is given in App.\,\ref{matrices}. This method was first described in Ref.\,\cite{Ballesteros:2022hjk}. The Eq.\,\eq{matrix_eq} comes from applying to \eq{moments} the Fokker-Planck equation that gives the time evolution of $P(\bm \Phi	,\bm \Phi^\star,N)$: 
\begin{equation} \label{FPe}
\frac{\partial P}{\partial N}=\sum_{k\ell}\bigg[{\bm A}_{k\ell}\frac{\partial}{\partial \bm \Phi	_k}(\bm \Phi_{\ell} P)+{\bm A}_{k\ell}\frac{\partial}{\partial \bm \Phi^\star_k}(\bm \Phi^\star_{\ell} P)+({\bm B}{\bm B}^{\rm T})_{k\ell}\frac{\partial^2P}{\partial \bm \Phi_k\partial \bm \Phi	^\star_\ell}\bigg].
\end{equation}

\subsubsection{Tensor modes}
The evolution equation for tensor modes in warm inflation is the same as in cold inflation. There is no stochastic source. In Fourier space:
\begin{equation}
v_k'' + \left( 1 - \epsilon \right)v_k' + \left( \dfrac{k^2}{a^2 H^2} - 2 + \epsilon \right)  v_k = 0\,,
    \label{eq:eqtensormodes}
\end{equation}
where $h_k = 2 v_k /a$ and $h_k$ is the mode function of any of the two possible independent polarizations. One can easily solve numerically Eq.\,(\ref{eq:eqtensormodes}) by assuming Bunch-Davies initial conditions and get the value of the tensor perturbation a few e-folds after horizon crossing. The dimensionless power spectrum for tensor modes (including the two polarizations) is then given by 
\begin{equation}
    \mathcal{P}_t \, = \, 2 \dfrac{k^3}{2 \pi^2} |h_k|^2 \, .
    \label{eq:Pt}
\end{equation}

\section{Phenomenological analysis and constraints}
\label{ss:pheno}

\subsection{General setup}

\textbf{Models considered.} We focus on the following inflaton potential and dissipation rate
\begin{align}
\Gamma(\phi, T) = C\,\mn\,\left(\frac{\phi}{\mn}\right)^\alpha \left(\frac{T}{\mn}\right)^\beta\,,  \,\,  \beta\geq 1 \quad {\rm and}\quad
V(\phi) = \frac{\lambda}{n} \mm^{4} \left(\frac{\phi}{\mn}\right)^n\,, \,\,  n> 1\,.
\label{eq:GammaandV}
\end{align} 
Each model is defined by a set of integer exponents $n, \alpha, \beta$. For each model the dynamics of inflation is uniquely determined by the dimensionless parameters $\lambda$ and $C$ as well as the effective number of relativistic thermalized species $g_\star$. We consider three cases for the dissipation coefficient, motivated by existing models that appear to be the most popular realizations of warm inflation. We list them below with some concrete examples from the literature:
\begin{itemize}
 \item[$\bullet$] $\Gamma \propto T$: The distributed mass model~\cite{Berera:1998px}, the warm little inflation scenario~\cite{Bastero-Gil:2016qru}, a supersymmetric version of the distributed mass model~\cite{Bastero-Gil:2018yen}, Nambu-Goldstone inflaton~\cite{Kitazawa:2022gzk}.
 \item[$\bullet$] $\Gamma \propto T^3$: The heavy mediator mass limit in the two-stage mechanism~\cite{Berera:2008ar},  via sphaleron processes in minimal warm inflation~\cite{Berghaus:2019whh,Berghaus:2020ekh}.
 \item[$\bullet$] $\Gamma \propto T^3/\phi^2$: Small mass limit in the two-stage mechanism~\cite{Bastero-Gil:2012akf}, in the context of extra-dimensions~\cite{Bastero-Gil:2011zxb}.
\end{itemize} 
The physics of these models is nicely summarized in a recent review \cite{Kamali:2023lzq}. We refer the reader to this reference or to the original works for details about the field theory implementation in each case. In what follows we discard any potential thermal corrections to the inflaton potential as well as thermal contributions to the occupation number of inflaton fluctuations. Such model-dependent effects have been argued to play a role (see e.g.\ \cite{Kamali:2023lzq} and references therein). We do not consider these effects in the main part of our work. However, we aim to clarify in Sec.\,\ref{sec:stoch} that a commonly applied correction to the primordial spectrum of curvature fluctuations –that may arise from high occupation numbers of inflaton fluctuations modes thermalized with the radiation plasma– is not required by the stochastic inflation formalism.

 \par \medskip

\noindent
\textbf{Computational procedure with the matrix formalism.} We scan the parameter space of each model using the matrix formalism presented in the previous section. That is, we solve Eq.\,(\ref{matrix_eq}) to compute the scalar primordial spectrum in a densely populated grid in parameter space. The tensor power spectrum is obtained for those same points in parameter space by simply solving the ordinary differential equation (\ref{eq:eqtensormodes}) for the tensor modes. For each point, the background evolution is obtained solving Equations \eq{fried}-\eq{eq:KGequation} at least until the end of inflation $N_{\rm end}$, and then fed into the computation of the fluctuations. 

We determine the time $N_{k_*}$ corresponding to Hubble-radius crossing for the CMB fiducial scale $k_*=0.05~\text{Mpc}^{-1}$ using Eq.\,(\ref{eq:forNk}), choosing adequate values of $N_\text{tr}\geq N_\text{end}$ and $\Delta N$, as defined in Sec.\,\ref{sec:durationinflation}. And we solve for both scalar and tensor modes until $\sim 10$ $e$-folds after horizon crossing for several scales around $k=k_*$. We determine the amplitude of the power spectrum $A_s$ for the scale $k_*$ and the scalar spectral index $n_s$ by fitting the resulting power spectrum with the usual relation
\begin{equation} \label{plaw}
\mathcal{P}_\mathcal{R}(k) \,= \, A_s \left( \dfrac{k}{k_*} \right)^{n_s-1} \, .
\end{equation}
The tensor-to-scalar ratio $r$ is derived computing the ratio of Eq.\,(\ref{eq:Pt}) and Eq.\,(\ref{eq:pspec_projection}). \par \medskip

\noindent
\textbf{Self-consistency check for the matrix formalism.} We have tested the precision of our results by developing two independent codes implementing the matrix formalism. For example, we have tested both codes with the model described by the potential $V(\phi)=(\lambda/4)\,\phi^4$ and $\Gamma =C\, {T}$ with $\lambda=1.74 \times 10^{-15}, C= 0.012$, which is compatible with BICEP/Keck-Planck constraints (see below). The relative difference between the outputs from two codes for the relevant CMB parameters (amplitude of the scalar power spectrum, scalar spectral index, tensor-to-scalar ratio and duration of inflation $N_{k_*}$) is found to be below $0.3\%$. 
\par \medskip

\noindent
\textbf{Effective number of relativistic species during inflation.} 
We often consider as a reference the value $g_\star=12$. Typically, several fields in addition to the inflaton are required to achieve the desired form of $\Gamma$ and $g_\star=12$ is a commonly used value in the literature. Notice that this value is smaller than the Standard Model (SM) expected value for temperatures above the electroweak scale, $g_\star(T>T_\text{EW})=106.75$ for $T_\text{EW}\sim 200~\text{GeV}$. We remain agnostic about the transition between the early thermal plasma and the SM plasma after inflation. The only effect of this transition on our results would be to affect the determination of $N_{k_*}$, which can be encapsulated in the variable $\Delta N$ defined in Section~\ref{sec:durationinflation}. We will discuss the effects of $g_\star$ and $\Delta N$ further on through our analysis.
\par \medskip

\noindent
\textbf{Reheating.} The transition to a radiation dominated universe after the end of inflation can sometimes be achieved automatically in warm inflation without introducing any further coupling between the inflaton and other species. App.\,\ref{app:reheating} explores this possibility for a subset of the inflaton potentials and dissipation rates parametrized in Eq.\,(\ref{eq:GammaandV}). Two distinct cases are discussed, corresponding to strong or weak dissipative regimes according to the value of the dissipation rate at the end of inflation, respectively $Q_\text{end}\gg1$ and $Q_\text{end}\ll1$. For each model, we use the possibility to achieve a smooth transition into radiation domination (or lack thereof) to motivate our choices for $N_\text{tr}$ and $\Delta N$, discussed in Sec.~\ref{sec:durationinflation}, for the determination of the CMB fiducial scale crossing time $N_{k_*}$. \par \medskip

\noindent
\textbf{Constraints.} We confront the parameter space of each model with CMB and other data. Specifically, we consider the bounds on the tensor-to-scalar ratio $r$ and scalar spectral index $n_s$ from an analysis of BAO~{\cite{eBOSS:2020yzd}, BICEP/Keck 2018 (BK18) and Planck (PR4) data~\cite{Tristram:2021tvh} (including CMB lensing data~\cite{Planck:2018lbu}). This work found $n_s=0.9668\pm 0.0037$ at 68\% confidence level (CL) $r<0.032$ at 95\% CL. The amplitude of the scalar power spectrum is measured to be $\log(10^{10}A_s)=3.044 \pm 0.014$ by Planck (TT,TE,EE+lowE+lensing)~\cite{Planck:2018vyg}. Given the fact that the measurements of $n_s$ and $A_s$ reach an $\mathcal{O}(1\%)$ precision, we must ensure our numerical predictions to be accurate to at least the same level of precision in order to derive meaningful results. 

We also compare our results to the planned sensitives of upcoming CMB-dedicated experiments and large scale surveys. The Simons Observatory should reach $r<6\times 10^{-3}$ ~\cite{SimonsObservatory:2018koc}, LiteBIRD should reach $r<2\times 10^{-3}$~\cite{LiteBIRD:2022cnt} and CMB Stage-4 (CMB-S4) $r<10^{-3}$ ~\cite{SO-CMBS4}. In addition, we consider a forecast for a CMB-Euclid joined analysis~\cite{Euclid:2021qvm} with an expected accuracy of $\sigma(n_s)\simeq 0.002$. We will use this prediction assuming as central value for $n_s$ the current one determined by Ref.\,\cite{Tristram:2021tvh} and will refer to it as (Euclid+CMB). We consider as well the possibility for the scalar index to vary with the scale $k$ by computing the running $\diff n_s/\diff \log k$ around the pivot scale $k_*$ in the models we study. We find values of the running that are generically much smaller than the central value allowed by Planck constraints $\diff n_s/\diff \log k = 0.0011 \pm 0.0099$ (TT,TE,EE+lowE+lensing+BAO)~\cite{Planck:2018vyg}. This justifies the parametrization \eq{plaw} and using only constraints on the scalar index assuming zero running, which are stronger.
\par \medskip

We start immediately below with a detailed analysis for $V=(\lambda/4)\,\phi^4$ and $\Gamma = C\,T$, studying the impact of each parameter on the predictions for the various CMB observables. We use this example to comment in Sec.\,\ref{secapp} on the accuracy of our results by comparing  the matrix formalism and the Langevin approach (see Section~\ref{sec:computationspectrum}). The subsequent parts of this section are dedicated to exploring additional models in a systematic way, covering all the possibilities listed in Table~\ref{tab:my_table}.

\subsection{A detailed case: linear dissipation rate ($\Gamma \propto T$)}
\label{sec:linear}

\subsubsection{Quartic potential}
\label{sec:linearquartic}

We perform a scan in the 2-dimensional parameter space $[C,\lambda]$ and select the parameters satisfying constraints on the amplitude of the scalar power spectrum from~\cite{Planck:2018vyg}. We choose the parameters $N_\text{tr}=5$ and $\Delta N=0$ to quantify the number of e-folds of inflation, as explained later on. We represent on the top panel of Fig.~\ref{fig:lineargammag} the parameter space, allowed at the 2$\sigma$ (or 3$\sigma$) level\footnote{Here and in the rest of this work, in order to obtain 2$\sigma$ intervals from CMB bounds at 68\% C.L. we simply assume Gaussianity for the posterior distribution functions whenever necessary. This approximation is sufficient for our purposes.}  with $g_\star$ chosen to be $4$ (left), $12$ (center) and $120$ (right). In addition, we represent constraints from the BICEP/Keck-Planck analysis of Ref.\,\cite{Tristram:2021tvh} at $2\sigma$ ($1\sigma$) as light (dark) blue regions. Forecasts for constraints from the Simons Observatory (SO)~\cite{SimonsObservatory:2018koc}, LiteBIRD~\cite{LiteBIRD:2022cnt} and CMB Stage-4 (CMB-S4)~\cite{SO-CMBS4} are represented as purple dashed lines in addition to predictions for a joined Euclid+CMB analysis~\cite{Tristram:2021tvh} in green (assuming the central value for $n_s$ to remain identical to the current one determined in Ref.\,\cite{Tristram:2021tvh}). The bottom panel of Fig.~\ref{fig:lineargammag} shows the parameter space corresponding to values of $A_s$ allowed at the 3$\sigma$ level in between two solid red lines.\footnote{We chose 3$\sigma$ (bottom panel), different from 2$\sigma$ (top pannel), for readibility as the two red lines are very close and would become almost undistinguishable if 2$\sigma$ were chosen.} The color of the allowed parameter space codes values of $\log_{10}r$. The blue region in the bottom panel of Fig.~\ref{fig:lineargammag} represents the allowed parameter space from the BICEP/Keck-Planck analysis of Ref.\,\cite{Tristram:2021tvh} at the $2\sigma$ level. The light-orange dashed lines represent isocountours of the duration $N_{k_*}$ between the time the scale $k_*$ crosses outside the horizon during inflation and the end of inflation.\par \medskip

\begin{figure}[t!]
    \centering
  \includegraphics[width=0.32\textwidth]{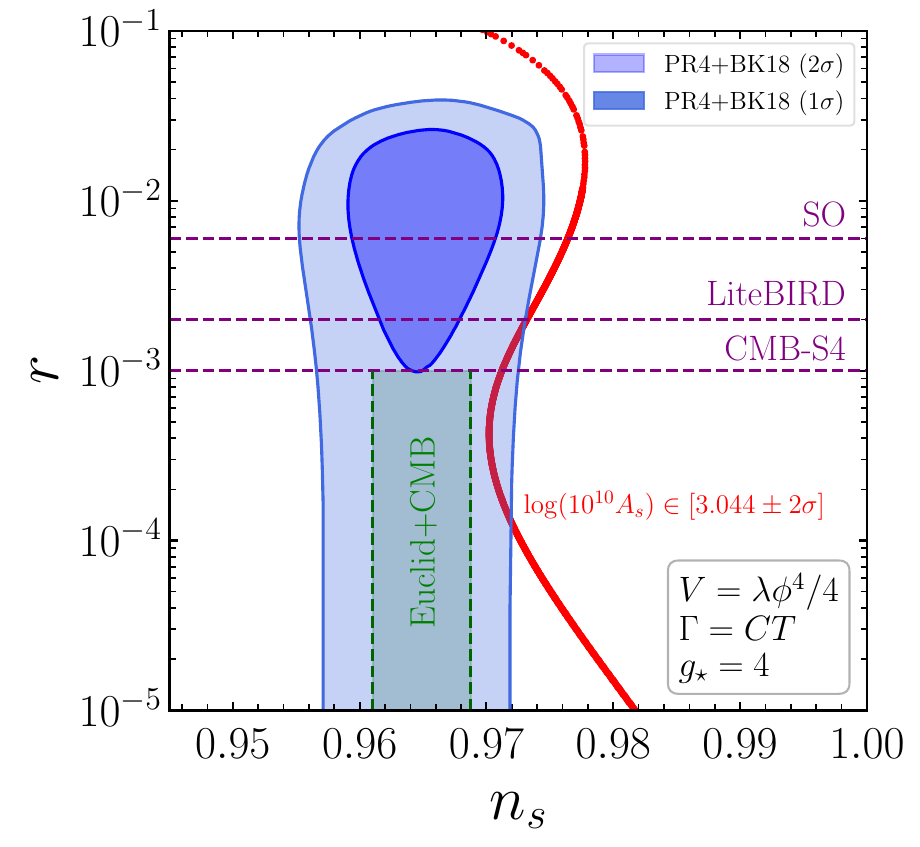}
  \includegraphics[width=0.32\textwidth]{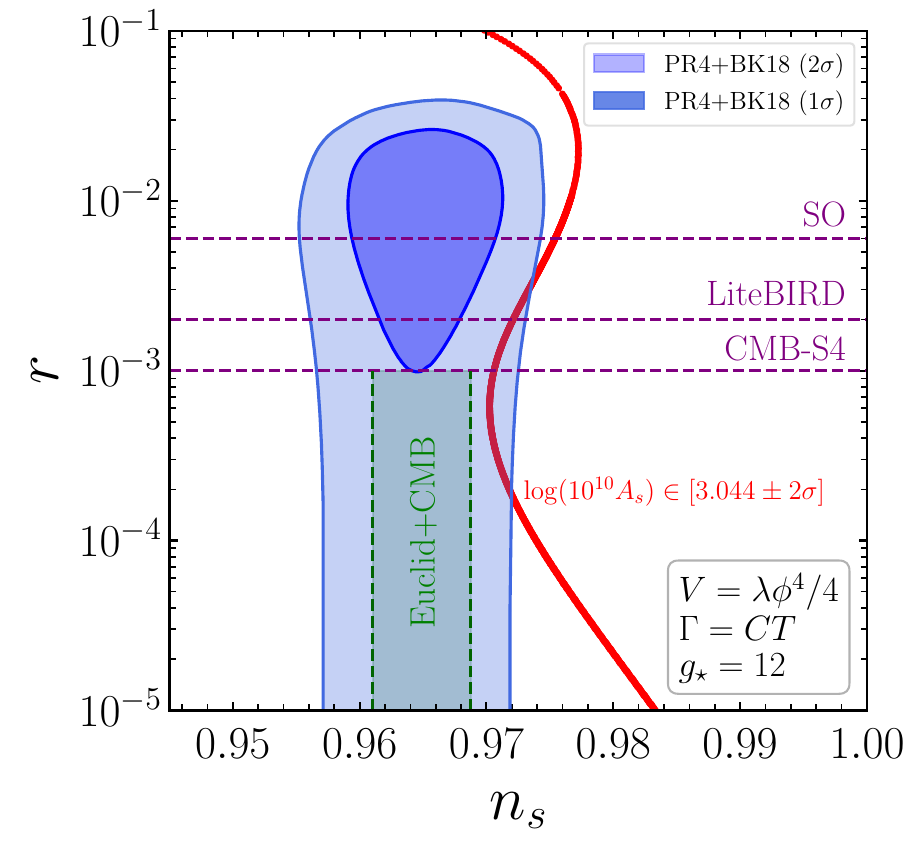}
  \includegraphics[width=0.32\textwidth]{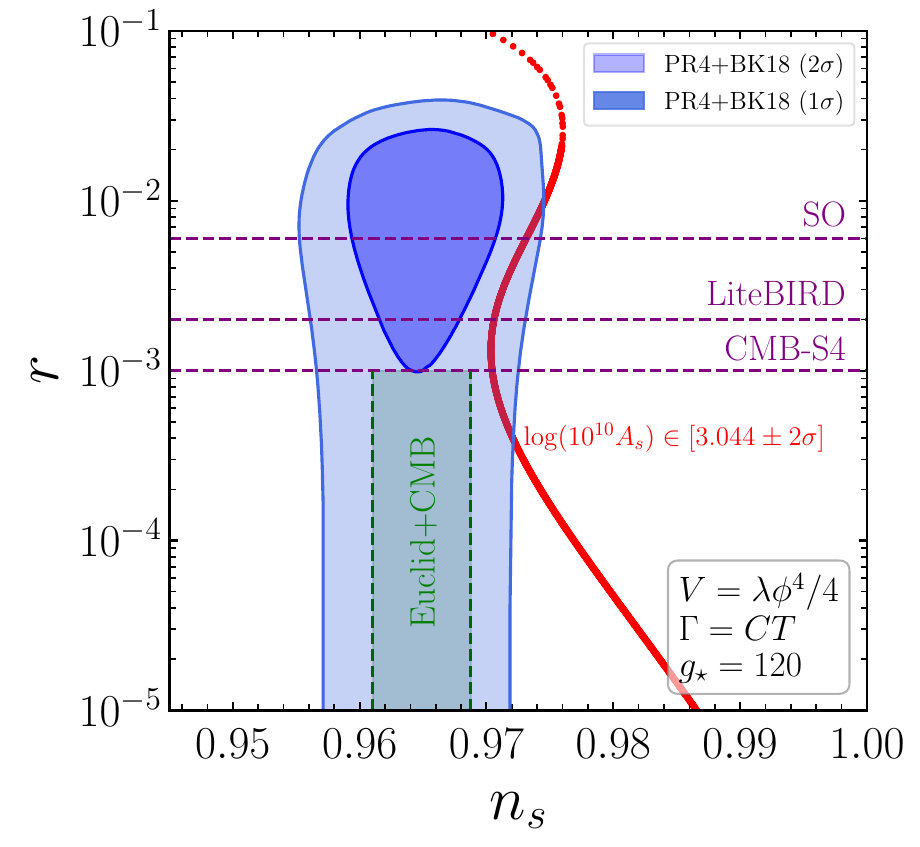}
  \includegraphics[width=0.32\textwidth]{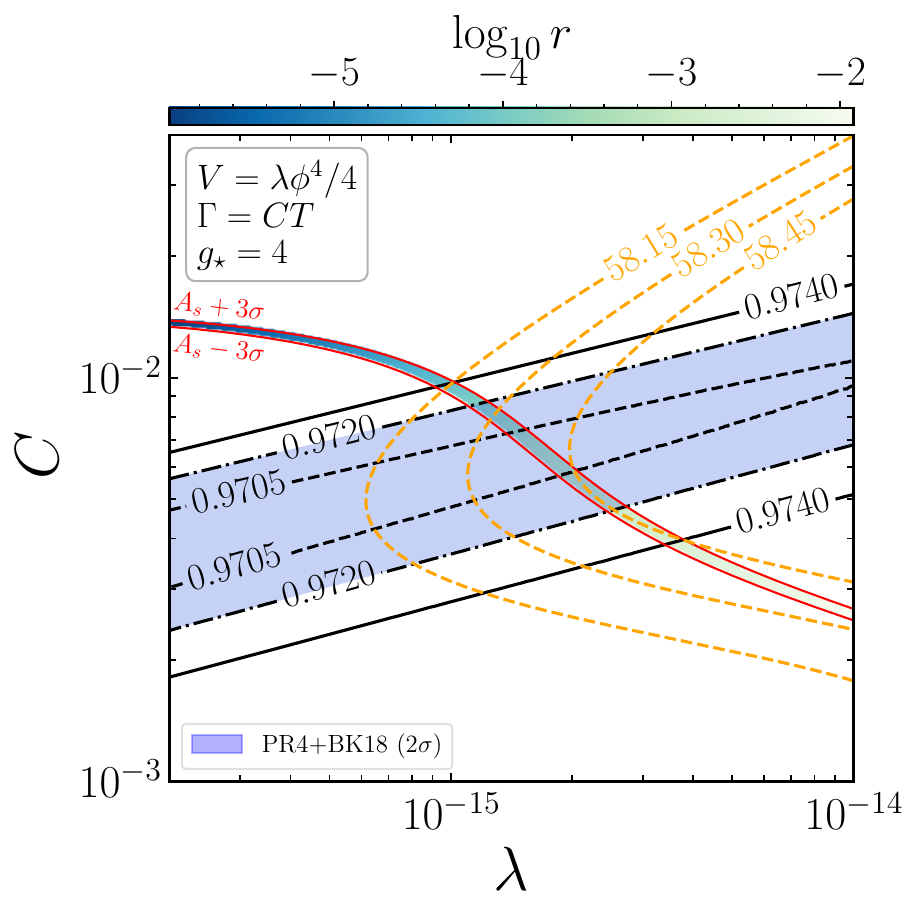}
  \includegraphics[width=0.32\textwidth]{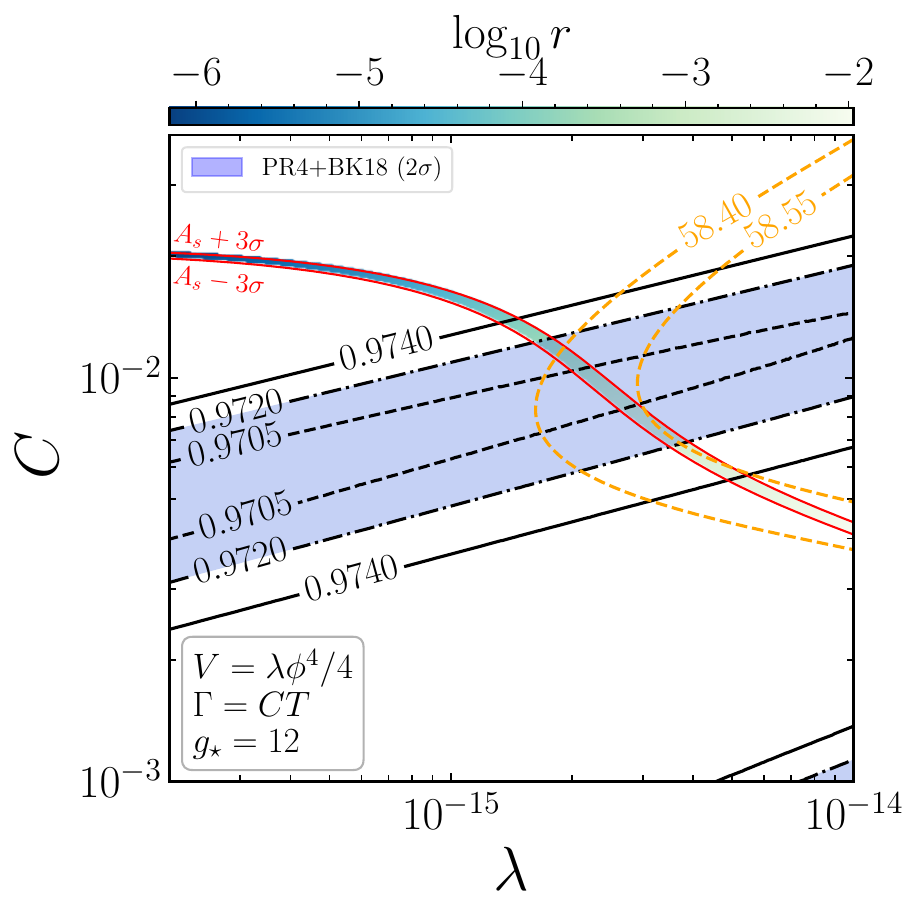}
  \includegraphics[width=0.32\textwidth]{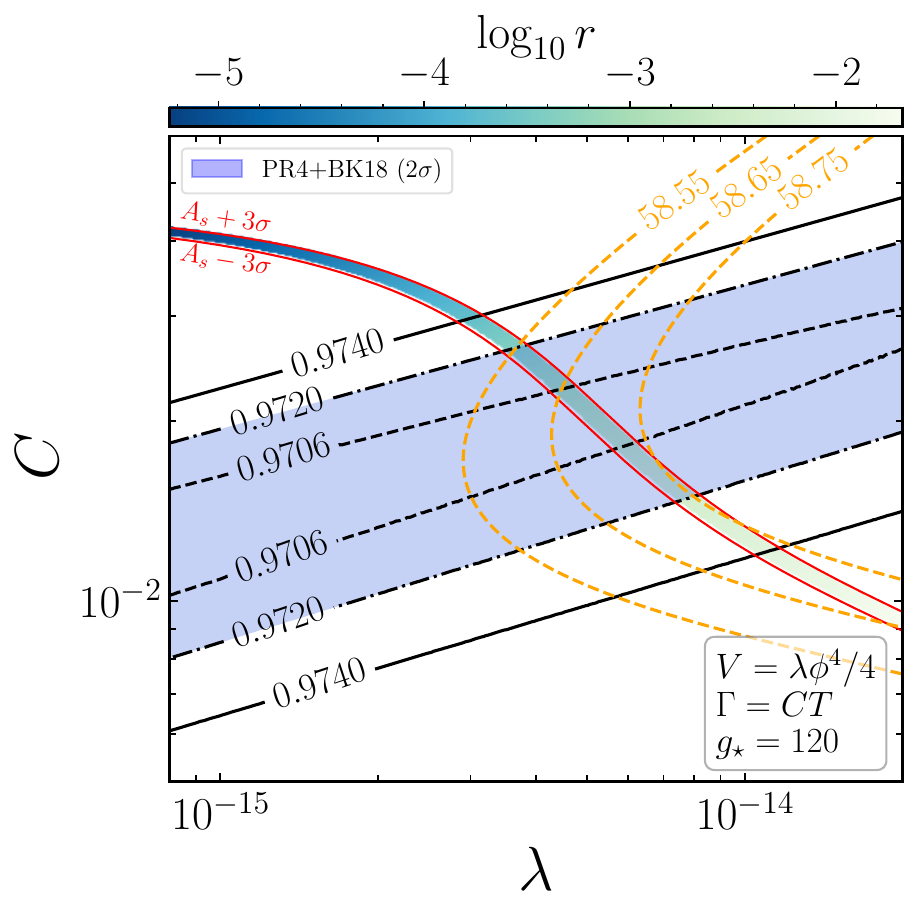}
\caption{\em \label{fig:lineargammag} {Parameter space for $V = (\lambda/4)\, \phi^4$, $\Gamma = C\,T$, described in Section~\ref{sec:linearquartic}, compatible with constraints on the amplitude of the scalar power spectrum $A_s$ from~\cite{Planck:2018vyg}. {\bf Top:} The parameter space, allowed at the 2$\sigma$ level, is represented in the plane $\{r,n_s\}$ in red. $g_\star$ is chosen to be $4$ (left), $12$ (center) and $120$ (right). Constraints from the BICEP/Keck-Planck analysis of Ref.\,\cite{Tristram:2021tvh} at $2\sigma$ ($1\sigma$) are represented as light (dark) blue regions. Forecasts for constraints from the Simons Observatory (SO)~\cite{SimonsObservatory:2018koc}, LiteBIRD~\cite{LiteBIRD:2022cnt} and CMB Stage-4 (CMB-S4)~\cite{SO-CMBS4} are represented as purple dashed lines. The sensitivity estimate for a joint Euclid+CMB analysis~\cite{Tristram:2021tvh} is represented in green, assuming the central value for $n_s$ to remain identical to the current one determined in \cite{Tristram:2021tvh}. {\bf Bottom:} The parameter space, allowed at 3$\sigma$, is represented in the plane $\{C,\lambda\}$ between the 2 solid curves in red. The colored region represented in each panel in between these two lines shows the value of $\log_{10}r$. The light-orange dashed lines represent isocontours of the duration $N_{k_*}$ between the time the CMB fiducial scale $k_*=0.05$~{\rm Mpc}$^{-1}$ crosses outside the horizon during inflation and the end of it. Dashed, dot-dashed and solid black lines represent isocontours of the scalar index $n_s$. The region in blue represents the allowed values of $n_s$ from Ref.\,\cite{Tristram:2021tvh} at $2\sigma$. Further details can be found in Sec.\,\ref{ss:pheno}.} We stress that the values of the spectral parameters $A_s, n_s$ and $r$ are always relative to the scale $k_*$, in this figure and through all this work.}
\end{figure}

\noindent
\textbf{Parameter space.} The allowed parameter space for $g_\star=12$ is projected in Fig.~\ref{fig:more_linear} (left panel) in a plane where the spectral index is represented on the y-axis and the dissipation coefficient $Q=\Gamma/(3H)$ evaluated at Hubble-radius crossing of the comoving scale $k_*$ on the x-axis. In this plane, going from large values of $Q_*$ to smaller values decreases the scalar index $n_s$ until reaching a local minimum $n_s\simeq 0.97$  for $Q_* \simeq 0.5$. For lower values of $Q_*$, $n_s$ increases to reach a local maximum $n_s\simeq 0.977$ for $Q_* \simeq 0.04$ before decreasing and reaching an asymptotic value $n_s \simeq 0.9485$  at small $Q_*$ in the limit where inflation is essentially cold. Such local $n_s$ maximum and minimum features appear in the plane $\{r,n_s\}$, in the top panels of Fig.~\ref{fig:lineargammag}, in the form of turning shapes for the allowed parameter space. Such a shape appears to be a rather universal feature in the various models we consider in this work. From Fig.~\ref{fig:lineargammag}, the parameter space for $g_\star=12$  compatible with constraints on $A_s$, $n_s$ and $r$ points towards a dimensionless coupling $C\simeq 7 \times 10^{-3}$ and $\lambda \simeq 2 \times 10^{-15}$. The allowed window for the duration of inflation between the crossing of $k_*$ and the end of inflation is also relatively narrow $N_{k_*}\in [58.40,58.55]$. \\

\begin{figure}[t!]
\begin{center}
$\includegraphics[width=.48\textwidth]{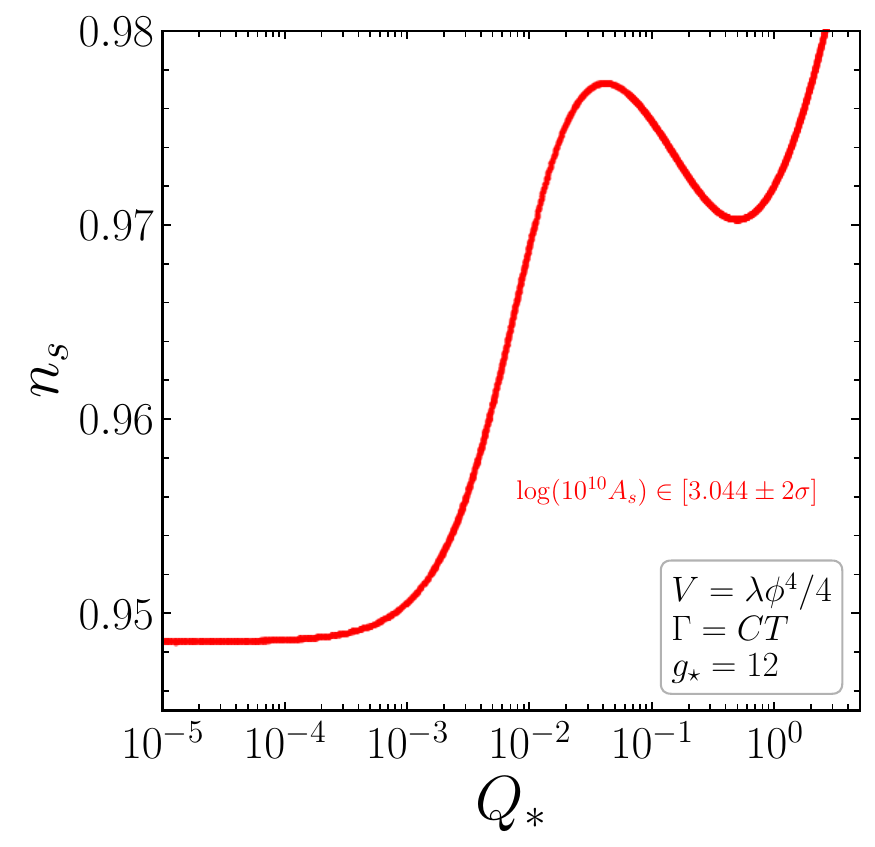}$
$\includegraphics[width=.49\textwidth]{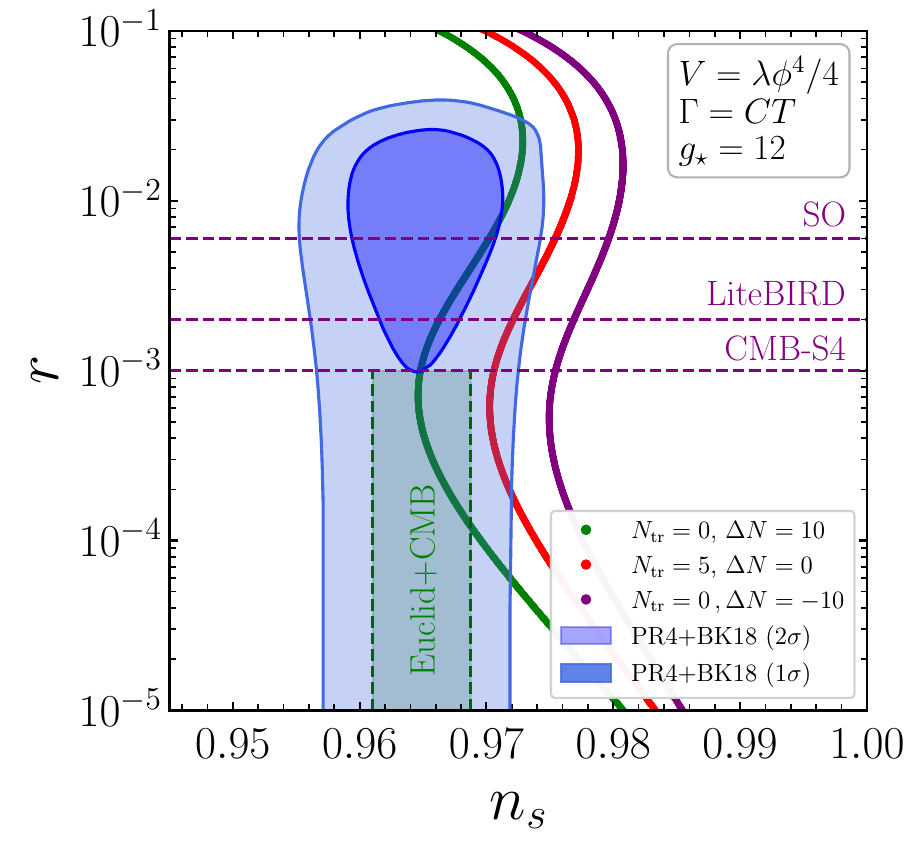}$
\caption{\em \label{fig:more_linear} {Parameter space for $V = (\lambda/4)\, \phi^4$, $\Gamma = C\,T$, described in Section~\ref{sec:linearquartic},  compatible with the constraints on the amplitude of the scalar power spectrum $A_s$ from~\cite{Planck:2018vyg}. {\bf Left:} The allowed parameter space projected in the plane $\{n_s,Q_*\}$ is represented in red assuming $N_\text{tr}=5$ and  $\Delta N=10$. See Sec.\,\ref{sec:durationinflation} for the meaning of these parameters and the discussion in Sec.\,\ref{sec:linear}. {\bf Right:} The allowed parameter space is projected in the plane $\{r,n_s\}$ as green, red and purple dots (from left to right) for different choices of the parameters $N_\text{tr}$ and  $\Delta N$, as indicated in the legend of the plot. Constraints from PR4+BK18 are represented in blue. Sensitivities on $r$ expected from SO, LiteBIRD and CMB-S4 are represented in dashed-purple lines. The green region represents forecast for a joint Euclid+CMB analysis. For more details, see the caption of Fig.~\ref{fig:lineargammag} top-panel and discussion in Sec.\,\ref{ss:pheno}.}}
\end{center}
\end{figure}

One can also observe in Fig.~\ref{fig:lineargammag} that varying the parameter $g_\star$ results in an approximate vertical shift of the allowed parameter space in the plane $\{r,n_s\}$. Therefore, the value of $g_\star$ does not affect the value of $n_s$ at the local minimum but increasing the value of $g_\star$ by a factor of $\mathcal{O}(10)$ tends to increase the predicted tensor-to-scalar ratio by a factor of $\mathcal{O}(1)$. As shown in Section~\ref{ss:cos}, the power spectrum for values of $0.2<Q_*<1$ (which are those relevant to fit the CMB data) scales as
\begin{equation}
\mathcal{P}_\mathcal{R} \, \sim \, \dfrac{C^{5/3} \lambda^{1/3}}{g_\star^{2/3}} \, .
\end{equation}
This expression agrees qualitatively with the results that can be observed comparing the bottom panels of Fig.~\ref{fig:lineargammag}: the allowed parameter space requires larger values of the dimensionless parameters $C$ and $\lambda$ as $g_\star$ increases, in order to maintain an amplitude of the spectrum at the same level. \par \medskip

\noindent
\textbf{Reheating.} In the strong dissipation regime, in which the value of $Q$ at the end of inflation is $Q_\text{end}\gg1$, as detailed in App.\,\ref{app:reheating} and summarized in Tab.~\ref{tab:1} and Tab.~\ref{tab:3}, radiation domination is achieved a few e-folds after the end of inflation. This is also illustrated in the right panel of Fig.~\ref{fig:efolds}. The choice $N_\text{tr}=5$ ensures that the transition from inflation to radiation domination is properly accounted for. 

An inflaton field oscillating about a quartic potential, in the weak dissipative regime ($Q_\text{end}\ll1$) redshifts as radiation (see Tab.~\ref{tab:1} and Tab.~\ref{tab:3} on the left panel). However, in the weak dissipative regime after inflation, $\Gamma$ increases (and so does $Q$) until reaching the strong dissipative regime where the ratio of the inflaton energy density over radiation decays exponentially fast and the transition to radiation is eventually achieved. The effect of choosing different values of the equation of state after inflation is illustrated in Fig.~\ref{fig:more_linear} on the right panel. We consider two extreme cases where the universe instantanesouly transitions after inflation to the longest possible phase of matter domination ($w=0$) or kination ($w=1$), which correspond to $N_\text{tr}=0$, $\Delta N=10$ ($w=0$) or $\Delta N=-10$ ($w=1$). The effect of these parameters on the determination of the CMB fiducial scale crossing is illustrated in Fig.~\ref{fig:efolds}. 
While a phase of matter domination tends to decrease the value of $n_s$ and ease the agreement with the central value of the BK-PR4 preferred region, a phase of kination would imply the opposite and is excluded by those results. 

An interesting aspect about this model is that there is  a lower bound on the tensor-to-scalar ratio ($r\gtrsim 7 \times 10^{-5}$) that is achievable in the region allowed by constraints on $n_s$. In addition, unless a phase of matter domination is achieved directly after inflation, the future Euclid+CMB might be able to rule out this scenario. Our results appear to be qualitatively consistent with Ref.\,\cite{Bastero-Gil:2016qru}, which considered a wide range of duration of inflation $N_{k_*}\simeq 50-60$. Ref.\,\cite{Bastero-Gil:2016qru} considered also the effect of a thermal distribution for the occupation number of inflaton fluctuations, which is a model dependent correction and, as we argue in Sec.\,\ref{sec:stoch}, is not required by the stochastic inflation formalism. Further on, we show the phenomenological impact of such term on the {allowed} parameter space for $\Gamma = C T$ and $V=(\lambda/4)\phi^4$ .

\subsubsection{A second approach: solving the Langevin equations.} \label{secapp}

\begin{figure}[t!]
\begin{center}
$\includegraphics[width=.43\textwidth]{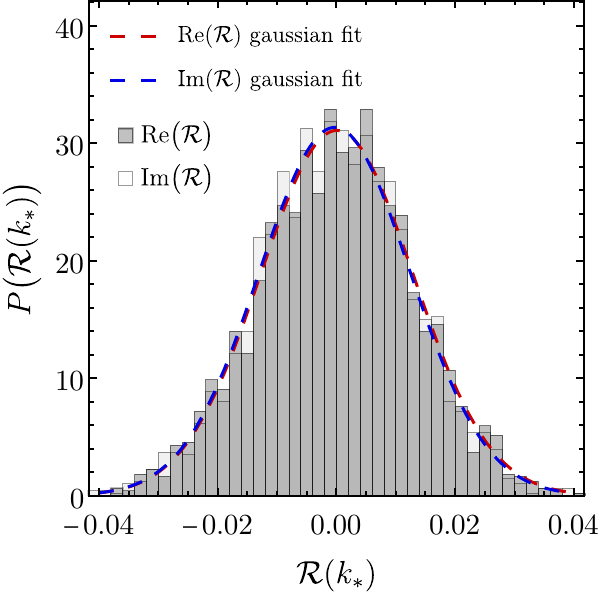}
\qquad \includegraphics[width=.43\textwidth]{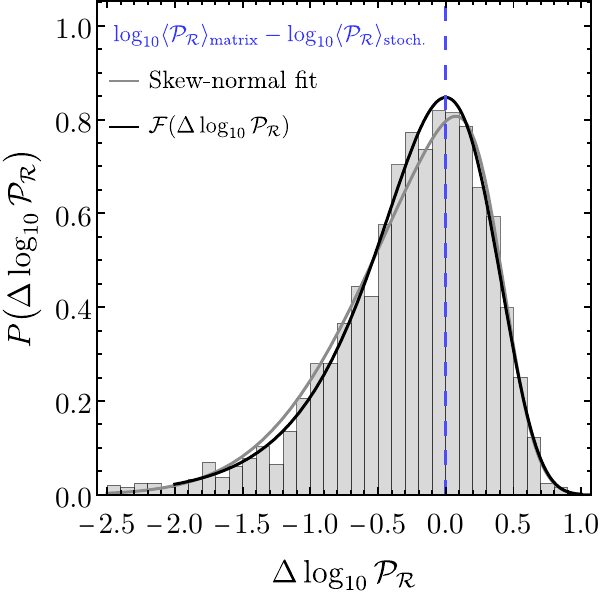}$
\caption{\em \label{fig:histogram1} Histograms and probabilty distribution functions for $2400$  solutions of the Langevin equations in the case $V=(\lambda/4)\, \phi^4$ and $\Gamma = C\, T$. {\bf Left:} Generated data for the real and imaginary parts of the curvature perturbation $\mathcal{R}$ for the CMB fiducial scale $k_*$ are shown respectively in dark-grey and light-grey. The red-dashed  and blue-dashed lines represent, respectively, the best (normalized) Gaussian fits of the real and imaginary parts. {\bf Right:} The same data is used to generate a (normalized) histogram in grey for the variable defined in Eq.\,(\ref{eq:kindeppower}) for the CMB fiducial scale $k_*$. The solid grey line represents a skew-normal distribution (Eq.\,(\ref{eq:skewnormal})) fit. The black solid line corresponds to the function defined in Eq.\,(\ref{eq:TheFfunction}). The vertical blue-dashed line represents the difference between evaluting the variable $\log_{10} \langle \mathcal{P}_\mathcal{R}  \rangle$ using the matrix approach and from averaging over stochastic realizations.}
\end{center}
\end{figure}

As discussed in Sec.\,\ref{sec:computationspectrum}, in order to determine the power spectrum of curvature pertubations, one can alternatively solve the Langevin equations (\ref{eq:Langevin}). We have used this approach, setting $g_\star=12$,  to check the accuracy and consistency of the results derived from the matrix approach. \par \medskip

\noindent
\textbf{Statistics for the curvature perturbation and power spectrum.} 
We have solved the system (\ref{eq:Langevin}) for 2400 stochastic realizations of the benchmark model $\lambda=1.74 \times 10^{-15}$ and $C=0.012$, compatible with current constraints. We have extracted the probability distribution functions for the real and imaginary parts of the curvature perturbation $\mathcal{R}$ as well as for its power spectrum. Values for the real and imaginary part of $\mathcal{R}$ generated from those stochastic realizations are represented respectively in dark-grey and light-grey in the histogram of Fig.\ \ref{fig:histogram1}. Both distribution functions  can be well fitted by independent Gaussian distributions whose relative difference in average and standard deviation are found to be at most $0.8\%$. The best (Gaussian) fits are represented for the real and imaginary parts in dashed-red and dashed-blue, respectively, in Fig.~\ref{fig:histogram1}. \par \medskip

As discussed in App.\,\ref{app:distributionforR}, since both the real and imaginary parts of the curvature perturbation $\mathcal{R}$ are Gaussian distributed, if one assumes them to be uncorrelated it is possible to show that there is a universal, i.e. $k$-{\em independent}, probability distribution function for the variable
\begin{equation}
\label{eq:kindeppower}
\Delta \log_{10} \mathcal{P}_\mathcal{R} \equiv  \log_{10} \mathcal{P}_\mathcal{R} -  \log_{10} \langle \mathcal{P}_\mathcal{R}  \rangle \,,
\end{equation}
where $\langle \cdots \rangle$ represents an average over stochastic realizations (see also Eq.\ \eq{eq:pspec_projection}). In Ref.\,\cite{Ballesteros:2022hjk}, it was found that the distribution $\mathcal{F}(\Delta \log_{10} \mathcal{P}_\mathcal{R})$, defined via
\begin{equation}
 \int_{-\infty}^\infty \mathcal{F}(x)\diff x \, = \,1 \,,
\end{equation}
could be well fitted by a skew-normal distribution:
\begin{equation}
P_\text{skew-normal}(x\,|\,\mu, \sigma, \alpha)=\dfrac{1}{\sqrt{2\pi} \sigma} e^{-\frac{(x-\mu)^2}{2 \sigma^2} } \text{erfc}\left[ -\dfrac{\alpha(x-\mu)}{\sqrt{2}\sigma} \right]\,,
\label{eq:skewnormal}
\end{equation}
where erfc is the complementary error function. In App.\,\ref{app:distributionforR}, we show that such (normalized) distribution is indeed universal (scale-independent) and follows
\begin{equation}
\mathcal{F}(x) \, =\,  \log(10) \, 10^x \exp(-10^x) \,  \,.
\label{eq:TheFfunction}
\end{equation}
Remarkably, this distribution is independent of any model parameter. The agreement between this distribution and the results from the 2400 independent stochastic realizations is illustrated on the right panel of Fig.~\ref{fig:histogram1}. The figures show,  in addition, a fit to the corresponding data with the skew-normal distribution of Eq.\,(\ref{eq:skewnormal}). Both the function of Eq.\,(\ref{eq:TheFfunction}) and the skew-normal fit can describe the distribution of the data given the size of our statistical sample. \par \medskip

\noindent
\textbf{Accuracy of the stochastic code.}
In App.\,\ref{app:limitationLangevin} we estimate the relative error on the expected value of $\mathcal{P}_\mathcal{R}$ one can achieve by averaging over $n$ realizations with the stochastic approach as compared to employing the matrix formalism:
\begin{equation}
\dfrac{\langle \mathcal{P}_\mathcal{R}(k) \rangle_\text{stoch.}-\langle \mathcal{P}_\mathcal{R} (k) \rangle_\text{matrix}}{\langle \mathcal{P}_\mathcal{R}(k) \rangle_\text{matrix}} \simeq \, 1\,\% \,\left(\dfrac{n}{10^4} \right)^{-1/2} \,.
\label{eq:precisionstochastic}
\end{equation}
 Fig.\,\ref{fig:histogram1} shows (in dashed-blue) the difference between the quantity $\log_{10} \langle \mathcal{P}_\mathcal{R}  \rangle$ computed with the matrix approach and from averaging over 2400 stochastic realizations of the benchmark model $\lambda=1.74 \times 10^{-15}$ and $C=0.012$. The agreement between both methods is found to be $\sim 0.03 \%$ which is consistent (and in fact even better) with an order of magnitude estimate of the precision that one would expect with $\mathcal{O}(10^3)$ realizations from Eq.\,(\ref{eq:precisionstochastic}) (see the discussion in App.\,\ref{app:limitationLangevin}). 
 
In the top panel of Fig.\,\ref{fig:stochasticandmatrixwithQstar}, we represented 1080 stochastic realizations of the power spectrum for each of five selected values of $C$ ($C=\{2\times 10^{-5},10^{-4},8 \times 10^{-4},6 \times 10^{-3},4 \times 10^{-2}\}$) and $\lambda=10^{-15}$, corresponding to different values of the dissipation coefficient $Q_*$. The dark blue dots represent the arithmetic average of all the 1080 realizations for each $Q_*$. The continuous black curve, corresponds to the numerical evaluation of the power spectrum with the matrix formalism. The bottom panel of Fig.\,\ref{fig:stochasticandmatrixwithQstar} shows the relative difference between the stochastic approach and the matrix formalism solution. The relative difference is found to be at most $\sim 5 \%$, in good agreement with expectations from Eq.\,(\ref{eq:precisionstochastic}).

\begin{figure}[t!]
    \begin{center}
  \includegraphics[width=.8\textwidth]{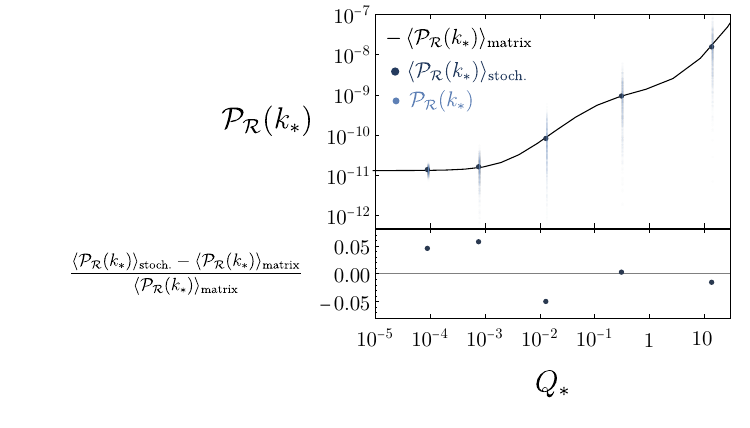} 
\caption{\it \textbf{Top:} Stochastic average of the power spectrum of the curvature perturbation $\mathcal{P}_\mathcal{R}$ for $V= \lambda/4 \phi^4$ and $\Gamma = C\,T$ for five different values of the dissipative coefficient $Q_*$ ($C=\{2\times 10^{-5},10^{-4},8 \times 10^{-4},6 \times 10^{-3},4 \times 10^{-2}\}$ from left to right and $\lambda=10^{-15}$ for all points), evaluated at the CMB fiducial scale $k_*$, (dark blue dots). The number of realizations for each value of $Q_*$ is 1080. Each realization is represented as light blue dot. The solid black line represents the average of the power spectrum obtained from the matrix approach. \textbf{Bottom:} Relative difference between
the stochastic average and the matrix approach determination of $\mathcal{P}_\mathcal{R}$. The agreement for each value of $Q_*$ is at most $\sim 5\%$. }
\label{fig:stochasticandmatrixwithQstar}
\end{center}
\end{figure}

\subsubsection{Parameter space for quadratic and sextic potentials}

We remind the reader that we are considering $\Gamma\propto T$.

\label{sec:linear_quadraticandsextic}

\par \medskip
\noindent
\textbf{Reheating.} Similarly to the $V\propto \phi^4$ case, both for $V\propto \phi^2$ and $V\propto \phi^6$, in the large dissipation regime ($Q_\text{end}\gg 1$), radiation is dominant and decays less rapidly than the inflaton (see Table~\ref{tab:2} and Table~\ref{tab:4}). If inflation ends in the weak dissipative regime ($Q_\text{end}\ll 1$), radiation is subdominant at the end of inflation. In this regime, after the end of inflation the inflaton dominates the energy budget for a while but the dissipation rate increases and eventually reaches the strong dissipative regime ($Q\gg1$) where the ratio of the inflaton to radiation energy density becomes exponentially or superexponentially suppressed for $V\propto \phi^6$ or $V\propto \phi^2$ respectively. A smooth transition into radiation domination can therefore always be achieved in these cases without invoking extra fields or couplings. \par \medskip

\noindent
\textbf{Parameter space.} The parameter space satisfying constraints on the amplitude of the power spectrum is represented in Figure~\ref{fig:linear_quadratic_sextic_r_VS_ns} on the left and center panels for $V\propto \phi^2$ and $V\propto \phi^6$ for different values of $N_\text{tr}$ and  $\Delta N$. In both cases, the characteristic S-like shape for the allowed parameter space can be seen. From the left panel of that figure, one can directly conclude that the quadratic potential is excluded for this model of dissipation ($\Gamma\propto T$) from constraints on $n_s$. Considering a maximal (i.e.\ until BBN) phase of matter domination after inflation ($\Delta N=10$) would ease the tension but cannot make to agree the model with the constraints on the scalar index.  

The sextic potential accommodates a wider range of values for $n_s$, from $0.92$ up to $0.98$. Two distinct region of parameter space are compatible with constraints on $n_s$, as illustrated in the right panel of Fig.~\ref{fig:linear_quadratic_sextic_r_VS_ns}. A first region predicts values of $n_s$ of the order of the central value, $n_s\simeq 0.965$. This corresponds to $r\sim  10^{-2}-3\times 10^{-2}$ for $\Delta N=0$ and $N_\text{tr}=5$ and $r\sim 6 \times 10^{-3}-3\times 10^{-2}$ assuming a maximal (until BBN) phase of kination ($\Delta N=-10$). This region of parameter space will be in the reach of SO, LiteBIRD and CMB-S4. An extended phase of matter domination ($\Delta N=10$) with such large values of $r$ is not compatible with constraints on $n_s$ and is therefore excluded. A second region of parameter space predicts $r\sim 10^{-5}$, beyond the reach of future planed experiments, but could accommodate any value of $n_s$ irrespectively of the value of $\Delta N=\pm 10$. In this case, the allowed range of $N_{k_*}$ is much larger than in the quartic case, with values $N_{k_*}\in[61,64]$ for $\Delta N=0$ and $N_\text{tr}=5$. The allowed dimensionless couplings are $C\in[10^{-3},4\times 10^{-3}]$ and $\lambda \in [4\times 10^{-17},1.5\times 10^{-16}]$ and a narrow region for $C\simeq 1.5 \times 10^{-2}$ and $\lambda\simeq 8 \times 10^{-15}$. \par \medskip

\begin{figure}[t]
\begin{center}
\includegraphics[width=.32\textwidth]{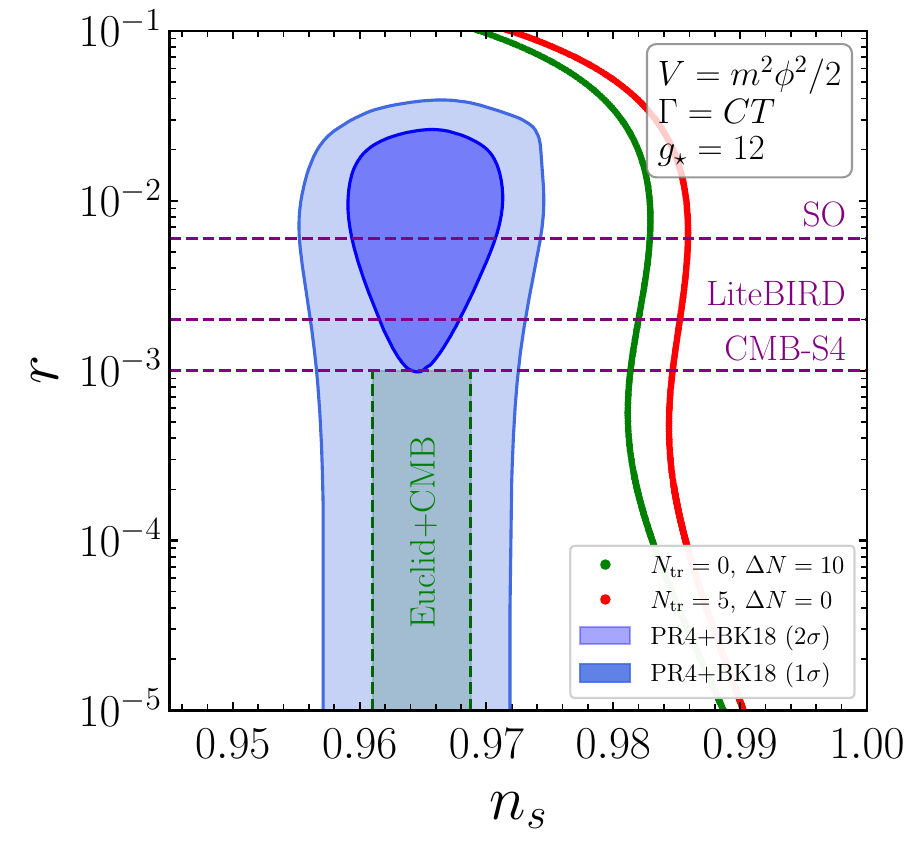}
\includegraphics[width=.32\textwidth]{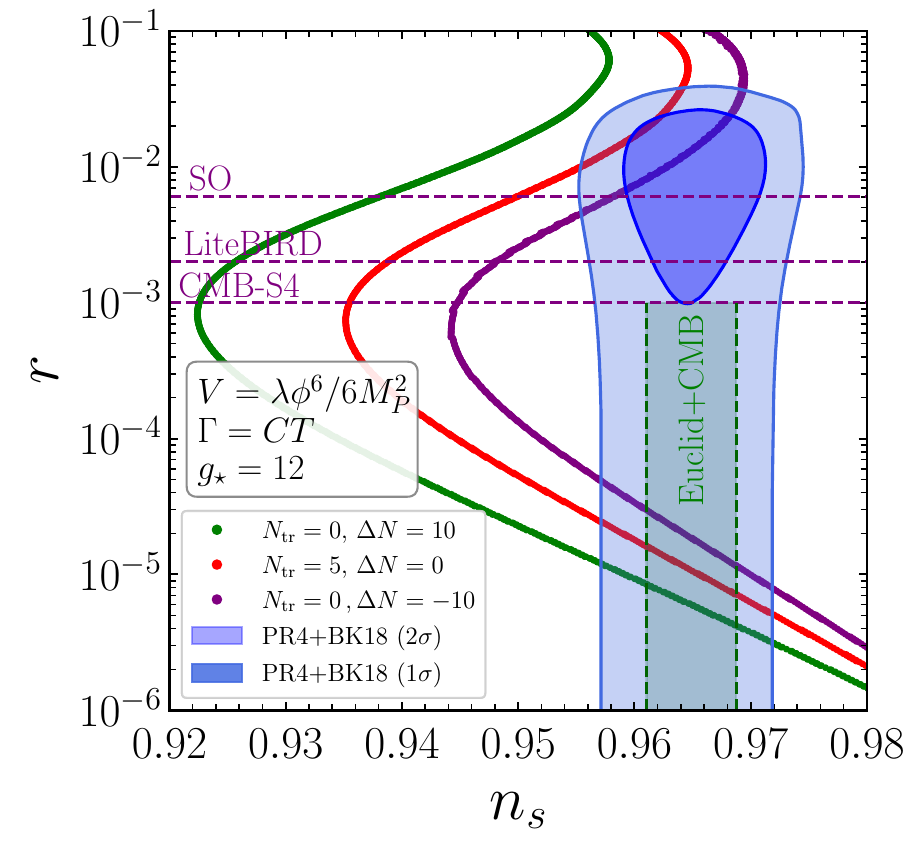}
\includegraphics[width=.32\textwidth]{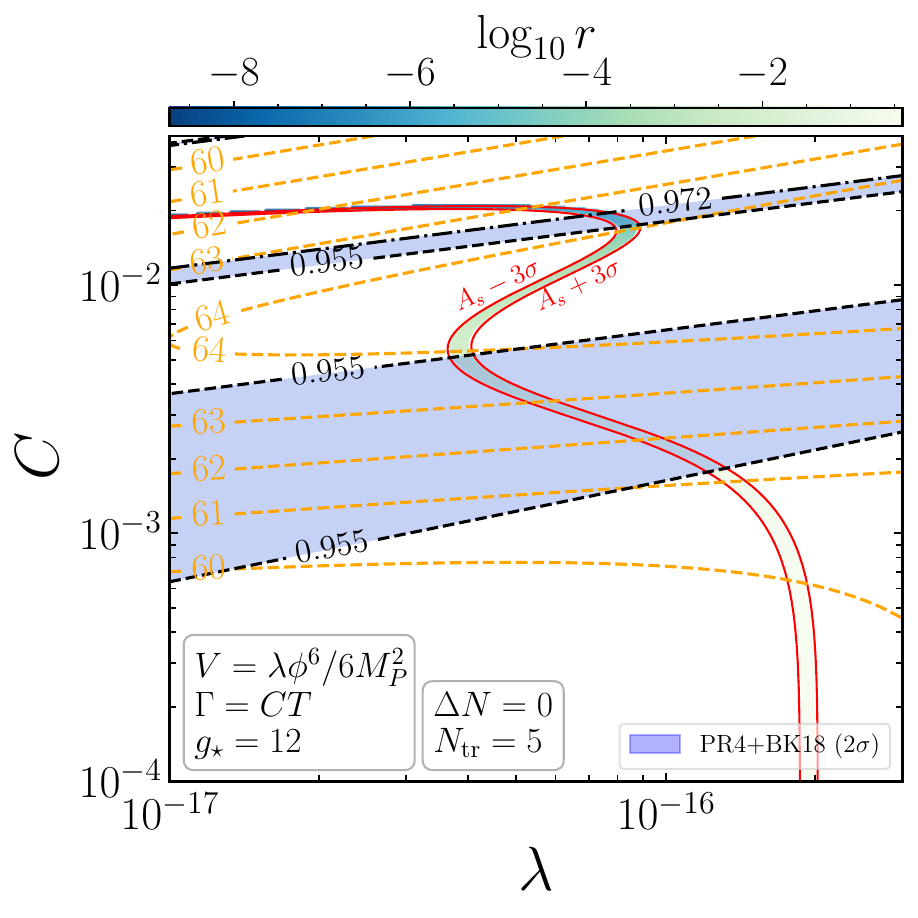}
\caption{\em \label{fig:linear_quadratic_sextic_r_VS_ns} {Parameter space for the models described in Sec.\,\ref{sec:linear_quadraticandsextic} compatible with constraints on the amplitude of the scalar power spectrum $A_s$ from~\cite{Planck:2018vyg}.  Constraints from PR4+BK18 are represented in blue. The expected sensitivity reaches on $r$ from SO, LiteBIRD and CMB-S4 are represented in dashed-purple lines. The green region represents the expected sensitivity of a joint Euclid+CMB analysis.  {\bf Left:} The allowed parameter space for a quadratic potential projected onto the plane $\{r,n_s\}$ is represented in red (assuming $N_\text{tr}=5$ and  $\Delta N=0$) and in green (assuming $N_\text{tr}=0$ and  $\Delta N=10$). See Sec.\,\ref{sec:durationinflation} for the meaning of these parameters and the discussion in Sec.\,\ref{sec:linear}. {\bf Center:} The allowed parameter space for $V\propto \phi^6$ is projected in the plane $\{r,n_s\}$ as green, red and purple dots (from left to right) for different choices of the parameters $N_\text{tr}$ and  $\Delta N$ as indicated in the plot legend. {\bf Right:} The allowed parameter space for $V\propto \phi^6$ projected onto the plane $\{C,\lambda\}$ for $N_\text{tr}=5$ and  $\Delta N=0$ (corresponding to the red dots in the center panel) between two solid red lines corresponding to constraints on $A_s$. The color of the allowed region codes the value of $\log_{10} r$. The light-orange dashed lines represent isocontours of the duration $N_{k_*}$ between the CMB fiducial-scale $k_*=0.05$~{\rm Mpc$^{-1}$} crosses outside the horizon during inflation and the end of it. Dashed, dot-dashed and solid black lines represent isocontours of the scalar spectral index $n_s$. For comparison and further details, see the caption of Fig.~\ref{fig:lineargammag} and the discussion in Sec.\,\ref{ss:pheno}. }}
\end{center}
\end{figure}

\subsection{$\Gamma \propto T^3$}
\label{sec:trilinear}

\begin{figure}[t]
\begin{center}
$\includegraphics[width=.49\textwidth]{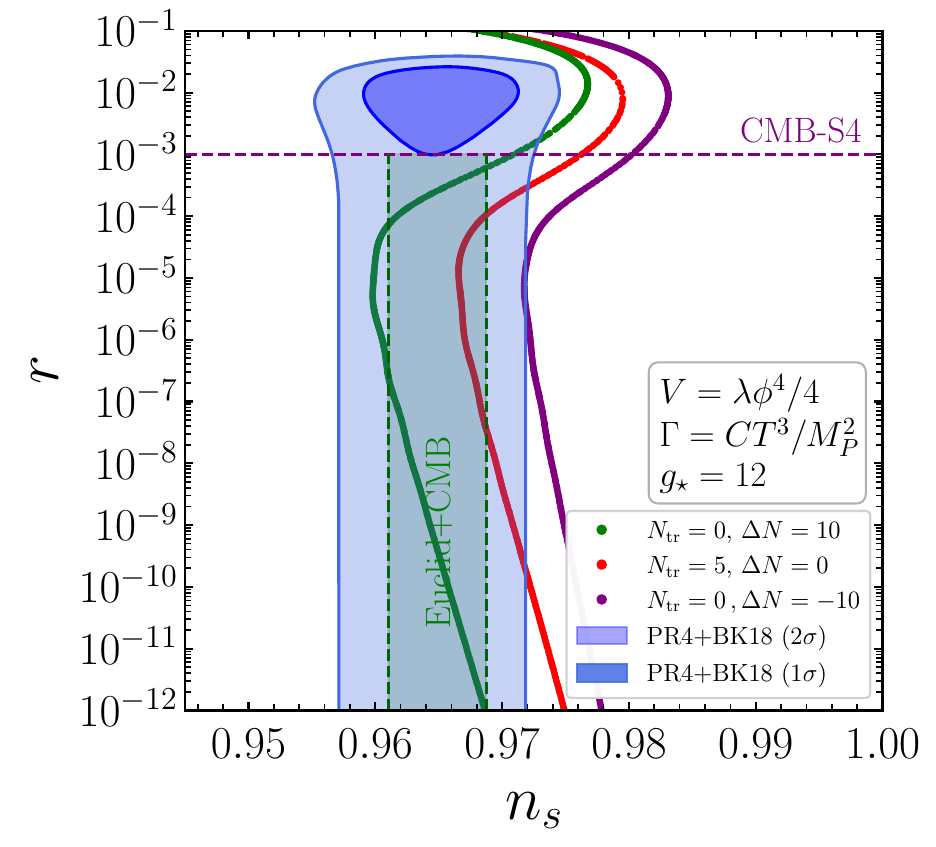}
\qquad\includegraphics[width=.49\textwidth]{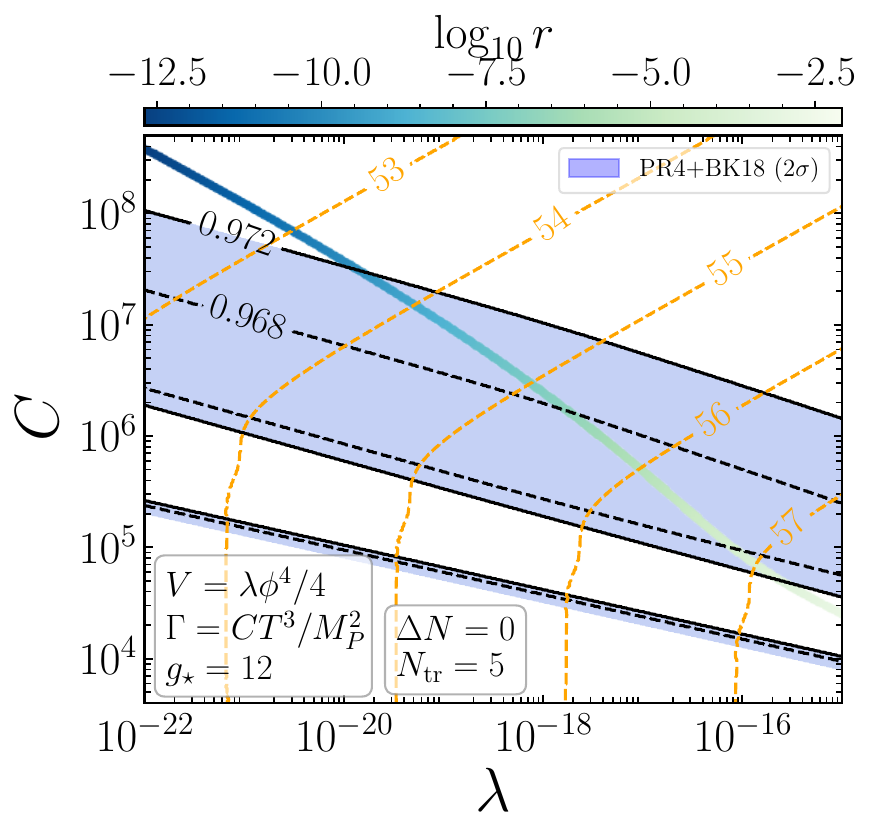}$
\caption{\em \label{fig:trilineargamma} {Parameter space for the model described in Sec.\,\ref{sec:trilinear} (cubic dissipation rate and $V\sim \phi^4$) compatible with the constraints on the amplitude of the scalar power spectrum $A_s$ from~\cite{Planck:2018vyg}.  Constraints from PR4+BK18 are represented in blue. The expected sensitivities on $r$ from SO, LiteBIRD and CMB-S4 are represented in dashed-purple lines. The green region represents the expected sensitivity for a joint Euclid+CMB analysis.   {\bf Left:} The allowed parameter space is projected onto the plane $\{r,n_s\}$ as green, red and purples dots (from left to right) for different choices of the parameters $N_\text{tr}$ and  $\Delta N$ as indicated in the legend. {\bf Right:} The allowed parameter space is projected onto the plane $\{C,\lambda\}$ for $N_\text{tr}=5$ and  $\Delta N=0$ (corresponding to the red dots in the center panel) between two solid red lines corresponding to $2\sigma$ constraints on $A_s$. The color of the allowed region codes the value of $\log_{10} r$. The light-orange dashed lines represent isocontours of the duration $N_{k_*}$ between the time the  CMB fiducial-scale $k_*$ crosses outside the horizon and the end of inflation. Dashed, dot-dashed and solid black lines represent isocontours of the scalar index $n_s$. For more details and for the purpose of comparison, see the caption of Fig.~\ref{fig:lineargammag} and the discussion in Sec.\,\ref{ss:pheno}. }}
\end{center}
\end{figure}

\textbf{Reheating.} The fate of the post-inflationary universe is summarized in the right panels of Table~\ref{tab:2}, Table~\ref{tab:3} and Table~\ref{tab:4} for $\Gamma\propto T^3$. For a quadratic potential  in the weak dissipative regime at the end of inflation, $Q_\text{end}\ll 1$, radiation redshifts faster than the inflaton component which behaves as non-relativisitic matter oscillating around its minimum. The dissipation rate decays exponentially and therefore the weak dissipative regime would essentially persist and a smooth transition into radiation domination cannot be obtained in this case unless extra physics is invoked. In the strong dissipative regime, $Q_\text{end}\gg 1$, the inflaton density becomes superexponentially suppressed after the end of inflation but the dissipation rate still decays exponentially with time. Therefore a weak dissipation regime follows the strong dissipation phase and, again, there is no transition into radiation domination (and some reheating mechanism is required). 

For a quartic potential  in the weak dissipative regime $Q_\text{end}\ll 1$, the inflaton density redshifts as radiation. The ratio of radiation to inflaton energy-density is therefore frozen to a (small) value and the dissipation rate decays exponentially, which perpetuates the weak dissipative regime. Reheating cannot be achieved in this case (unless, again, extra couplings are included). However in the strong dissipative regime, $Q_\text{end}\gg 1$, the inflaton density redshifts faster than radiation and the radiation dominates the energy budget. The dissipation rate still exponentially drops to reach the weak dissipation regime where the ratio of radiation to inflaton energy-density becomes frozen but to much larger values than unity. A transition into radiation domination after inflation is therefore achieved in this case. 

For a sextic potential, $V\sim \phi^6$, the inflaton energy density redshifts faster than radiation for both weak and strong dissipation regimes. The dissipation rate exponentially decays ensuring that the weak dissipation regime would always be achieved. As the ratio of radiation to inflaton energy-density increases with time, the transition to radiation domination is always achieved. \par \medskip

\noindent
\textbf{Parameter space.} The parameter space compatible with the bounds on the amplitude of the scalar power spectrum is shown in Figure~\ref{fig:trilineargamma} for a quartic potential. Given that whether a transition into radiation domination occurs in this case depends on the value of $Q_\text{end}$ at end of inflation, we represent the allowed parameter spaces corresponding to several choices of $N_\text{tr}$ and $\Delta N$. The left panel of the figure shows that current CMB data constrain the available parameter space to values of $r<10^{-3}$, smaller than the sensitivity expected from CMB-S4. A maximal phase of kination ($N_\text{tr}=0$ and $\Delta N=-10$) is on the edge of being excluded by current data. For smaller post-inflationary equation of states and for $r<10^{-3}$, the predicted values of $n_s$ are compatible with the current preferred range. However, values of the tensor-to-scalar ratio could possibly reach extremely small values beyond any foreseeable near-future detection reach. The right panel of Fig.~\ref{fig:trilineargamma} shows the large parameter space available in this case, where values of $C$ and $\lambda$ respectively span three and four orders of magnitude. The typical duration of inflation after the CMB fiducial scale crossing $N_{k_*}$ can cover values from $53.5$ up to $57$, a range much larger than the  one we find in the $\Gamma \propto T$ case.\\
The parameter space for $V \propto \phi^2$ and $V \propto \phi^6$ potentials is represented in Fig.~\ref{fig:trilinear_sextic_and_quadratic}, respectively on the left and right panels. An interesting feature specific to the sextic potential, is that the value of $Q$ is actually constant up to slow-roll corrections during inflation. This is discussed in Section~\ref{sec:analyticalestimates}. However, the quadratic and sextic potentials are obviously excluded from the results of Fig.\,\ref{fig:trilinear_sextic_and_quadratic}.

\begin{figure}[t]
\begin{center}
$\includegraphics[width=.48\textwidth]{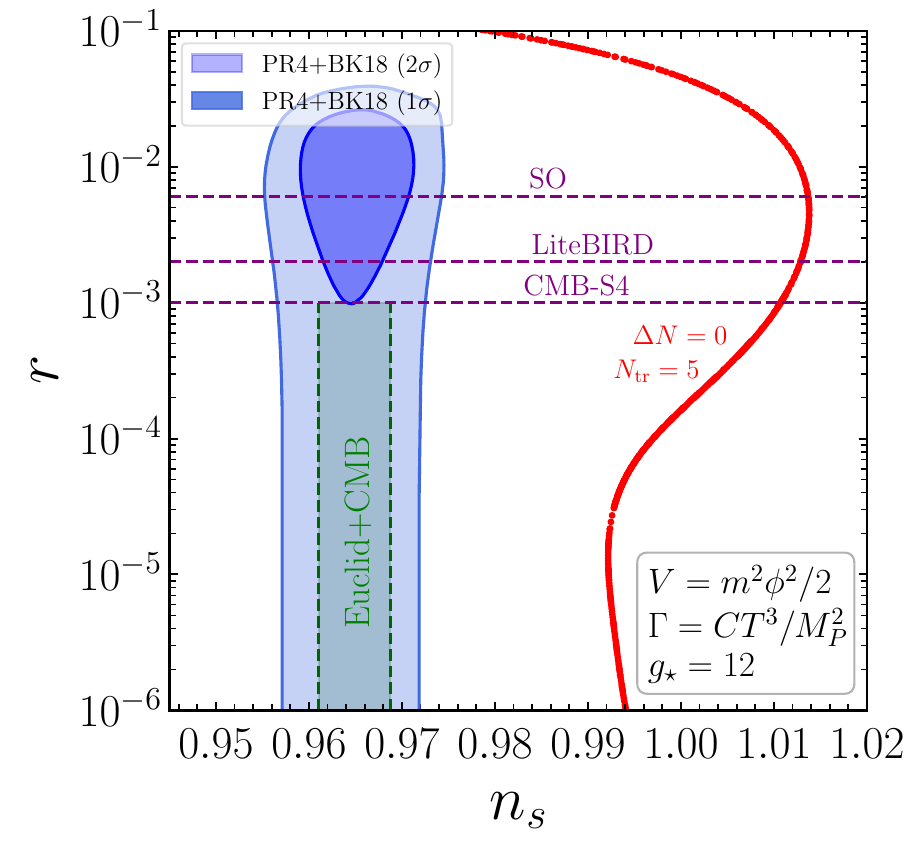}\includegraphics[width=.49\textwidth]{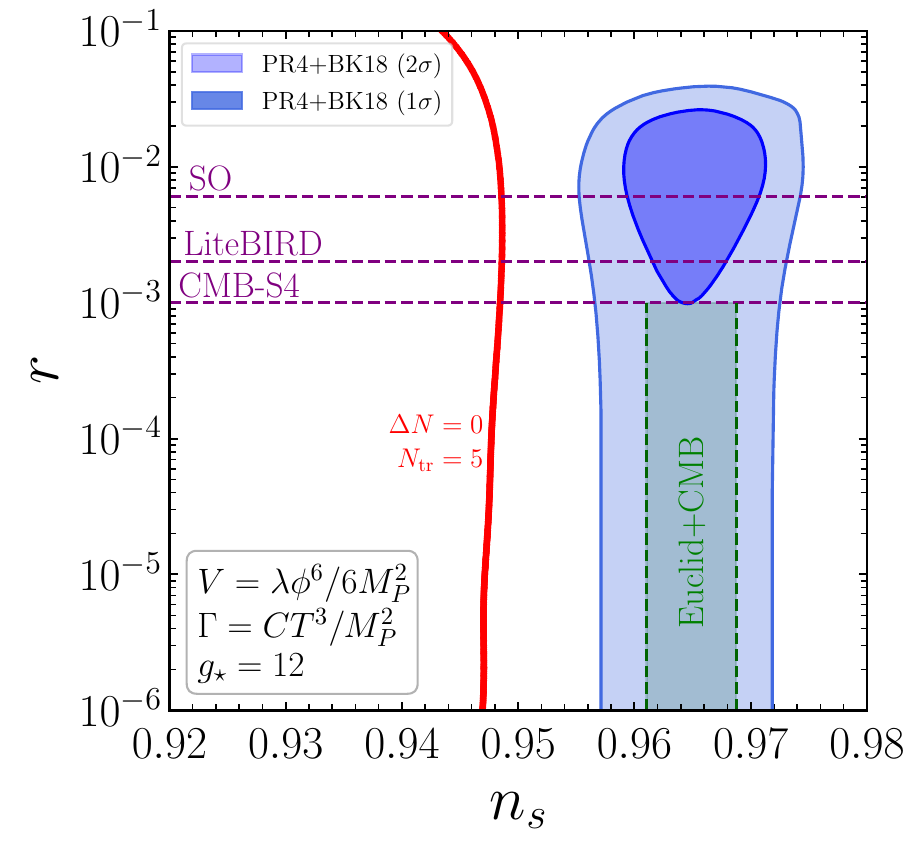} 
$
\caption{\em \label{fig:trilinear_sextic_and_quadratic} {The parameter space for the model described in Sec.\,\ref{sec:trilinear}, compatible with the constraints on the amplitude of the scalar power spectrum $A_s$ from~\cite{Planck:2018vyg}, is represented in red assuming $N_\text{tr}=5$ and  $\Delta N=0$. Constraints from PR4+BK18 are represented in blue. The expected sensitivities on $r$ from SO, LiteBIRD and CMB-S4 are represented in dashed-purple lines. The green region represents a sensitivity forecast for a joint Euclid+CMB analysis.  {\bf Left:} The allowed parameter space for a quadratic potential $V\sim \phi^2$. {\bf Right:} The allowed parameter space for sextic potential. For more details, see the caption of Fig.~\ref{fig:lineargammag} and the discussion in Sec.\,\ref{ss:pheno}.}}
\end{center}
\end{figure}

\subsection{$\Gamma\propto T^3/\phi^2$}
\label{sec:T3phi-2}

\textbf{Reheating.} This is case is more problematic as once the inflaton field reaches the minimum of his potential, the dissipation rate diverges, which implies this simple form of $\Gamma$ stops being valid. For this reason we solve the various equations up to the end of inflation (corresponding to $N_{tr}=0$) and do not extraplote past this time. Going beyond this point would require extra assumptions on the model. \par \medskip

\noindent
\textbf{Parameter space.} The regions of parameter space compatible with constraints on the amplitude of the scalar power spectrum for quadratic, quartic and sextic potential are respectively represented in the left, center and right panels of Fig.~\ref{fig:T3phi-2_quartic_r_VS_ns}. The predicted values for $n_s$ in the three cases are significantly larger than the  constraints on $n_s$ from BICEP/Keck-Planck. As a result, the $\Gamma \sim T^3/\phi^2$ is ruled out for these potentials. This conclusion remains even if we add kination or matter domination phases immediately after inflation.

\begin{figure}[t!]
    \centering
  \includegraphics[width=0.33\textwidth]{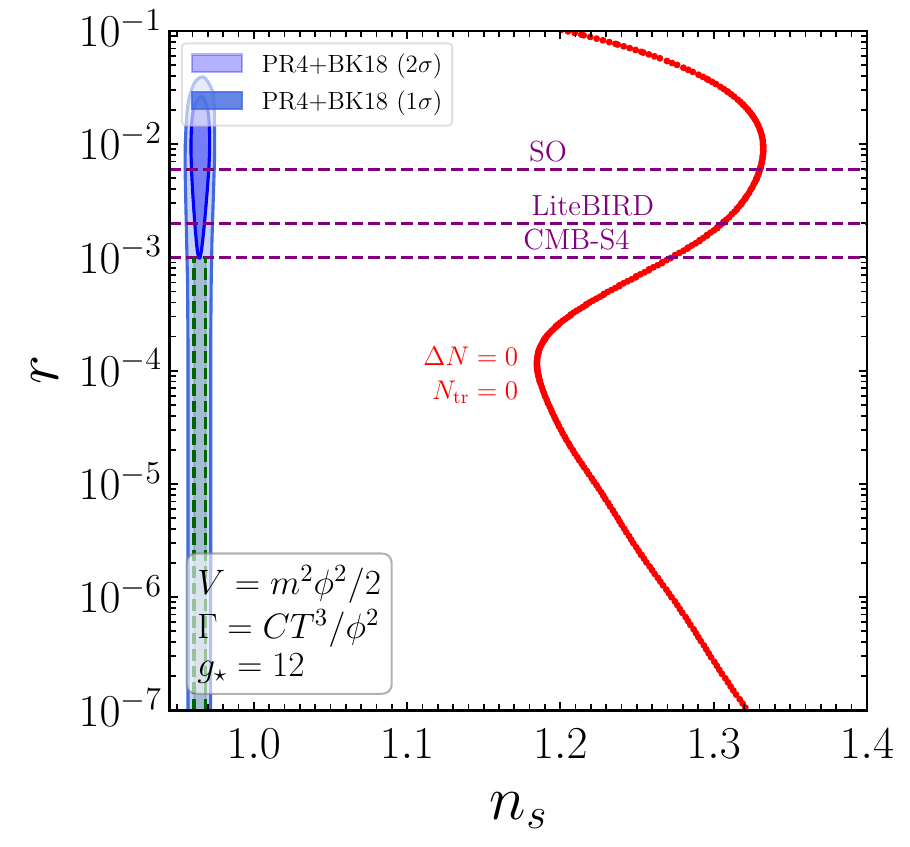}
  \includegraphics[width=0.32\textwidth]{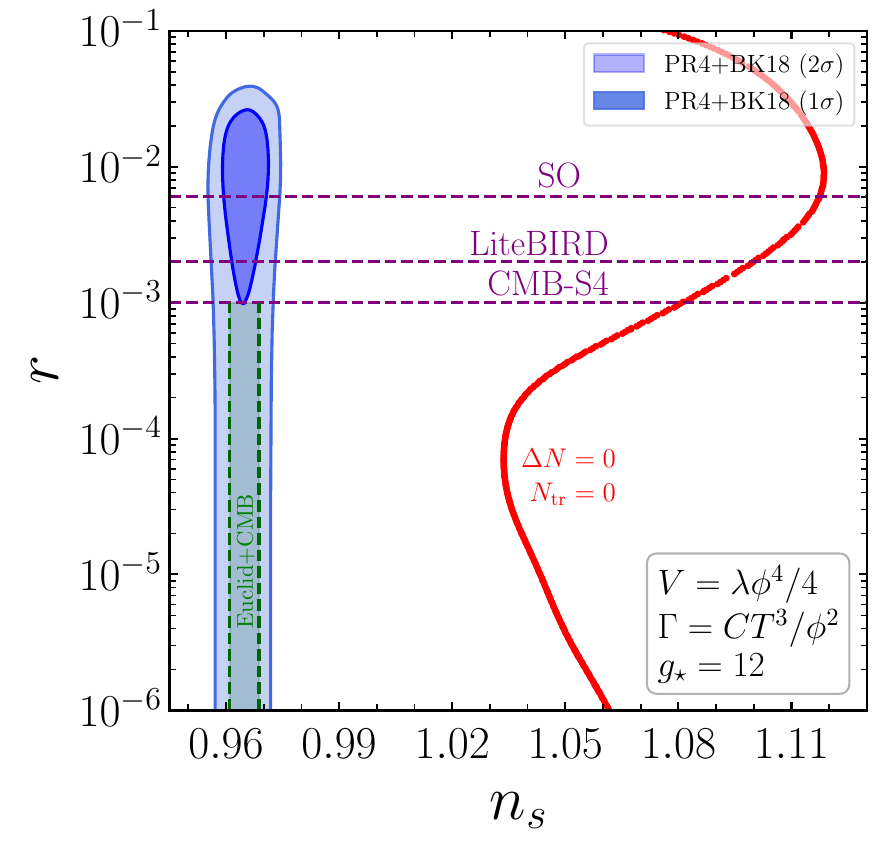}
  \includegraphics[width=0.32\textwidth]{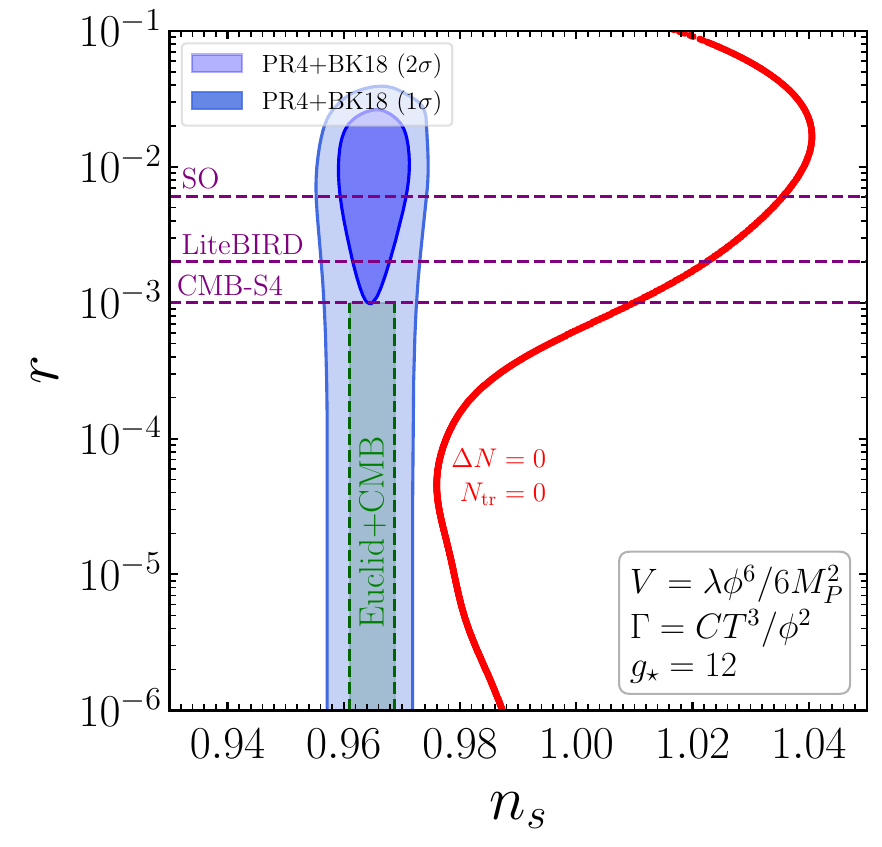}
\caption{\em \label{fig:T3phi-2_quartic_r_VS_ns} {The parameter space for the model described in Sec.\,\ref{sec:T3phi-2}, compatible with the CMB limits on the amplitude of the scalar power spectrum $A_s$ from~\cite{Planck:2018vyg}, is represented in red assuming $N_\text{tr}=0$ and  $\Delta N=0$. Constraints from PR4+BK18 are represented in blue. Expected sensitivities on $r$ from SO, LiteBIRD and CMB-S4 are represented in dashed-purple lines. The green region represents a sensitivity forecast for a joint Euclid+CMB analysis.  {\bf Left:} Parameter space for a quadratic potential. {\bf Center:} Parameter space for a quartic potential.  {\bf Right:} Parameter space for sextic potential. For more details, see the caption of Fig.~\ref{fig:lineargammag} and the discussion in Sec.\,\ref{ss:pheno} and Sec.\,\ref{ss:pheno} .}}
\end{figure}

\subsection{Comparison to earlier works}

\label{sec:comparison}

Some of the combinations of $\Gamma(\phi,T)$ and $V(\phi)$ that we have considered have also been studied in previous works. In this section we discuss how our procedure and results compare to some of those works, highlighting the most important differences.

Reference \cite{Benetti:2016jhf} considered the four possible combinations of $V\propto \phi^n$, $n=4,6$ and $\Gamma$ proportional to $T$ or $T^3/\phi^2$. The quartic potential was considered with $\Gamma\propto T$ in \cite{Bastero-Gil:2017wwl} and with $\Gamma\propto T^3/\phi^2$ in \cite{Arya:2017zlb}. The latter work was complemented in \cite{Arya:2018sgw} with the possibilities previously studied in \cite{Benetti:2016jhf} that we have just mentioned. These four papers \cite{Benetti:2016jhf,Bastero-Gil:2017wwl,Arya:2017zlb,Arya:2018sgw} used Planck 2015 data \cite{Planck:2015fie} to do Monte Carlo samplings of (some) of the parameters of each model. The case $V\propto\phi^4$, $\Gamma\propto T$ had been considered earlier in \cite{Bastero-Gil:2016qru}, where predictions for the plane $\{r,n_s\}$ were obtained and compared to Planck 2015 bounds. A subsequent analysis taking into account theoretical constraints in a specific field theory implementation was done in \cite{Bastero-Gil:2018uep}. 

The amplitude of the primordial scalar spectrum was obtained in these papers using fitting formulas that originate from \cite{Ramos:2013nsa}, \cite{Bastero-Gil:2014jsa} and \cite{Bastero-Gil:2016qru}. This type of semi-analytical formula, that relies on a numerical fit, has been widely used in the literature on warm inflation after the year 2013.\footnote{See e.g.\ \cite{Visinelli:2016rhn} for a previous study of monomial potentials. In that reference analytical limits of Eq.\ \eqref{ec:frompapers} with $G(Q)=1$ were used.} Its form is the following:
\begin{align}\label{ec:frompapers}
\mathcal{P_R} = \overbrace{\underbrace{\left(\frac{H}{\dot\phi}\right)^2\left(\frac{H}{2\pi}\right)^2}_{\text{cold-like}}\left(\frac{T}{H}\frac{2\pi\,Q}{\sqrt{1+4\pi\, Q/3}}\right.}^{\text{ purely analytical inhomogeneous}}+\left.\underbrace{1+2\,n_{\rm BE}}_{\Theta}\right)\times \underbrace{G(Q)}_{\text{numerically-fitted correction}}\,.
\end{align}
This approximate formula combines four pieces:
\begin{enumerate} 
\item An additive cold-like piece of the form $H^4/(2\pi\dot\phi)^2$. We call it cold-like because although it has the same form as in standard cold inflation, $H$ and $\dot\phi$ are evaluated on the background solution of warm inflation.
\item A purely analytical approximation for the inhomogeneous solution of the inflaton fluctuations (which, like our own analytical approximation, fails to reproduce the numerical spectrum as $Q$ grows, see Sec.\,\ref{ss:cos} below). It is induced by the thermal noise and dominates over the cold-like solution for large enough $Q$.
\item A correction, $2\,n_{\rm BE}$, to the cold-like piece. This correction has been argued to come from inflaton modes with high occupation number, thermalized with the radiation plasma, see e.g.\
\cite{Ramos:2013nsa}.
\item A function $G(Q)$ implementing a fit to numerical solutions of the spectrum\footnote{{See \cite{Graham:2009bf, Bastero-Gil:2011rva} for early works in this direction}.} that accounts for the effect of radiation fluctuations on the growth of curvature perturbations and therefore corrects the failure at large $Q$ of the purely analytical solution; see point 2 above. This numerical correction depends on the specific form of $\Gamma$ and $V$. For instance, in the case of $V \propto \phi^4$ and $\Gamma \propto T$ the function $G(Q)$ from \cite{Bastero-Gil:2016qru} reads $G(Q) = 1+0.0185\,Q^{2.315}+0.335\,Q^{1.364}$.
\end{enumerate}
All the time-dependent (background) quantities appearing in \eq{ec:frompapers} have been customarily evaluated in the literature at horizon crossing. We discuss next some aspects of  \eq{ec:frompapers}. 
The correction $n_{\rm BE}$, where $\rm BE$  stands for Bose-Einstein, is $n_{\rm BE} = (\exp(H/T)-1)^{-1}$ and therefore vanishes for $T/H\rightarrow 0$. In the oposite limit it is potentially relevant, but in practice the inhomogeneous solution dominates, regardless of $n_{\rm BE}$.  This means that the expression \eq{ec:frompapers}  is a good approximation  in the $Q_*\ll 1$ limit because the homogeneous solution tends to the cold one, and it is also a good approximation in the $Q_*\gg 1$ limit {({see} 
Eq.\ \eqref{ec:correctedbythermal} and {the} 
discussion {around it})}. For intermediate values of $Q_*\lesssim 1$, $n_{\rm BE}$ can be relevant, as we discuss below.

In Sec.\,\ref{sec:analyticalestimates} we discuss in detail the derivation of an analytic approximation for the power spectrum that performs better than \eq{ec:frompapers}  given the function $G(Q)$ proposed in the literature, see the green curve in Fig.\,\ref{fig:comparewithprevious} (left). 
\begin{figure}[t!]
\begin{center}
$\includegraphics[width=.485\textwidth]{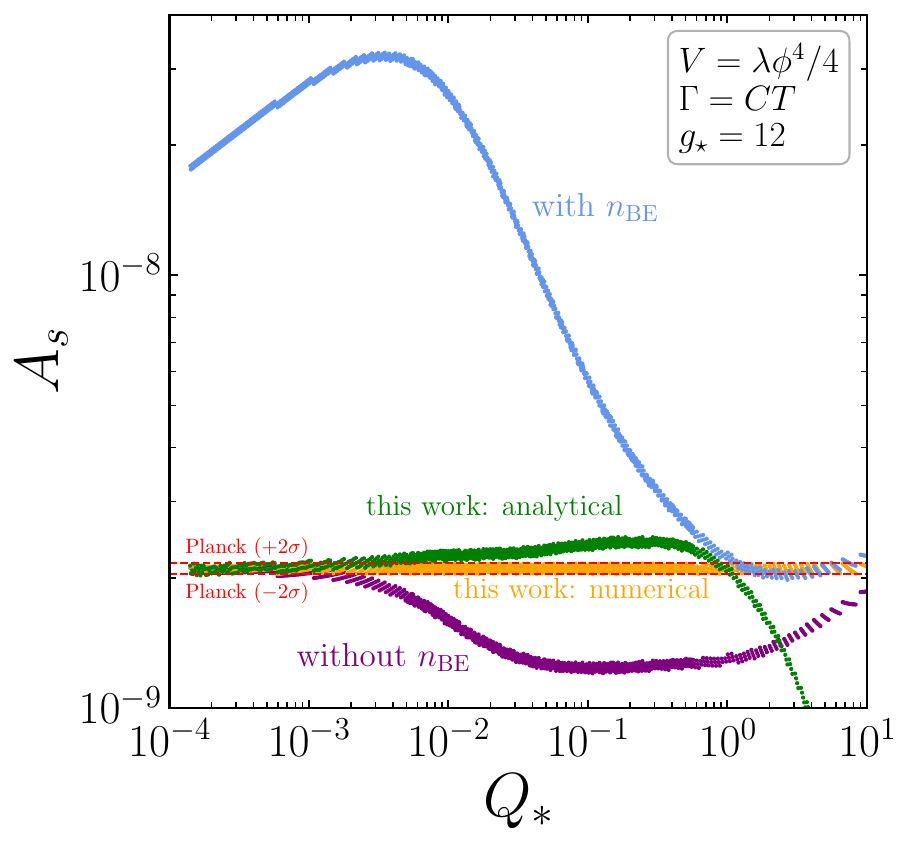}\includegraphics[width=.47\textwidth]{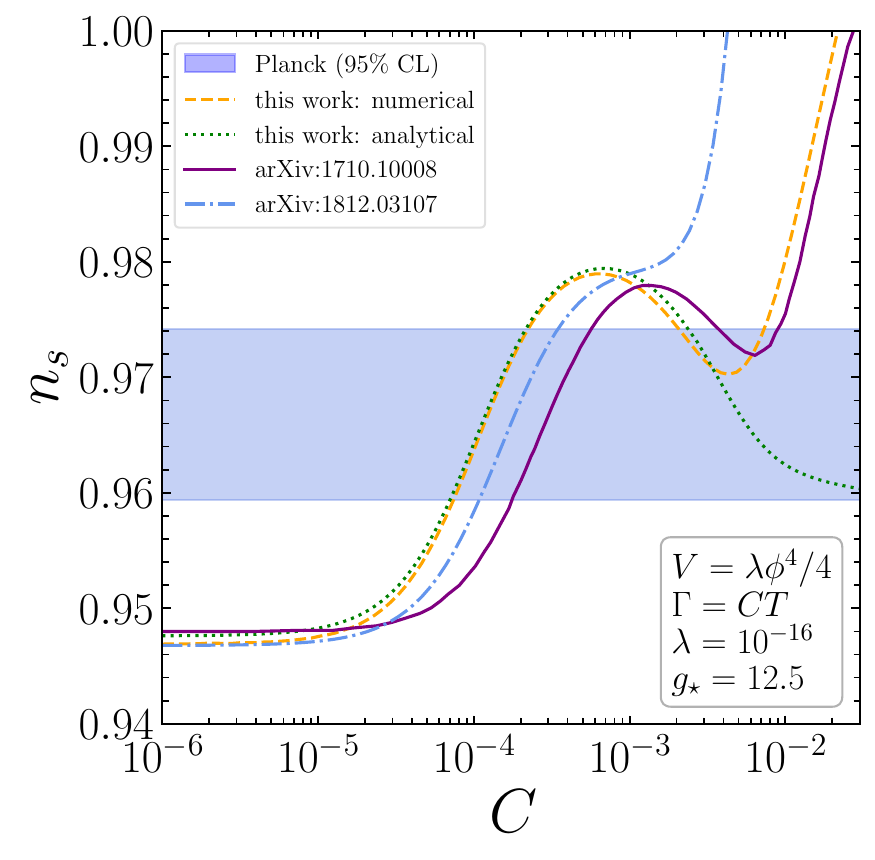} 
$
\caption{\em \label{fig:comparewithprevious} { Comparison between our numerical and analytical results with earlier works.  {\bf Left:} Amplitude of the scalar power spectrum, $A_s$, for $k=k_*$ as a function of $Q_*$. The orange dots are the result of a scan in parameter space \{$\lambda$, $C$\} giving rise to $A_s$ in agreement with the bounds from Ref.\,\cite{Planck:2018vyg}, within 2$\sigma$ (represented in red dashed lines). The figure illustrates the performance of our  purely analytical approach as computed in Sec.\,\ref{sec:analyticalestimates} (green dots) and the semi-analytical formula Eq.\,\eqref{ec:frompapers} without the $n_\text{BE}$ term (purple). To obtain these two sets of points we have used the same set of parameters \{$\lambda$, $C$\} as for the orange dots. In addition, to illustrate the size of the effect of including the $n_\text{BE}$ term, we also apply Eq.\,\eqref{ec:frompapers} with it (blue dots), using again the same set of \{$\lambda$, $C$\}. Clearly, this last set of points cannot be interpreted as the analytical and semi-analytical approximations we are comparing (purple and green), since it assumes different physics than the other sets of points; it just illustrates the magnitude of the effect the $n_{BE}$ term would have. {\bf Right:} Scalar spectral index as a function of the dimensionless parameter $C$ (of $\Gamma = C\,T$). The orange dashed line and green dotted line correspond respectively to our numerical and analytical results. The purple solid line labeled ``arXiv:1710.10008" corresponds to the results of Ref.\,\cite{Bastero-Gil:2017wwl} and the blue dotted-dashed line to Eq.\,(B.1) of \cite{Arya:2018sgw}. The blue region corresponds to the allowed parameter space by Planck constraints on $n_s$ at $95 \%$ confidence level. For details, see the discussion in Sec.\,\ref{sec:comparison}.}}
\end{center}
\end{figure}
The left panel of Fig.\,\ref{fig:comparewithprevious} compares the power spectrum as a function of $Q_*$ computed using the approximation \eq{ec:frompapers} by taking the function $G(Q)$ provided in \cite{Bastero-Gil:2016qru}. The blue and purple curves are obtained (with and without $n_{\rm BE}$) by applying \eq{ec:frompapers} to a set of multiple choices for the pair \{$\lambda$, $C$\} ($g_\star=12$ for all pairs) that give a spectrum (computed with our numerical matrix formalism, orange horizontal band) falling within 2$\sigma$ of current observational bounds {$\log(10^{10}A_s)=3.044 \pm 0.014$ by Planck (TT,TE,EE+lowE+lensing)~\cite{Planck:2018vyg}}. In order to make the most meaningful comparison to previous literature we have taken $N_{\rm tr}=0$ in our numerical approach to produce our prediction for this figure (see Sec.\,\ref{sec:durationinflation} for our discussion of the determination of the number of e-folds). Clearly, the inclusion of $n_{\rm BE}$ leads to a very large discrepancy from our numerical result for any $Q_*\lesssim 1$. The semi-analytical approximation \eq{ec:frompapers} without $n_{\rm BE}$ works well only for sufficiently small $Q_*$, as expected. For the larger values of $Q_*$ necessary to fit $n_s$ (see right panel on the same figure), the approximation \eq{ec:frompapers} (with the form of $G(Q)$ typically assumed, see point 4 above) underestimates the power spectrum by a factor of $\sim 1/2$, which amounts to a discrepancy from the correct result as large as $\sim 5\sigma$. This is clearly insufficient to use \eq{ec:frompapers} for precision cosmology, regardless of whether $n_{\rm BE}$ is included.  We advocate instead the use of our numerical matrix formalism, which is not only accurate and precise but also fast. 

In Fig.\,\ref{fig:comparewithprevious} (right) we also compare the spectral index $n_s$ as a function of $C$ (from $\Gamma = C\,T$) that we obtain numerically (orange dashed line) to the corresponding result from \cite{Bastero-Gil:2017wwl} and the analytic fitting formula of App.\,B (Eq. (B.1)) of \cite{Arya:2018sgw}. For a given $C$ between $10^{-5}$ and $10^{-3}$ the error committed by these approximations is $\mathcal{O}(1\%)$, which is significantly larger than the precision with which the marginalized value of $n_s$  is determined with current CMB (and other) data: $n_s = 0.9668 \pm 0.0037$ at 68\% c.l.\ for Planck (TT,TE,EE+lowE+lensing+BK15+BAO), assuming non-vanishing $r$.

Whereas $n_{\rm BE}$ was included in \eq{ec:frompapers} in  the analysis in \cite{Benetti:2016jhf,Arya:2017zlb,Arya:2018sgw}, the results of \cite{Bastero-Gil:2016qru,Bastero-Gil:2017wwl} were obtained with and without it. As we discuss in Sec.\,\ref{quantumn}, a similar correction can arise when quantizing inflaton perturbations, assuming they acquire a Bose-Einstein distribution. Whether this term is actually required is a model-dependent question. 

In Sec.\,\ref{sec:stoch} we argue that the application of the stochastic inflation formalism \cite{Starobinsky:1986fx} to warm inflation does not imply by itself the necessity of implementing it.

 It would be necessary to focus on a specific implementation of warm inflation to determine whether $\Theta = 1+2 n_{\rm BE}$ is equal to $1$ or not. We assume $\Theta =1$ since we perform a phenomenological analysis without deriving $\Gamma$ from a specific Lagrangian and, to the best of our knowledge, the literature on warm inflation does not show $n_{\rm BE}$ to arise in general. Choosing any other value for $\Theta$ would mean adding an extra assumption, which does not appear to be justified.

Moreover, we find that if an occupation number correction, $\Theta$, arises at all, it does so multiplying the actual homogeneous solution for inflaton fluctuations in warm inflation, instead of the cold-like spectrum (as it is the case in the semi-analytical approximation \eqref{ec:frompapers}). As we argue in Sec.\,\ref{quantumn}, this conclusion follows from a consistent quantization scheme for a scalar field with a classical (stochastic) source. As we explicitly  show in App.~\ref{app:stochasticv2}, the origin of this difference arises from terms neglected in Ref.\,\cite{Ramos:2013nsa}. The difference is not too large for values of $Q_*\gg 1$ or $Q_* \ll 1$. However, for intermediate values of $Q_*\lesssim 1$, the difference becomes important (see Fig.\,\ref{fig:comparison}, left panel). The latter range of $Q_*$ is the relevant one to fit CMB constraints. We illustrate the difference in the plane \{$n_s$,$r$\} (for a quartic potential with a linear dissipation coefficient) between implementing $\Theta$ on the cold-like solution (as it has been done in previous works, see \cite{Bastero-Gil:2016qru}) or in the homogeneous solution--as our quantization procedure indicates--in Fig.\,\ref{fig:comparison_nBE_older}. 
\begin{figure}[t!]
    \begin{center}
  \includegraphics[width=.40\textwidth]{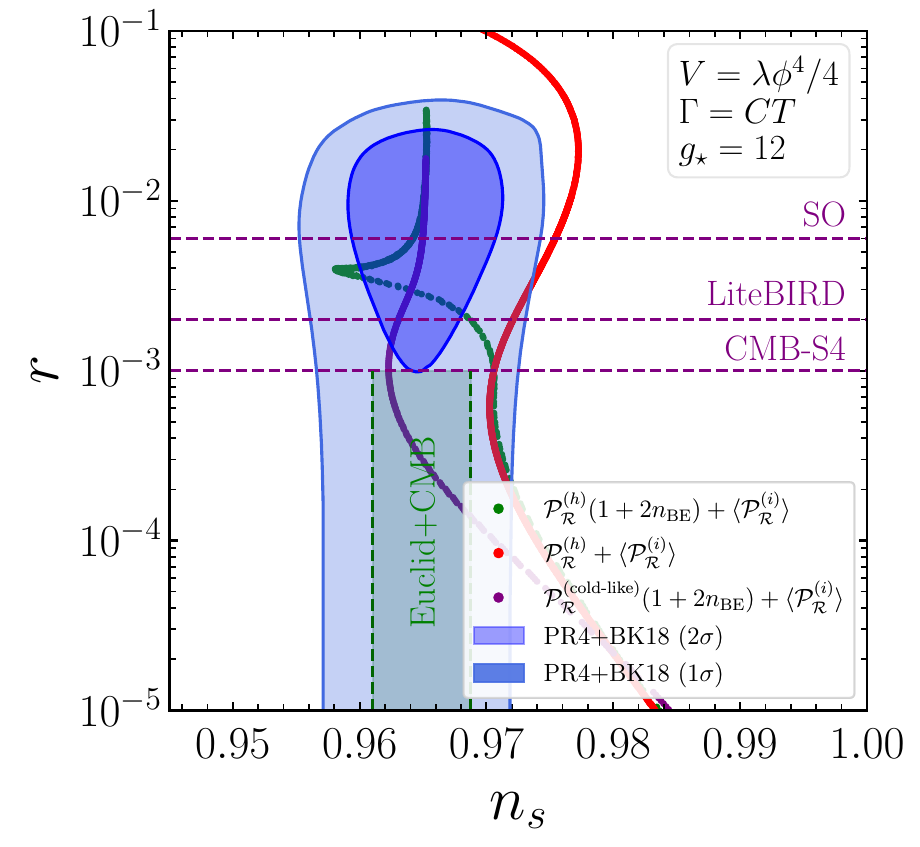}
\caption{\it {Scalar spectral index and tensor-to-scalar ratio for a quartic potential with linear dissipative coefficient obtained by computing $\langle\mathcal{P}_\mathcal{R}\rangle$ in three different ways. Each point corresponds to a combination of $(C, \lambda)$ which satisfies the CMB constraints on $A_s$. The power spectrum for purple points is computed assuming a Bose-Einstein enhancement of the cold-like solution, i.e.\, ${\langle\mathcal{P}_\mathcal{R}\rangle = \mathcal{P}_\mathcal{R}^{\text{(cold-like)}}(1+2n_{\text{BE}})+\mathcal{P}_\mathcal{R}^{(i)}}$, see Fig.\ 2 of \cite{Bastero-Gil:2016qru}. The power spectrum for green points is computed assuming an analogous enhancement in the homogeneous solution, ${\langle\mathcal{P}_\mathcal{R}\rangle = \mathcal{P}_\mathcal{R}^{(h)}(1+2n_{\text{BE}})+\mathcal{P}_\mathcal{R}^{(i)}}$, as our quantization procedure would require, if inflaton perturbations actually thermalize. If no Bose-Einstein enhancement is present (as we assume through this work for the numerical analysis of the various models we consider) the corresponding result is shown with red dots, which is the same curve as the one in Fig.\,\ref{fig:lineargammag} (upper center panel).}}
\label{fig:comparison_nBE_older}
\end{center}
\end{figure}
\begin{figure}[t!]
    \begin{center}
  \includegraphics[width=1.0\textwidth]{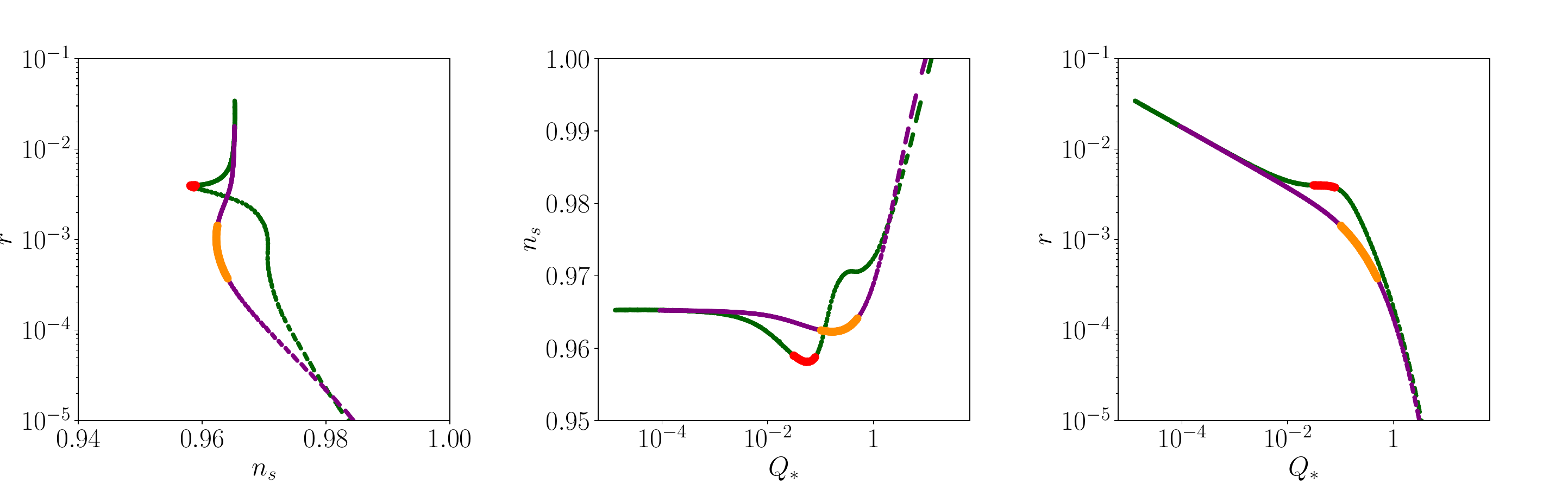}
\caption{\it {\textbf{Left}: Each point corresponds to a combination of $(C, \lambda)$ which satisfies the CMB constraints on $A_s$. Like in Fig.\,\ref{fig:comparison_nBE_older}, the power spectrum for the purple points is computed as ${\langle\mathcal{P}_\mathcal{R}\rangle = \mathcal{P}_\mathcal{R}^{\text{(cold-like)}}(1+2n_{\text{BE}})+\mathcal{P}_\mathcal{R}^{(i)}}$, while the power spectrum for the green points is computed as ${\langle\mathcal{P}_\mathcal{R}\rangle = \mathcal{P}_\mathcal{R}^{(h)}(1+2n_{\text{BE}})+\mathcal{P}_\mathcal{R}^{(i)}}$. The same convention holds for the rest of the panels. The regions in which the behaviour of $n_s$ with $r$ changes are indicated in red and orange. \textbf{Center}: projection of the points in the left panel onto the {$\{Q_*, n_s\}$} plane. \textbf{Right}: Analogous, but projecting onto the {$\{Q_*,r\}$} plane. }}
\label{fig:parmetric_curves_ns_r}
\end{center}
\end{figure}
The different shape in the two cases can be understood with the aid of 
Fig.\,\ref{fig:parmetric_curves_ns_r}. The right and center panels of the latter show how $r$ barely changes with $Q_*$ while $n_s$ goes through its minimum if $\Theta$ is implemented on the homogeneous solution. This translates into a ``peaky'' turnaround at $r\approx 4\times 10^{-3}$ in the \{$n_s$,$r$\}  plane. Computing the curvature power spectrum as ${\langle\mathcal{P}_\mathcal{R}\rangle = \mathcal{P}_\mathcal{R}^{\text{(cold-like)}}(1+2n_{\text{BE}})+\mathcal{P}_\mathcal{R}^{(i)}}$ makes it a growing function of $Q_*$, and the corresponding $r$ decreases as $Q_*$ grows. However, if we compute the scalar power spectrum as ${\langle\mathcal{P}_\mathcal{R}\rangle = \mathcal{P}_\mathcal{R}^{{(h)}}(1+2n_{\text{BE}})+\mathcal{P}_\mathcal{R}^{(i)}}$, it is not necessarily a growing function of $Q_*$ because $\mathcal{P}_\mathcal{R}^{{(h)}}$ decreases with $Q_*$.  The exact behaviour of $\langle\mathcal{P}_\mathcal{R}\rangle$ as a function of $Q_*$ depends, as explained in detail in Sec.\,\ref{quantumn} and illustrated in the right panel of Fig.\,\ref{fig:comparison}, on the initial conditions for $\delta\phi_k$. Setting Bunch-Davies initial conditions 5 e-folds before CMB scales cross the horizon, $\langle\mathcal{P}_\mathcal{R}\rangle$ is nearly constant for $Q\approx \mathcal{O}(0.1)$, and so is $r$.  \par \medskip

}

\section{Analytic estimates}
\label{sec:analyticalestimates}

In this section, we consider a simplified version of the equations for the background (Sec.\,\ref{s:bck}) and the perturbations (Sec.\,\ref{s:pert}) which can be treated fully analytically. We solve them to obtain a purely analytical estimate of $A_s$ and $n_s$ which can be used to understand the main features of the numerical results shown in Sec.\,\ref{ss:pheno}. We also discuss the quantization of inflaton perturbations sourced by the thermal noise.

\subsection{Background in slow-roll}

We define the usual (potential) slow-roll parameters for convenience:
\begin{align}
\epsilon_V=\frac{\mm^2}{2}\left(\frac{V_{\phi}}{V}\right)^2\,,\quad \eta_V = \mn^2\frac{V_{\phi\phi}}{V}\,.
\end{align}
In the slow-roll approximation, the background equations of motion \eq{fried} and \eq{eq:KGequation} are
\begin{align} \label{srA}
\phi' = -\frac{\sqrt{2\epsilon_V}}{1+Q}\mm\,,\quad
\rr = \frac{Q\,\epsilon_V}{2(1+Q)^2}V\,.
\end{align}
As in the previous section, we consider a model of warm inflation described by the dissipation rate and potential
\begin{align}
\Gamma(\phi, T) = C\,\mn\,\left(\frac{\phi}{\mn}\right)^\alpha \left(\frac{T}{\mn}\right)^\beta\,,  \,\,  \beta\geq 1 \quad {\rm and}\quad
V(\phi) = \frac{\lambda}{n} \mm^{4} \left(\frac{\phi}{\mn}\right)^n\,, \,\,  n> 1\,.
\end{align} 
Using the slow-roll attractor \eq{srA}, we obtain the following two identities:
\begin{align}\label{ec:n22}
\left(\frac{\phi}{\mn}\right)^\vartheta=\frac{2\pi^2g_\star}{15\,n\,\lambda}\left(\frac{3\lambda}{n\,C^2}\right)^{2/\beta} \frac{(1+Q)^2}{Q^{1-4/\beta}	}\,, \quad {\rm where} \quad
\vartheta \equiv n\left(1-\frac{2}{\beta}\right)-2\left(1-2\frac{\alpha}{\beta}\right)\,,
\end{align}
and
\begin{align}
\frac{4-\beta+(4+\beta)Q}{2Q} \frac{\diff Q}{\diff N}=2\ev-2\sqrt{2\epsilon_V}\alpha \frac{\mn}{\phi} - \beta(\etv-\ev)	\,.\label{ec:master2}
\end{align}
If $\vartheta = 0$, then $Q$ is a constant. If instead $\vartheta\neq 0$, we can combine \eq{ec:n22} and \eq{ec:master2} to get a differential equation for $Q$:
\begin{equation}\label{ec:phiasq}
\frac{\diff Q}{\diff N} = \left(\frac{15\,n\,\lambda}{2\pi^2\,g_\star}\left(\frac{n\, C^2}{3\lambda\,Q^2}\right)^{2/\beta}\frac{Q}{(1+Q)^2}\right)^{2/\vartheta}    \frac{n\,\beta\,\vartheta\,Q}{\beta-4-(4+\beta)Q}\,.
\end{equation}
In the following, we estimate the average over realizations of the thermal noise (stochastic average) of the power spectrum $\langle\PP_\RR\rangle$ by solving the system of Eqs.\,\eqref{ec:b01}-\eqref{ec:b03} with these slow-roll approximations. To do so, we will first estimate $\ddr$. Then, we will estimate $\ddp$, which is partially sourced by $\ddr$, and compute $\langle\mathcal{P}_{\delp}\rangle$. Finally, we will relate $\ddp$ and $\mathcal{R}$ to compute $\langle\mathcal{P}_\mathcal{R}\rangle$.

\subsection{An approximation for $\delta \rho_r$}\label{ss:dr}

Neglecting in \eqref{ec:b02} the slow-roll suppressed terms (i.e. terms suppressed by $\epsilon$ or $\epsilon^{1/2}$) and taking the super-horizon limit ($k/(aH)\rightarrow 0$) we get\footnote{Notice that we introduce subscripts to indicate the comoving momentum dependence of perturbation variables.}
\begin{equation}\label{ec:b071}
\ddr'+(4-\beta)\ddr = -\sqrt{\frac{2\Gamma\, T\, H}{a^3}}\phi'\xi_{\bm k}\,.
\end{equation}

\subsubsection{The linear dissipation case}
In the linear dissipation case ($\beta=1$), the solution of \eqref{ec:b071} is
\begin{equation}\label{ec:b13}
\ddr(N) \simeq c_1e^{-3N} - a_0^{-3/2}e^{-3N}\sqrt{2 \Gamma\, T\, H}\ \phi'\int_{-\infty}^{N} e^{3\tilde N/2}\,\xi_{\bk}(\tilde N)\, \diff \tilde N\,,
\end{equation}
where $c_1$ is an integration constant. To get this solution we have approximated $a(N)=a_0e^N$ (constant $H$) and used that $\sqrt{2\Gamma\, T\, H}\phi'$ is approximately constant during slow-roll (evaluating it at horizon crossing). The equal-time correlator is then
\begin{equation}\label{ec:b15}
\begin{split}
\langle \ddr(N) \ddrp(N)\rangle =\left(c_1^2e^{-3N} + \frac{2}{3}\Gamma\, T\, H\,\phi'^2 a_0^{-3} \right)e^{-3N}\delta(\bk+\bk')\,,
\end{split}
\end{equation}
where we have used $\langle\xi_{\bk}(N)\xi_{\bk'}(\tilde N)\rangle=\delta(N-\tilde N)\delta(\bk+\bk')$ and $\langle\xi_{\bk}(N)\rangle=0$ with $\langle... \rangle$ representing the average over stochastic realizations of the noise. The homogeneous solution is exponentially suppressed with respect to the inhomogeneous one and we neglect it, so that:
\begin{align}
\ddr(N) &= - a^{-3}\sqrt{2 \Gamma\, T\, H}\phi'\int_{-\infty}^{N} a^{3/2}(\tilde N)\,	 \xi_{\bk}(\tilde N)\, \diff \tilde N\,, \label{ec:b19}\\
\langle \ddr(N) \ddrp(N)\rangle	 &= \frac{8}{3} \rr\, T\, a^{-3}(N)\ \delta(\bk+\bk')\,.\label{ec:b20}
\end{align}

\subsubsection{The cubic dissipation case}

The analytical description of the cubic dissipation case ($\beta=3$) is slightly more complicated. The solution of \eqref{ec:b071} is
\begin{equation}\label{ec:w2}
\ddr(N) \simeq c_1a^{-1}{(N)} + a_0^{-3/2}e^{-N}\sqrt{2\Gamma\, T\, H}\phi'\int_N^\infty e^{-\tilde N/2}\,\xi_{\bk}(\tilde N)\, \diff \tilde N\,.
\end{equation}
The equal-time correlator reads
\begin{equation}\label{ec:w5}
\begin{split}
\langle\ddr(N)\ddrp(N)\rangle = \left[c_1^2a^{-2}(N) +  2\Gamma\, T\,H\,\phi'^2a^{-3}(N)\right]\delta(\bk+\bk')\,.
\end{split}
\end{equation}
Unlike in the linear case, the inhomogeneous solution is suppressed for large $a$. Hence, asymptotically $\langle\ddr(N)\ddr(N)\rangle \sim a^{-2}(N)$. In order to approximate the full solution by the homogeneous term, we need to determine $c_1$. However, our simplified description tells us nothing about the boundary conditions of $\ddr(N)$. From numerical calculations, we notice that the inhomogeneous contribution to the solution appears to be a reasonable approximation for the solution up to $N_{\text{cut}}=N_{\text{horizon}}-\ddnc$, for some (model-dependent) $\ddnc>0$.\footnote{A similar observation and reasoning was made in \cite{Ballesteros:2022hjk}, where the $T$ dependence of $\Gamma$ was also cubic.} Matching the homogeneous and inhomogeneous terms at $N_{\text{cut}}$, we have
\begin{equation}\label{ec:w6}
c_1^2 \simeq
2\Gamma\,T\,H\,\phi'^2\frac{H}{k}e^{\ddnc}\,,
\end{equation}
where we have again used that background quantities are approximately constant during slow-roll. We therefore write $\ddr$ for $N>N_{\text{cut}}$ as
\begin{equation}
\ddr(N)= \sqrt{\frac{2\Gamma\, T\, H^2\phi'^2}{k}}e^{\ddnc/2}a^{-1}(N)= \sqrt{\frac{8H\,\rho_r\, T}{k}}e^{\ddnc/2}a^{-1}(N)\,,\label{ec:b23.9}
\end{equation}
and the equal-time correlator as
\begin{align}
\langle\ddr(N)\ddrp(N)\rangle  = \frac{8H\,\rho_r\, T}{k}e^{\ddnc}a^{-2}(N)\delta(\bk+\bk')\,.
\end{align}

\subsection{An approximation for $\PP_{\delta\phi}$}\label{ss:approx_deltaphi}

Neglecting slow-roll suppressed terms, Eq.\,\eqref{ec:b01} reads
\begin{equation}\label{ec:b24}
\ddp''+\left(3+\frac{\Gamma}{H}\right)\ddp'+\frac{k^2}{(aH)^2}\ddp+\Gamma_{T}\frac{\phi'\,T}{4H\,\rho_r}\ddr=\sqrt{\frac{2\Gamma\, T}{a^3H^3}}\xi_\bk\,.
\end{equation}
The fourth term in the left-hand side can be rewritten by evaluating the prefactor in the slow-roll attractor and substituting $\ddr$ by \eqref{ec:b19}, \eqref{ec:b23.9}. We then have
\begin{align}\label{ec:b25}
\Gamma_{T}\frac{\phi'T}{4H\rho_r}\ddr &=  -\sqrt{\frac{2\Gamma\, T}{H^3}}a^{-3}\int_{-\infty}^Na^{3/2}(\tilde N)\, \xi_{\bm{k}}(\tilde N)\, \diff \tilde N \quad \text{(linear dissipation)}\,,\\
\Gamma_{T}\frac{\phi'T}{4H\rho_r}\ddr &=  \frac{3}{aH}\sqrt{\frac{2\Gamma\, T}{k}}e^{\ddnc/2} \quad \text{(cubic dissipation)}\,,
\end{align}
Substituting in \eqref{ec:b24} and defining a new time variable $z=k/(aH)$ we have\footnote{Recall that, by virtue of Ito's rule, changes of variables act on the noise as $\xi(N)=\sqrt{\diff z / \diff N }\xi(z)$, see \cite{Ballesteros:2022hjk}.}
\begin{align}
z^2\frac{\diff^2 \ddp}{\diff z^2} + z(1-2\nu)\frac{\diff\ddp}{\diff z} + z^2  \ddp &= \sqrt{\frac{2\Gamma T}{k^3}}\left(z^2\xi_{\bm k} +  z^3\int_z^\infty d\tilde{z}\,\tilde{z}^{-2}\, \xi_{\bm{k}} (\tilde z)\right) \quad \text{(linear)}\,, \label{ec:b32}\\
z^2\frac{\diff^2 \ddp}{\diff z^2} + z(1-2\nu)\frac{\diff \ddp}{\diff z}  + z^2  \ddp &= \sqrt{\frac{2\Gamma T}{k^3}}\left(z^{2}\xi_{\bm k} -3e^{\ddnc/2}z\right)  \quad \text{(cubic)}\,, \label{ec:b32.1}
\end{align}
where 
\begin{align} \label{defnu}
\nu\equiv\frac{3}{2}(1+Q)\,.
\end{align} 
The solutions to these equations can be written as $\ddp = \delp_\bk^{(h)} + \delp_\bk^{(i)}$, where $\delp_\bk^{(h)}$ is the general solution of the reduced (homogeneous) equation, and $\delp_\bk^{(i)}$ is a particular solution to the full (inhomogeneous) equation constructed with the retarded Green's function. As we shall proof in Sec.\,\ref{ss:quan}, the time-dependent power spectrum (averaged over stochastic realizations) of inflaton perturbations is given by the sum $
\langle\PP_{\delp}(z) \rangle = 
\PP_{\delp}^{(h)}(z) + 
\langle\PP_{\delp}^{(i)}(z)\rangle$, where
\begin{equation}\label{ec:b34a}
 \PP_{\delp}^{(h)}(k,z) = \frac{k^3}{2\pi^2}|\delp_k^{(h)}(z)|^2 \quad \text{and} \quad \delta(\bk+\bk')\langle \PP_{\delp}^{(i)}(k,z) \rangle =\frac{k^3}{2\pi^2}\langle\ddp^{(i)}(z)\ddpp^{(i)}(z) \rangle\,.
\end{equation}
The power spectrum at the end of inflation is obtained by taking the $z\to 0$ limit in \eqref{ec:b34a}.

\subsubsection{The homogeneous solution}\label{ss:homo}
The reduced equation of both \eqref{ec:b32} and \eqref{ec:b32.1} reads
\begin{equation}\label{ec:b35}
z^2\frac{\diff^2\delp_k^{(h)}}{\diff z^2} + z(1-2\nu)\frac{\diff \delp_k^{(h)}}{\diff z} + z^2  \delp_k^{(h)}= 0\,.
\end{equation}
Notice that the homogeneous solution only depends on $k=|\bk|$. Introducing a more convenient variable $y_k(z) \equiv \delp_k(z)/z^\nu$ we can rewrite \eqref{ec:b35} as canonical Bessel equation. Any solution to this equation can be expressed as
\begin{equation}\label{ec:b42}
y_k(z)=A_k J_\nu(z)+B_k Y_\nu(z), \quad \text{and} \quad \delp_k^{(h)}(z)=z^{\nu}\left[A_k J_\nu(z)+B_k Y_\nu(z)\right]\,,
\end{equation}
where $J_\nu$, $Y_\nu$ are the Bessel functions of the first and second kind, respectively. The constants $A_k$ and $B_k$ ought to be determined from the boundary conditions, which we discuss next. From \eqref{ec:b34a}, we have that the homogeneous contribution to the inflaton power spectrum  at the end of inflation is
\begin{equation}
\PP_{\delp}^{(h)} = \lim_{z\to 0}\frac{k^3}{2\pi^2}|\delp_k^{(h)}(z)|^2 = \frac{k^3}{2\pi^2}|B_k|^2\frac{2^{2\nu}\Gamma_E(\nu)^2}{\pi^2}\, ,
\label{eq:homoegenoust}
\end{equation}
where $\Gamma_E$ is the Euler gamma function.
\paragraph*{Boundary conditions for the homogeneous solution.}
Deep inside the horizon (i.e.\ in the $z\to\infty$ limit), the two independent solutions to $\delp_k^{(h)}$ behave as \cite{abramowitz1948handbook}
\begin{equation}
{z^\nu}J_\nu(z) \sim {z^\nu}\sqrt{\frac{2}{\pi z}}\cos\left(z-\frac{\nu\pi}{2}-\frac{\pi}{4}\right), \quad
{z^\nu}Y_\nu(z) \sim {z^\nu}\sqrt{\frac{2}{\pi z}}\sin\left(z-\frac{\nu\pi}{2}-\frac{\pi}{4}\right)\,.
\end{equation} 
Specific boundary conditions are implemented by the choice of the constants $A_k$ and $B_k$ in \eqref{ec:b42}. For instance, in the cold limit ($\nu=3/2$, see the definition \eq{defnu}), the usual Bunch-Davies boundary conditions are given by $A_k=\sqrt{\frac{H^2\pi}{4k^3}}$, $B_k=iA_k$. Let us consider a generic warm case with $\nu=3/2(1+Q)$, $Q>0$. In the asymptotic past $(z\rightarrow \infty)$, the solution behaves as 
\begin{equation}\label{ec:b64}
\delp_k^{(h)}(z) \sim z^{3/2(1+Q)}
{\frac{e^{iz}}{\sqrt{z}}
}
\sim z^{1+3Q/2}e^{iz}\,,
\end{equation}
while in the asymptotic future $(z\rightarrow 0)$, one has \cite{abramowitz1948handbook}
\begin{equation}\label{ec:b641}
\delp_k^{(h)}(z) \sim {z^\nu}Y_\nu(z) \sim  2^\nu\Gamma_E(\nu)\,.
\end{equation}
We observe that:
\begin{enumerate}
\item The early-time limit is incompatible with Bunch-Davies boundary conditions. Indeed, Eq.\,\eqref{ec:b64} is a factor $z^{3Q/2} \neq 1$ away from it.
\item The late-time perturbations (and therefore, the power spectrum imprinted on observable perturbations) grow super-exponentially with $\nu$ as per \eqref{ec:b641}, compromising the perturbative character of the perturbations.
\end{enumerate}
Both puzzles are solved if we assume that inflation was cold in the asymptotic past, and interactions with the thermal bath appeared at some time scale $z_0$. This solves, by construction, point 1 (since Eq.\,\eqref{ec:b64} for $Q=0$ corresponds to Bunch-Davies boundary conditions). Regarding point 2, let us assume the coefficients of the solution were those imposed by Bunch-Davies in cold inflation, $A_k^-=\sqrt{\frac{H^2\pi}{4k^3}}$, $B_k^-=i\sqrt{\frac{H^2\pi}{4k^3}}$. For $z<z_0$, continuity of $\delp_k$ and $\delp_k'$ imposes that the coefficients of $\delp_k^{(h)}(z)$ are
\begin{align}
A_k^+ &= -\sqrt{\frac{H^2\pi^2}{8k^3}}e^{iz_0} z_0^{1-\nu}\left[iz_0Y_\nu(z_0)+(i+z_0)Y_{\nu-1}(z_0)\right]\,, \label{ec:joint1}\\ 
B_k^+ &= \sqrt{\frac{H^2\pi^2}{8k^3}}e^{iz_0}z_0^{1-\nu}\left[-iz_0J_\nu(z_0)+(i+z_0)J_{\nu-1}(z_0)\right]\,.\label{ec:joint2}
\end{align}
In the asymptotic future ($z\to 0$), this matching has the following effects:
\begin{itemize}
\item For strong dissipation ($\nu\gg3/2$),
\begin{equation}\label{ec:b68}
\PP_{\delp}^{(h)}(k) = \frac{k^3}{2\pi^2}|B_k^+|^2\frac{2^{2\nu}\Gamma_E(\nu)^2}{\pi^2}\simeq \frac{k^3}{2\pi^2}|B_k^+|^2\frac{2^{2\nu}}{\pi^2}\frac{2\pi}{\nu}\left(\frac{\nu}{e}\right)^{2\nu} \sim \frac{{z_0}^{3}}{\nu}\left(\frac{2\nu}{z_0 e}\right)^{2\nu}\,,
\end{equation}
where we have used the Stirling approximation for $\Gamma_E$ and the asymptotic expansion of Bessel functions for small $z$ and large $z_0$. We see that the homogeneous solution is now exponentially suppressed as long as $z_0>\frac{2\nu}{e}$ (which is a reasonable condition, since $z_0$ is a time in the far past, i.e. $z_0 \gg1$). Not only does this remove the super-exponential divergence in $\nu$, but it also suppresses the homogeneous solution with respect to the inhomogeneous one, ensuring that perturbations are brought to a thermal attractor (this was already noticed in \cite{LopezNacir:2011kk} and discussed in \cite{Ballesteros:2022hjk}).
\item For weak dissipation ($\nu\gtrsim 3/2$), $3Q/2 = \nu-3/2\ll 1$. As seen in \eqref{ec:b68}, $B_k^+$ is the only coefficient contributing to the power spectrum. Expanding $B_k^+$ in powers of $Q$, we obtain:
\begin{equation}\label{ec:b70}
B_k^+ = B_k^{-}(1- \mathcal{O}(\log z_0)Q + \mathcal{O}(Q^2))\,.
\end{equation}
Therefore, the power spectrum for small $Q$ is perturbatively close to that in cold inflation (as expected), and the dependence on the transition scale $z_0$ is suppressed by both $Q$ and the presence of the logarithm. 
\end{itemize}
In practice, following \eqref{ec:b70} we can pick $B_k=i\sqrt{\frac{H^2\pi}{4k^3}}$ when $\nu \gtrsim 3/2$, making
\begin{equation}
\PP_{\delp}^{(h)}(k)  = \frac{k^3}{2\pi^2}\frac{H^2\pi}{4k^3}\frac{2^{2\nu}\Gamma_E(\nu)^2}{\pi^2}=\frac{H^2\Gamma_E(\nu)^22^{2\nu}}{8\pi^3}\,,
\end{equation}
and set $\PP_{\delp}^{(h)}(k)  =0$ by fiat when $\nu \gg 3/2$. This ensures that we neglect the homogeneous solution in the limit in which it is exponentially suppressed, and that we naturally recover the cold inflation result in the $\nu\to 3/2$ limit.
\subsubsection{The inhomogeneous solution}

\paragraph*{Linear dissipation.}

The retarded Green's function associated to \eqref{ec:b32} is
\begin{equation}\label{ec:b44}
\tilde{G}(z,w) = \frac{\pi}{2}G(z,w), \quad G(z,w)= z^\nu w^{-(1+\nu)}\left[J_\nu(z)Y_\nu(w)-J_\nu(w)Y_\nu(z)\right]\Theta(z-w)\,.
\end{equation}
A particular solution to \eqref{ec:b32} is therefore 
\begin{equation}\label{ec:b46}
 \delp_\bk^{(i)}(z) = \frac{\pi}{2}\sqrt{\frac{2\Gamma T}{k^3}}\int_z^\infty \diff w \ G(z,w)\left[w^2\xi_{\bm k}(w)+w^3\int_w^\infty \diff z_1 \ z_1^{-2}\xi_{\bm k}(z_1)\right]\,.
\end{equation}
Notice that $ \delp_\bk^{(i)}(z)$ depends of $\bk$ through the noise; the Green's function only depends on $k$ (through the time variable $z$). We now use Eq.\,\eqref{ec:b34a} to compute $\PP_{\delta\phi}^{(i)}(k,z)$. Notice that the dependence on $\bk$ of the equal-time correlator only appears in $\delta(\bk+\bk')$; the dimensionless power spectrum only depends on its modulus $k$. Indeed,
\begin{multline}\label{ec:b461}
\langle \PP_{\delta\phi}^{(i)}(k,z) \rangle=\frac{k^3}{2\pi^2}\times \frac{\pi^2}{4}\frac{2\Gamma T}{k^3}\int_z^\infty \diff w_1\int_z^\infty \diff w_2 G(z,w_1) G(z,w_2) \\\times
\left[w_1^2w_2^2\delta(w_1-w_2)
+ w_1^2w_2^3\int_{w_2}^\infty \diff z_1 \ \frac{1}{z_1^2}\delta(w_1-z_1)+w_2^2w_1^3\int_{w_1}^\infty \diff z_2\frac{1}{z_2^2}\delta(w_2-z_2)\right.\\
\quad \left.+w_1^3w_2^3 \int_{w_1}^\infty \diff z_1\int_{w_2}^\infty \diff z_2 \frac{1}{z_1^2}\frac{1}{z_2^2} \delta(z_1-z_2)\right]\,,
\end{multline}
where the Dirac deltas in time variables arise from the correlators of the noise. The second term in brackets can be rewritten as
\begin{equation}\label{ec:b48}
 w_1^2w_2^3\int_{w_2}^\infty \ \diff z_1\frac{1}{z_1^2} \delta(z_1-w_1) 
= w_1^2w_2^3 \times \frac{1}{w_1^2}\Theta(w_1-w_2) = w_2^3\Theta(w_1-w_2)\,,
\end{equation}
and analogously for the third term, where $\Theta$ is the Heaviside step function. If we now define
\begin{align}
F_1(z)&\equiv\int_z^\infty \diff w_1 \int_z^\infty \diff w_2 \ G(z,w_1)G(z,w_2)w_1^2w_2^2
\delta(w_1-w_2) = \int_z^\infty \diff w_1 \ G(z,w_1)^2 w_1^4 \\
F_2(z)&\equiv\int_z^\infty \diff w_2 \ G(z,w_2)w_2^3\int_{w_2}^\infty \diff w_1 G(z,w_1)\,,
\end{align}
Eq.\,\eqref{ec:b461} can be written as
\begin{equation}\label{ec:b49}
\langle \PP_{\delta\phi}^{(i)}(k,z) \rangle = \frac{\Gamma T}{4}\left(F_1(z) +\frac{8}{3}F_2(z)\right) \implies \langle \PP_{\delta\phi}^{(i)}(k) \rangle=\frac{\Gamma T}{4}\left [F_1(0) + \frac{8}{3}F_2(0)\right]\,.
\end{equation}

\paragraph*{Cubic dissipation.}

Since the difference between the linear and cubic dissipation cases lies only in the source term for $\ddp$, the Green's function is the same in both cases. Using again \eqref{ec:b34a},
\begin{equation}\label{ec:i2}
\langle \PP_{\delta\phi}^{(i)}(k,z) \rangle = \frac{k^3}{2\pi^2}\frac{\pi^2}{4}\frac{2\Gamma T}{k^3}\left[\int_z^{\infty}\diff  w \ G(z,w)^2w^{4} +9e^{\ddnc}\left(\int_{z}^{\infty}{ \diff  w \ G(z,w)w}\right)^2\right]\,.
\end{equation}
Defining 
\begin{equation}
F_3(z)\equiv\left(\int_{z}^{\infty}{ \diff w \ G(z,w)w}\right)^2\,,
\end{equation}
we can write
\begin{equation}\label{ec:b58}
\langle \PP_{\delta\phi}^{(i)}(k) \rangle= \lim_{z\to 0} \frac{\Gamma T}{4}\left [F_1(z) + 9e^{\ddnc}F_3(z)\right]=\frac{\Gamma T}{4}\left [F_1(0) + 9e^{\ddnc}F_3(0)\right]\,.
\end{equation}

\subsection{Scalar power spectrum and spectral index}\label{ss:rr}

So far, we have computed the power spectrum for the inflaton perturbation. To obtain the curvature power spectrum, we start from the definition of $\RR$,
\begin{equation}\label{ec:b72}
\RR = \frac{H}{p+\rho}(\delta q_r - H\phi'\delta\phi) - \varphi.
\end{equation}
Writing $\delta q_r$ in terms of the other variables using \eqref{ec:b04} and keeping only terms proportional to $\delta\phi$ (which are found numerically to be dominant), we have $\delta q_r \simeq \left(V_{\phi}/(3H)+H\phi'\right)\delta\phi$. Substituting in \eqref{ec:b72} and neglecting the metric perturbation (which is subdominant)
\begin{equation}\label{ec:b74}
\RR \simeq  \frac{1}{\mm} \frac{1+Q}{\sqrt{2\epsilon_V}}\delta\phi,
\end{equation}
and therefore
\begin{equation}
\langle \mathcal{P}_\mathcal{R} \rangle  =  \frac{1}{\mm^2} \frac{(1+Q)^2}{2\epsilon_V}\left(\PP_{\delp}^{(h)}+\langle \PP_{\delp}^{(i)}  \rangle\right).
\end{equation}
where
\begin{equation}\label{ec:b76}
\PP_{\delp}^{(h)}=\begin{cases}
\frac{H^2}{8\pi}\frac{2^{2\nu}\Gamma_E(\nu)^2}{\pi^2}, \quad &Q_*<Q_0,\\
0, \quad &Q_*>Q_0,
\end{cases}
\end{equation}
and $\langle \PP_{\delp}^{(i)} \rangle$ is given in \eqref{ec:b49} (linear dissipation) and \eqref{ec:b58} (cubic dissipation). In \eqref{ec:b76}, $Q_0$ is the cutoff we impose on the homogeneous contribution as discussed at the end of Sec.\,\ref{ss:homo}. We find $Q_0 \simeq 0.2$ is a good choice. Notice how, in the cold limit $Q\to 0$, the inhomogeneous contribution to the spectrum vanishes and the homogeneous one reduces to the standard cold inflation result. Let us introduce the compact notation
\begin{equation}\label{ec:b76.0}
\langle \PP_\RR \rangle = (\mathcal{G} + \mathcal{W}F){\mathcal{Y}}\,,
\end{equation}
where
\begin{equation}\label{ec:b76.1}
\mathcal{W}=\frac{\Gamma T}{4\mn^2}\,,\quad {\mathcal{Y}}=\frac{(1+Q)^2}{2\epsilon_V}\,, \quad \mathcal{G} = \begin{cases}
\frac{2^{2\nu-3}\Gamma_E(\nu)^2}{\pi^3 }\frac{H^2}{\mn^2} \,, &\ Q<Q_0\\
0, &\ Q>Q_0
\end{cases}\,.
\end{equation}
and
\begin{equation}\label{ec:F}
F = \begin{cases} F_1(0)+\frac{8}{3}F_2(0) \quad \text{(linear dissipation)}\\
F_1(0) + 9e^{\ddnc}F_3(0) \quad \text{(cubic dissipation)} \end{cases}
\end{equation}
As it is usually done when computing spectra in slow-roll, all background quantities (i.e. the ones in \eqref{ec:b76.1}) are considered constant during the calculation, with their value fixed at horizon crossing for each mode. This implicitly fixes the scale dependence of the power spectrum and allows to derive the spectral index from the evolution of the background.

\paragraph*{Spectral index when $\vartheta \neq 0$.}

Since $\phi$ can be written as a function of $Q$ using \eqref{ec:phiasq}, every background quantity from \eqref{ec:b76.1} can be expressed as a function of $Q$ using the attractor equations:
\begin{align}
H^2 = \frac{V}{3\mm^2}\,,\quad {\mathcal{Y}} = \frac{(1+Q)^2}{2\epsilon_V}\,,\quad \Gamma = \frac{\sqrt{3\,VQ^2}}{\mm}\,,\quad
T=\left(\frac{30\,\rr}{\pi^2\,g_\star}\right)^{1/4}\,,\quad  \rr=\frac{Q\,\epsilon_V}{2(1+Q)^2}\,V\,.\label{ec:zz2}
\end{align}
We compute the spectral index using
\begin{equation}\label{ec:b86a}
n_s-1=\left.\frac{\diff\log \langle \PP_\RR \rangle}{\diff \log k}\right|_{k=k_*} = \left.\frac{\diff \log \langle \PP_\RR \rangle}{\diff Q}\right|_{Q=Q_*}\left.\frac{\diff Q}{\diff N}\right|_{Q=Q_*}\left.\frac{\diff N}{\diff \log k}\right|_{k=k_*}\,.
\end{equation}
The last factor is ${\diff N}/{\diff \log k}=  1/(1-\epsilon) \simeq 1$. The second factor is given by \eqref{ec:phiasq}. The first factor can be computed from \eqref{ec:b76.0}, and \eqref{ec:b76.1}. We have
\begin{equation}\label{ec:b911}
\frac{\diff \log \langle \PP_\RR \rangle}{\diff Q} = \frac{\diff\log{\mathcal{Y}}}{\diff Q} + { \frac{1}{\mathcal{G}+\mathcal{W}F} \left( \frac{\diff \mathcal{G}}{\diff Q} + \frac{\diff \mathcal{W}}{\diff Q}F +\mathcal{W}\frac{\diff F}{\diff Q} \right) }.
\end{equation}
The term $\diff F/ \diff Q$ is computed numerically. Every other term follows from \eqref{ec:zz2}:
\begin{align}
\frac{\diff V(Q)}{\diff Q} &= n(4-\beta + (4+\beta)Q)\mathcal{Z}(Q)V(Q)\\
\frac{\diff\log{\mathcal{Y}}}{\diff Q}  &= 2\left [4 - \beta + (4(1 +\alpha) + \beta (n-1 ) - 2 n) Q\right]\,\mathcal{Z}(Q)\,,\\
\frac{\diff \mathcal{W}}{\diff Q} &= 2\alpha(5+ 3Q) +(1+\beta)(n+ 3 n Q - 4 (1 + Q))\frac{\mathcal{W}(Q)}{2}\mathcal{Z}(Q)\,,\\
\frac{\diff \mathcal{G}}{\diff Q} &= \frac{2^{2\nu-3}\Gamma_E(\nu)^2}{\mm^4\pi^3}\left[\frac{1}{3}\frac{\diff V(Q)}{\diff Q}+\left(\log 2 + \frac{\diff \log \Gamma_E(\nu)}{\diff \nu}\right)V(Q)\right]\,,
\end{align}
where $\diff \log \Gamma_E(\nu) / \diff \nu$ is the polygamma function and $\left[\mathcal{Z}(Q)\right]^{-1}= \beta\,\vartheta\,Q(1+Q)\,$.

\paragraph*{Spectral index when $\vartheta = 0$.} The formulae for background quantities in \eqref{ec:zz2} still hold; however, the background value of $\phi$ is now an independent variable, and $Q$ is a constant given by solving \eqref{ec:n22}.  We compute the spectral index using
\begin{equation}\label{ec:b86}
n_s-1=\left.\frac{\diff \log \langle \PP_\RR \rangle}{\diff \log k}\right|_{k=k_*} = \left.\frac{\diff\log \langle \PP_\RR \rangle}{\diff \phi}\right|_{\phi=\phi_*}\left.\frac{\diff \phi}{\diff N}\right|_{\phi=\phi_*}\left.\frac{\diff N}{\diff \log k}\right|_{k=k_*}\,.
\end{equation}
The derivative $\diff \phi/\diff N$ is given by \eqref{srA}. The first factor in \eqref{ec:b86} is
\begin{equation}\label{ec:b86.1}
\frac{\diff\log \langle \PP_\RR \rangle}{\diff \phi} = \frac{\diff\log{ \mathcal{Y}}}{\diff\phi} + { \frac{1}{\mathcal{G}+\mathcal{W}F} \left(\frac{\diff \mathcal{G}}{\diff \phi} + \frac{\diff \mathcal{W}}{\diff \phi}F  \right) } \,.
\end{equation}
Notice that $Q$ is a constant in this case, and therefore there is no derivative of $F$ in the last expression. The terms in \eqref{ec:b86.1} are
\begin{align}
\frac{\diff \log{ \mathcal{Y}}}{\diff  \phi}&=\frac{2}{\phi}\,,\quad
\frac{\diff  \mathcal{G}}{\diff  \phi} =\frac{n}{\phi}\mathcal{G}\,, \quad
\frac{\diff  \mathcal{W}}{\diff  \phi} =\frac{3n-2}{4\phi}\mathcal{W}.
\end{align}

\begin{figure}[t!]
\begin{center}
$\includegraphics[width=0.7\textwidth]{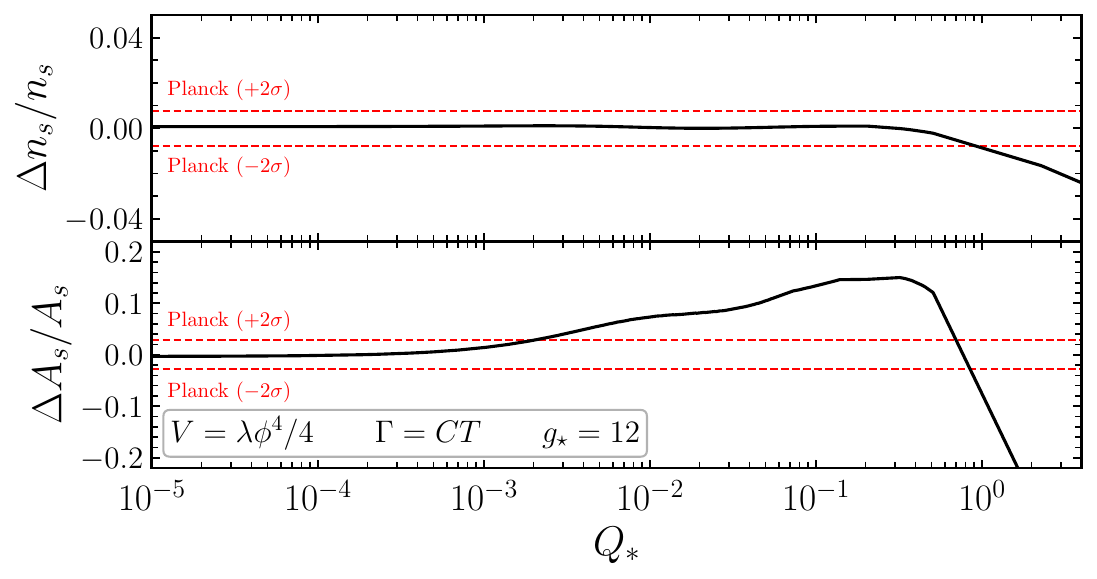}$
\caption{\em \label{fig:compareanalytical} Relative difference between analytical and numerical evaluations of the scalar spectral  index $n_s$ (top panel) and amplitude of scalar power spectrum $A_s$ (bottom panel), as a function of the dissipation coefficient $Q= \Gamma/(3H)$, evaluated at the time the CMB fiducial scale $k_* = 0.05$ Mpc$^{-1}$ crosses outside the horizon during inflation. The quantities on the vertical axis are defined in Eq.\,(\ref{eq:defBigO}). The red dashed lines represent relative uncertainties on $n_s$ and $A_s$, respectively  at $95 \%$ C.L. by Planck.}
\end{center}
\end{figure}

\subsection{Comparison with numerical solutions}
\label{ss:cos}

\noindent
\textbf{Precision of the analytical approach.} In order to quantify the precision of our analytical approach, we compute the relative difference between our analytical approximations for $n_s$ and $A_s$ and the matrix formalism:
\begin{equation}
\dfrac{\Delta \mathcal{O}}{\mathcal{O}} \, \equiv \, \dfrac{\mathcal{O}^\text{num}-\mathcal{O}^\text{ana}}{\mathcal{O}^\text{ana}},\, \qquad \text{for} \qquad \mathcal{O}=\{n_s,A_s\} \,.
\label{eq:defBigO}
\end{equation}
In this expression, $\mathcal{O}^\text{ana}$ is evaluted using our analytical approach, i.e.\ using Eq.\,(\ref{ec:b76.0}) for $A_s$ and Eq.\,(\ref{ec:b86a}) ($\vartheta \neq 0$) or Eq.\,(\ref{ec:b86}) ($\vartheta = 0$) for $n_s$. $\mathcal{O}^\text{num}$ is obtained numerically using the matrix formalism, which is exact, solving Eq.\,(\ref{matrix_eq}).

\begin{figure}[t!]
\begin{center}
$\includegraphics[width=0.7\textwidth]{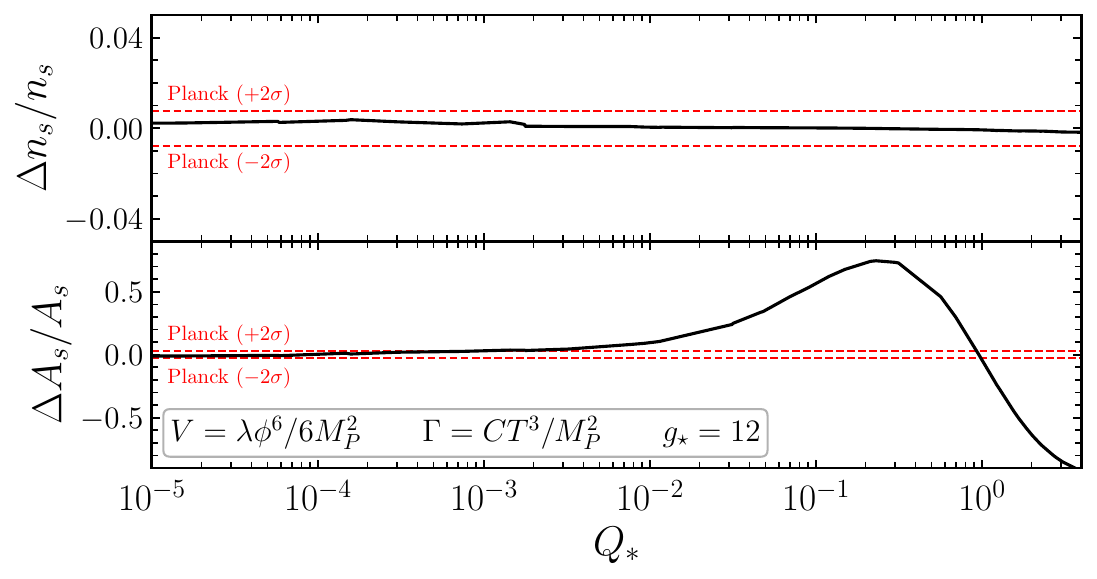}$
\caption{\em \label{fig:compareanalytical2}  {Same as Fig.\,\ref{fig:compareanalytical} but for $V \propto \phi^6$ and $\Gamma \propto T^3$. } }
\end{center}
\end{figure}

In Fig.\,\ref{fig:compareanalytical} and Fig.\,\ref{fig:compareanalytical2}, we represent the quantities defined in Eq.\,(\ref{eq:defBigO}) as a function of $Q_*$, for two examples  with $\vartheta\neq 0$ and $\vartheta= 0$. Our analytical approximations work reasonably well up to $Q_*\simeq \mathcal{O}\left(10^{-1}\right)$. The error incurred in the calculation of $n_s$ is smaller than the observational uncertainty at 95\% confidence level, although the approximation does not work so well for $A_s$ up to such values of $Q_*$. For $Q_*\gtrsim \mathcal{O}\left(10^{-1}\right)$, the couplings between different perturbations become too strong for the simplifications in the equations of $\delta\rho_{r,\bm{k}}$ and $\delta\phi_\bk$ that we have used to hold. \par \medskip

\noindent
\textbf{Parameter dependence for $V = \lambda\phi^4/4$ \& $\Gamma = C\, T$.} In Sec.\,\ref{sec:linear} we explore the parameter dependence of this model numerically. For comparison we use our analytical approximations to obtain a qualitative estimate for $\langle \PP_\RR \rangle$ in terms of $C$, $\lambda$ and $g_\star$ in the region allowed by cosmological data, i.e.\ for $Q_* \sim \mathcal{O}\left(10^{-1}\right)$.  For these values of $Q_*$ the quantity $F$ in \eqref{ec:F} is roughly constant and the homogeneous contribution to the spectrum is already subdominant. Therefore, using \eqref{ec:b76.0}, we can approximate the spectrum as
\begin{equation}
\langle \PP_\RR \rangle \propto { \mathcal{Y}} \frac{\Gamma\, T}{4\mm^2}\propto \frac{C\, \lambda^{1/2}}{g_\star^{1/2}}\left(\frac{\phi}{\mm}\right)^3 (1+Q_*)\ Q_*^{1/2}\propto \frac{C^3}{g_\star}\frac{1}{Q_*}\,,
\end{equation}
where we have used Eqs.\,\eqref{srA} and \eqref{ec:n22}\,. Notice that $Q_*$ itself depends on $\lambda$, $C$ and $g_\star$. The explicit dependence can be obtained by integrating \eqref{ec:phiasq}, which, for $Q_*\sim\mathcal{O}(10^{-1})$ gives approximately
\begin{equation}
N_{k_*} \propto	 \left(\frac{C^4}{g_\star\lambda}\right)^{1/3}  \frac{1}{Q_*}\,,
\end{equation}
which, taking into account that $N_{k_\star}$ is only mildly dependent on the specific combination of $C$, $\lambda$ and $g_\star$ (see Sec.\,\ref{sec:durationinflation}), yields
\begin{equation}
\langle \PP_\RR \rangle \propto \frac{C^{5/3}\lambda^{1/3}}{g_\star^{2/3}}\,.
\end{equation}

\subsection{Quantization of $\delta\phi$}\label{ss:quan}

In this section, we discuss how to quantize a scalar field sourced by white noise in de Sitter spacetime, as it is the case of $\delta\phi$ in warm inflation. As a conclusion of our discussion, we will argue that  the stochastic average of the power spectrum of the field is the sum of the power spectrum associated to the homogeneous solution and the stochastic average of the one associated to the inhomogeneous solution, with no mixing term among them, as stated in \eqref{ec:b34a}. We start by rewriting the equation of motion for $\ddp$ in conformal time $\eta$
\begin{equation}\label{ec:infquan}
\eta^2 \frac{\diff^2\ddp}{\diff \eta^2} + \eta(1-2\nu)\frac{\diff \ddp}{\diff\eta} + k^2\eta^2\ddp = \sigma(\eta)\xi_{\bm k}\,,
\end{equation} 
where $\sigma(\eta)$ is a function of $\eta$ whose explicit form is irrelevant for the argument (for instance, in \eqref{ec:b24} it would be $\sigma(\eta) = \sqrt{2\Gamma\, T/(a\,H)^3}$), and $\xi_{\bm k}$ is a white noise, i.e. $\langle\xi_{\bk}(\eta)\,\xi_{\bk'}(\eta')\rangle{\propto}\delta(\eta-\eta')\delta(\bk+\bk')$. Rescaling the field as 
\begin{equation}
\ddp = \chic\, \eta^{\nu-1/2}\,,
\end{equation}
we obtain 
\begin{equation}
\frac{\diff^2\chic}{\diff\eta^2} +\omega^2_k(\eta)\chic = S_\bk(\eta)\,,
\end{equation}
where
\begin{equation}\label{ec:q13}
S_\bk(\eta)= \eta^{-(\nu+3/2)}\sigma(\eta)\xi_{\bm k} \quad \text{and} \quad \omega^2_k(\eta) = k^2+\frac{1-4\nu^2}{4\eta^2}\,.
\end{equation}
In the cold limit, i.e. $S_\bk(\eta)\to 0$ and $\nu\to 3/2$ (up to slow-roll corrections), we recover the usual Mukhanov-Sasaki equation. 

Let us denote by $v_k$ a complexified solution of the homogeneous equation.\footnote{Given two real independent solutions of the homogeneous equation $\chi_{k,1}$, $\chi_{k,2}$, the complexified solution is defined as $v_k = \chi_{k,1}+i\chi_{k,2}$.} Together with its complex conjugate $v_k^*$, they form a basis of all the solutions to the homogeneous equation. As it corresponds to a harmonic oscillator (even if the frequency is time dependent), the Wronskian of the homogeneous equation $W[v_k, v_k^*]$ is a constant\footnote{Indeed, given $x'' -ax'-bx=0$ and $W$ the Wronskian of two independent solutions, one has $W'=aW$. Hence, if the equation is harmonic-oscillator-like, $W'=0$.} (whose exact value depends, of course, on the specific normalization of the solution). Let us illustrate this with a well-known example. If $\nu$ is constant, two linearly (real) independent solutions can be expressed in terms of Bessel functions of the first and second kind,
\begin{equation}
\chi_{k,1} = \eta^{1/2}J_\nu(k\eta), \quad \chi_{k,2} = \eta^{1/2}Y_\nu(k\eta)\, ,
\end{equation}
and the complexified solution and its complex conjugate are
\begin{equation}
v_k = \eta^{1/2} H^{(1)}_\nu(k\eta), \quad v_k^* = \eta^{1/2} H^{(2)}_\nu(k\eta)\,, 
\end{equation}
whose Wronskian $W[v_k,v_k^*] = -4i/\pi$ is indeed constant (here, $H^{(1)}_\nu$, $H^{(2)}_\nu$ are the Hankel functions of the first and second kind, respectively). 

Let us now consider a particular solution of the inhomogeneous equation, which we denote $s_\bk$. As we have seen, such a solution can be typically written using the retarded Green's function of the equation, which will be constructed with the homogeneous solution. Formally,
\begin{equation}
s_\bk (\eta) = \int \diff \eta'\,G^{(\text{ret})}(k\eta,k\eta')\,S_\bk(\eta')\,.
\end{equation}
We propose the following quantization for the Fourier modes of the field $\chi$:
\begin{equation}\label{ec:qu0}
\opchic =  v_k \,\hat{a}_\bk + v^*_k\, \hat{a}^\dagger_\bk + s_\bk \,\hat{\mathbb{I}}\,,
\end{equation}
where $\hat{\mathbb{I}}$ is the identity operator. $\hat{a}_{\bk}^{\dagger}$ and $\hat{a}_{\bk}$ are the creation and annihilation operators, respectively, acting on the vacuum state $\hat{a}_{\bk}|0\rangle=0$. They obey the standard commutation relations $[\hat{a}_{\bk},\hat{a}_{\bk'}^{\dagger}]=\delta(\bk+\bk')$ and $[\hat{a}_{\bk},\hat{a}_{\bk'}]=[\hat{a}_{\bk}^{\dagger},\hat{a}_{\bk'}^{\dagger}]=0$. Imposing the commutation relation 
\begin{equation}\label{ec:qu1}
\left[\hat{\chi}(\eta,\bm{x}),\frac{\diff \hat{\chi}}{\diff \eta}(\eta,\bm{x'})\right]=\delta(\bm{x}-\bm{x'}), \quad \text{which implies} \quad \left[\hat{\chi}_\bk(\eta),\frac{\diff \hat{\chi}_{\bk'}}{\diff \eta}(\eta)\right]=\delta(\bk+\bk')\,,
\end{equation}
and substituting \eqref{ec:qu0} in \eqref{ec:qu1}, we get
$\delta(\bk+\bk') = W[v_k,v_k^*][a_\bk,a^\dagger_{\bk'}]$. Considering that $W[v_k,v_k^*]$ is a constant --which we can set to {$i$} by appropriately normalizing the initial conditions for the homogeneous solution-- we see that \eqref{ec:qu0} is consistent with the usual commutation relation of the creation and annihilation operators. Notice that this is only the case because the inhomogeneous solution appears multiplying the identity operator in \eqref{ec:qu0}, which commutes with every other operator. Expanding the field in Fourier modes according to \eqref{ec:qu0}, we have
\begin{equation}
\hat{\chi}(\bx,\eta) = \int \frac{\diff^3 k }{(2\pi)^{3/2}}\left[v_k \,\hat{a}_\bk + v^*_k \,\hat{a}^\dagger_\bk + s_\bk\, \hat{\mathbb{I}}\right] e^{i\bk\cdot\bx }\,.
\end{equation} 
{This quantization scheme reduces to the usual one in the case of cold inflation, as $s_\bk\to 0$ and $v_k$ become the standard solutions to the Mukhanov-Sasaki equation. Furthermore, notice that the classical character of thermal fluctuations \cite{Berera:1999ws, Berera:2023liv} (encapsulated in $s_\bk$) is manifest because of the identity operator. In order to compute the variance of $\chi$, recall that $s_\bk$ is a stochastic function, while $v_k$ is a deterministic one. We thus have}
\begin{equation}\label{ec:infquan2}
\begin{split}
\langle 0| \hat{\chi}(\bx,\eta)\, \hat{\chi}(\bx,\eta) | 0\rangle  &= \int \frac{\diff^3 k \, \diff^3 k'}{(2\pi)^3} \left(v_k \,v^*_{k'}\,\langle 0|\hat{a}_\bk\,\hat{a}^\dagger_{\bk'}|0\rangle  + s_\bk\, s_{\bk'}\, \langle 0|\hat{\mathbb{I}}|0\rangle\right) e^{i(\bk+\bk')\cdot\bx}\\
&=\int \frac{\diff^3 k}{(2\pi)^3}\,|v_k|^2 + \int\frac{\diff^3 k\, \diff^3 k'}{(2\pi)^3} s_\bk\, s_{\bk'}e^{i(\bk+\bk')\cdot\bx}\,.
\end{split}
\end{equation}
There are no cross-terms involving the homogeneous and inhomogeneous solutions. This is because those terms involve only one creation or annihilation operator, whose vacuum expectation value vanishes. Taking the stochastic average of the above, we get
\begin{equation}\label{ec:q3}
\langle\langle 0| \hat{\chi}(\bx,\eta)\, \hat{\chi}(\bx,\eta) | 0\rangle\rangle =\int \frac{\diff^3 k}{(2\pi)^3}|v_k|^2 + \int\frac{\diff^3 k\, \diff^3 k'}{(2\pi)^3} \langle s_\bk\, s_{\bk'}\rangle e^{i(\bk+\bk')\cdot\bx}\,.
\end{equation}
 We emphasize that the notation $\langle\langle 0| \cdots | 0\rangle \rangle$ implies evaluating the relevant quantity in the (quantum) vacuum state and averaging over thermal noise realizations and that these two operations commute. In particular, notice that the stochastic average does not affect the first addend in the right-hand side, since it is a deterministic quantity. Analogously to \eqref{ec:b34a}, we define
\begin{equation}\label{ec:qu7}
\PP_{\chi}^{(h)}(k,\eta)= \frac{k^3}{2\pi^2}|v_k(\eta)|^2 \quad \text{and} \quad \delta(\bk+\bk')\,\langle \PP_{\chi}^{(i)}(k,\eta)\rangle =\frac{k^3}{2\pi^2}\langle s_\bk(\eta)\,s_{\bk'}(\eta) \rangle\,,
\end{equation}
with which Eq.\,\eqref{ec:q3} becomes
\begin{equation}\label{ec:qu4}
\langle\langle 0| \hat{\chi}(\bx,\eta)\, \hat{\chi}(\bx,\eta) | 0\rangle\rangle = \int \diff(\log k)\, \left( \PP_{\chi}^{(h)}(k,\eta) + \langle \PP_{\chi}^{(i)}(k,\eta) \rangle\right)\,.
\end{equation}
If we now define the dimensionless thermally averaged power spectrum as the quantity $\langle \PP_\chi(k,\eta) \rangle$ satisfying
\begin{equation}\label{ec:qu5}
\langle\langle 0| \hat{\chi}(\bx,\eta)\, \hat{\chi}(\bx,\eta) | 0\rangle\rangle = \int \diff(\log k) \, \langle \PP_\chi(k,\eta) \rangle\,,
\end{equation}
comparison between \eqref{ec:qu4} and \eqref{ec:qu5} yields
\begin{equation}\label{ec:qu8}
\langle \PP_\chi(k,\eta) \rangle = \PP_{\chi}^{(h)}(k,\eta) + \langle \PP_{\chi}^{(i)}(k,\eta) \rangle\,,
\end{equation}
with $\PP_{\chi}^{(h)}(k,\eta)$ and $\langle \PP_{\chi}^{(i)}(k,\eta) \rangle$ defined in \eqref{ec:qu7}. This explicitly proves the absence of mixing between homogeneous and inhomogeneous solutions in the power spectrum. Since $\chi$ and $\delp$ are the same field up to a rescaling by background variables, the analogous result to \eqref{ec:qu8} holds for $\PP_{\delp}(k,\eta)$, which is precisely what we used in Sec.\,\ref{ss:approx_deltaphi}.\par

\subsection{Quantization of $\delta\phi$ assuming a significant occupation number} \label{quantumn}

When computing expectation values in Section \ref{ss:quan} (e.g. in Eq.\ \ref{ec:infquan2}), it is assumed that the quantum fluctuations of the inflaton (whose Fourier modes correspond to the homogeneous solution of \eqref{ec:infquan}) are in the vacuum state $|0\rangle$. This is necessary at very early times, before the inflaton couples to extra degrees of freedom (c.f. the discussion on initial conditions in Section \ref{ss:homo}). The time evolution does not change the vacuum state because we are working in the Heisenberg picture.

During warm inflation, the inflaton interacts (possibly through a mediator) with thermalised degrees of freedom. If these interactions are strong enough, the state of inflaton perturbations may depart from the vacuum one. The calculations presented in the above section are then modified as follows. The quantum expectation value in \eqref{ec:infquan2} becomes
\begin{equation}\label{ec:infquan3}
\begin{split}
\langle  \hat{\chi}(\bx,\eta)\, \hat{\chi}(\bx,\eta) \rangle  &= \int \frac{\diff^3 k \, \diff^3 k'}{(2\pi)^3} \left(v_k \,v^*_{k'}\,\langle \hat{a}_\bk\,\hat{a}^\dagger_{\bk'}\rangle +      \text{h.c.}        + s_\bk\, s_{\bk'}\, \langle \hat{\mathbb{I}}\rangle\right) e^{i(\bk+\bk')\cdot\bx}\\
&=\int \frac{\diff^3 k}{(2\pi)^3}\,|v_k|^2 \,\Theta+ \int\frac{\diff^3 k\, \diff^3 k'}{(2\pi)^3} s_\bk\, s_{\bk'}e^{i(\bk+\bk')\cdot\bx}\,,
\end{split}
\end{equation}
where we have defined
\begin{equation}\label{ec:thermal_correction}
\Theta \,\delta(\bm{k}+\bm{k'})= \langle \hat{a}_\bk\,\hat{a}^\dagger_{\bk'}\rangle + \langle \hat{a}^\dagger_\bk\,\hat{a}_{\bk'}\rangle.
\end{equation} 
We emphasize that now we are not necessarily evaluating the correlators in vacuum. In other words, $\langle \hat{a}_\bk\,\hat{a}^\dagger_{\bk'}\rangle$ need not be equal to  $\langle 0| \hat{a}_\bk\,\hat{a}^\dagger_{\bk'}|0\rangle$. Notice that the inhomogeneous contribution receives no correction due to the normalization of quantum states. Calculations are henceforth analogous to those in Sec.\ \ref{ss:quan}, reaching
\begin{equation}\label{ec:correctedbythermal}
\langle \PP_\chi(k,\eta) \rangle = \PP_{\chi}^{(h)}(k,\eta)\,\Theta + \langle \PP_{\chi}^{(i)}(k,\eta) \rangle\,.
\end{equation}
In practice, the effect of $\delta\hat{\phi}$ not being in a vacuum state is enhancing the spectrum for $Q_*< 1$ (where the homogeneous solution dominates) and pushing the regime in which the inhomogeneous solution dominates to larger values of $Q_*$. As a consistency check, notice that $\Theta =1$ if inflaton perturbations are in the vacuum state. In any other case, $\Theta$ is a model-dependent quantity that has to be derived from a particular choice of the Lagrangian. A limit which has often been assumed in the literature is that in which the inflaton perturbations themselves attain a thermal equilibrium distribution. This has been argued to be the case in certain regions of parameter space for natural inflation \cite{Ferreira:2017lnd,Ferreira:2017wlx}. It has been assumed as a possibility (alongside with vacuum perturbations) for some monomial potentials \cite{Bastero-Gil:2014jsa, Bastero-Gil:2016qru, Bastero-Gil:2017wwl, Bastero-Gil:2019rsp}. 

If $\delta\hat{\phi}$ is in a thermal distribution of states, Eq.\ \eqref{ec:thermal_correction} reads \cite{Bhattacharya:2005wn} 
\begin{equation}
\Theta = 1+2\,n_{\be}\,,
\end{equation}
where $n_{\be}=[\exp(H/T-1)-1]^{-1}$ is the Bose-Einstein distribution, and the quotient $T/H$ is evaluated at horizon crossing in the literature on warm inflation assuming this correction. Notice that $n_{\be}\to 0$ for $T/H\to 0$. Also, as already discussed, in the $Q\to 0$ limit the homogeneous solution tends to the cold one and the inhomogeneous solution tends to zero. Therefore, in the $T/H\to 0\,,\, Q\to 0$ limit, the cold power spectrum is recovered. This is necessarily the case for consistency. However, the physics in the intermediate regime $0<T/H<1$ does not necessarily conform to the warm inflation formalism (i.e.\ the equations described in Sections \ref{dissipsec} and \ref{s:pert}) as the radiation bath is not expected to be thermalised.

\begin{figure}[t!]
    \begin{center}
  \includegraphics[width=1.0\textwidth]{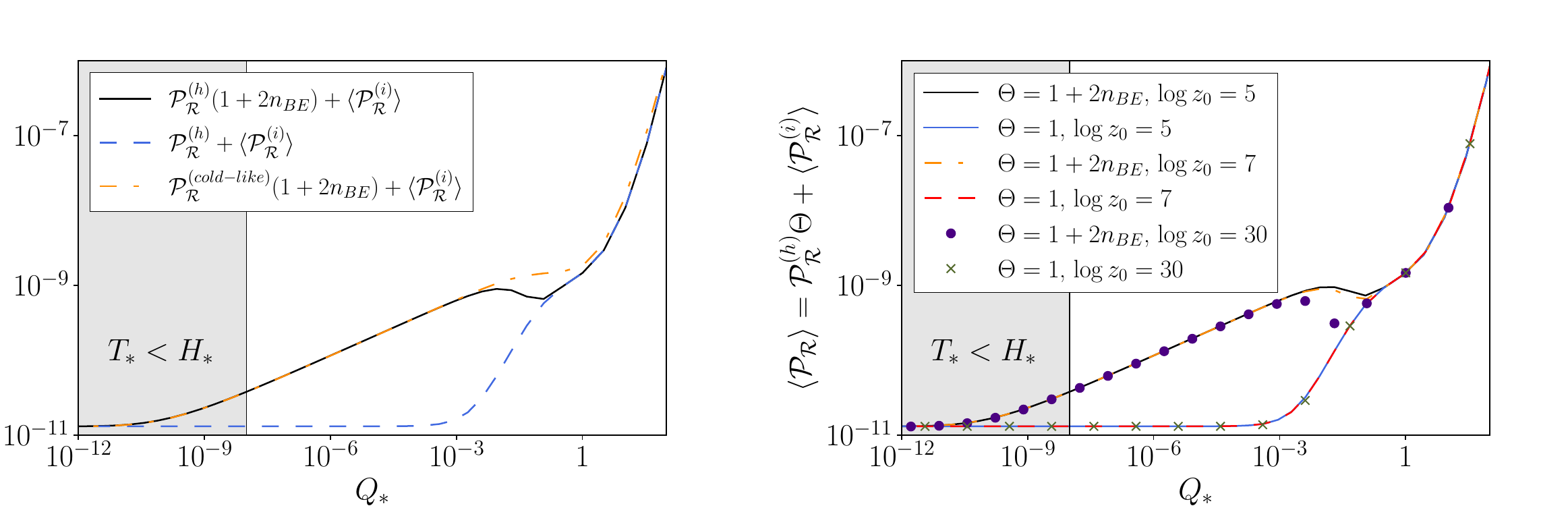}
\caption{\textbf{Left:} Comparison between the power spectrum computed with the matrix formalism with $\Theta=1+2n_{BE}$, implementing this correction using the cold-like power spectrum (yellow dot-dashed curve, analogous to \eqref{ec:frompapers}) and the homogeneous ($h$) spectrum (black continuous curve, Eq.\ \eqref{ec:correctedbythermal}). The spectrum with  $\Theta=1$, Eq.\ \eqref{ec:qu8}, is shown with a blue dashed curve. The second computation leads to a small dip around $Q \lesssim 1$, as a consequence of the suppression of the homogeneous solution competing with the Bose-Einstein enhancement.  \textbf{Right:} Influence of the value of the initial condition $z_0$ on the power spectrum computed as $\mathcal{P}_\mathcal{R}=\mathcal{P}_\mathcal{R}^{(h)}\Theta+\mathcal{P}_\mathcal{R}^{(i)}$. The larger is $z_0$, the further in the past the vacuum-to-thermal transition described in Sec.\ \ref{ss:homo} occurred, and therefore, the more suppressed the homogeneous solution is. In the case $\Theta=1$, the inhomogeneous solution quickly takes over, and the influence of the initial conditions is negligible. However, if the homogeneous solution is thermally enhanced, so is the influence of the initial conditions. For $T<H$ the radiation bath is not properly thermalised and the equations of warm inflation do not strictly apply (although, as one should expect, the equations yield the correct cold inflation limit). The curves in both plots have been computed for $V(\phi)=\lambda\phi^4/4$, $\Gamma=CT$, with $\lambda = 10^{-15}$, $g_\star = 12$ and varying $C$.}
\label{fig:comparison}
\end{center}
\end{figure}

An important comment is in order. As already anticipated in Sec.\,\ref{sec:comparison}, following our quantization approach, it is the homogeneous contribution to the power spectrum that is corrected by $\Theta$. This differs from the usual approach in the literature \cite{Bastero-Gil:2014jsa, Bastero-Gil:2016qru, Bastero-Gil:2017wwl,Arya:2019wck}, where the homogeneous solution is discarded and replaced by a cold-like power spectrum, which is then multiplied by $\Theta$. The latter approach yields the expression \eq{ec:frompapers} for the power spectrum. As it can be seen in Fig.\ \ref{fig:comparison}, left panel, the difference between both approaches is noticeable for $Q\lesssim 1$.\footnote{This difference depends slightly on the initial conditions for the homogeneous solution. As shown in Sec.\ \ref{ss:homo}, the larger is $z_0$ (as introduced in Eqs.\ \eqref{ec:joint1}, \eqref{ec:joint2}), the more suppressed is the homogeneous solution. This causes a small suppression of the spectrum for values of $Q_* \lesssim 1$ (just before the inhomogeneous solution starts dominating), see Fig.\ \ref{fig:comparison}, right panel. In the vacuum case ($\Theta=1$), the inhomogeneous solution starts dominating at smaller values of $Q_*$ and therefore this effect is smaller.} 

In any case, as we have already explained, we have not implemented an occupation number correction (in any form) in the main body of our numerical analyses, since the necessity of implementing this correction needs to be checked on a case by case basis. It is however worth stressing that the matrix formalism described in Sec.\ \ref{sec:computationspectrum} is still applicable in the presence of a $\Theta\neq 1$, with a slight modification. By construction, solving the matrix differential Eq.\ \eqref{matrix_eq} with a set of initial conditions (namely Bunch-Davies mode functions for inflaton perturbations in the asymptotic past) provides a solution which is the sum of a homogeneous solution and an inhomogeneous solution, given by the source term of the equation (which is stochastic). In order to apply the matrix formalism with $\Theta\neq 1$, we need the homogeneous and inhomogeneous solutions separately, because only the homogeneous one is enhanced a posteriori by $\Theta\neq 1$, as we have discussed above. To obtain them, we solve the matrix differential equations twice:
\begin{enumerate}
\item First, setting the source term (i.e. the noise) to zero, and the initial conditions to Bunch-Davies sufficiently back in time. This provides the homogeneous contribution to the power spectrum.
\item Second, keeping the source term and setting the initial conditions to zero. This provides the inhomogeneous contribution to the power spectrum.\footnote{Indeed, setting the initial conditions to zero imposes that the corresponding homogeneous solution is zero because of the uniqueness of the solution of the homogeneous equation.}
\end{enumerate}
This is the procedure we have followed to generate Fig.\ \ref{fig:comparison}.

\section{Stochastic inflation formalism} \label{sec:stoch}

So far, we have computed the power spectrum using the Langevin equations for the linear perturbations of the inflaton, metric and radiation energy density sourced by the thermal noise $\xi$. In this section, instead, we adapt the stochastic formalism of cold inflation \cite{Starobinsky:1986fx} to warm inflation. In cold inflation, the power spectrum obtained in the stochastic approach is, to first order, the same as the one obtained in linear perturbation theory, both for slow roll \cite{Vennin:2015hra} and ultra-slow roll \cite{Ballesteros:2020sre} inflation. As we will argue, this is also the case in warm inflation. 

In Ref.\ \cite{Ramos:2013nsa} the stochastic inflation formalism was used to show for the first time how the term $n_{\rm BE}$ may appear in the approximation \eqref{ec:frompapers}. The implementation of the stochastic formalism done in that reference differs from the more standard procedure we follow below. We revisit how $n_{\rm BE}$ would appear, under the assumption of large occupation number of inflaton fluctuations, taking into account the quantization procedure discussed in the previous section.

In App.\,\ref{app:stochasticv2}, we follow the approach to stochastic inflation introduced in Ref.\,\cite{Ramos:2013nsa}. Our derivation in App.\,\ref{app:stochasticv2} emphasizes the differences with the original treatment of Ref.\,\cite{Ramos:2013nsa} and agrees with the main conclusion of this section.

In order to obtain analytical results, we work in a simplified limit of the equations of motion in which the inflaton field with its thermal (stochastic) source is the only degree of freedom. We will therefore set $\rho_r$ to its background attractor and neglect both metric and radiation perturbations. 

\subsection{Effective description of the large wavelength modes}

We start by writting the equation of motion for the full (i.e. space-time dependent)  inflaton field $\phi(\bx,N)$ 
\begin{equation}\label{ec:a1}
\phi''(\bx,N)+(3+3Q-\epsilon)\phi'(\bx,N)-\frac{\nabla^2}{(aH)^2}\phi(\bx,N)+\frac{\dv}{H^2}=\prefactor\xi (\bx,N)\,.
\end{equation}
Following e.g.\ \cite{Ballesteros:2020sre}, Eq.\,\eqref{ec:a1} can be split into two first order coupled differential equations by introducing a new variable $\pi(\bx,t)$
\begin{align}
\phi'(\bx,N)&=\pi(\bx,N)\label{ec:a2}\,,\\
\pi'(\bx,N)&=-\left[(3+3Q-\epsilon)\pi(\bx,N)-\frac{\nabla^2}{(aH)^2}\phi(\bx,N)+\frac{\dv}{H^2}\right]+\prefactor\xi (\bx,N)\,.\label{ec:a3}
\end{align}
In the stochastic inflation formalism, we express the fields $\phi(\bx,t)$ and $\pi(\bx,t)$ as the sum of two components
\begin{equation}\label{ec:a3.5}
\phi(\bx,N)=\bar{\phi}(\bx,N)+\phi_q(\bx,N)\,, \qquad \pi(\bx,N)=\bar{\pi}(\bx,N)+\pi_q(\bx,N)\,,
\end{equation}
where $\bar{\phi},\bar{\pi}$ and $\phi_q,\pi_q$ respectively correspond to contributions from large and small wavelengths. Following the procedure presented in Sec.\,\ref{ss:quan}, we quantize the inflaton field $\phi(\bx,N)$ by expanding each component in terms of Fourier modes and creation/annihilation operators as
\begin{equation}
\begin{split}
\bar{\phi}(\bx,N)&=\int \dfrac{\diff^3 \bk}{(2 \pi)^{3/2}}  \, W(\sigmak-k) \left(\phih\ann + \phih^*\cre + \phii \hat \idd\right)e^{i\bk\cdot\bx},\\ 
\phi_q(\bx,N) &=\int \dfrac{\diff^3 \bk}{(2 \pi)^{3/2}} \, W(k-\sigmak) \left(\phih\ann + \phih^*\cre + \phii \hat \idd\right)e^{i\bk\cdot\bx}\,.\end{split}\label{ec:a4}
\end{equation}
We have introduced a window function $W$ selecting modes with wavenumber smaller than a comoving scale $\sigmak (N) = \sigma\, a(N)H(N)$\,, where $\sigma$ is a constant. The window function is taken to be a Heaviside function $W(\sigmak-k)=\Theta(\sigmak-k)$ for simplicity. In practice we consider the limit $\sigma \rightarrow 0$, in which the large wavelength contribution can reasonably be considered to be homogenous $\bar{\phi}(\bx,t)\simeq\bar{\phi}(t)$. The mode functions $\phih$ and $\phii$ respectively satisfy the following  homogeneous and inhomogenous equations
\begin{align}
\phih'&=\pi^{(h)}_k,\label{ec:a7.1}\\
{\pi^{(h)}_k}'&=-\left[(3+3Q-\epsilon){\pi^{(h)}_k}+\left(\frac{k^2}{a^2H^2}+\frac{\left.\ddv\bareval}{H^2}\right)\phih\right],\label{ec:a8.1}
\end{align}
and 
\begin{align}
\phii'&={\pi^{(i)}_\bk},\label{ec:a7.2}\\
{\pi^{(i)}_\bk}'&=-\left[(3+3Q-\epsilon){\pi^{(i)}_\bk}+\left(\frac{k^2}{a^2H^2}+\frac{\left.\ddv\bareval}{H^2}\right)\phii\right]+\prefactor\xi_\bk(N)\,.\label{ec:a8.2}
\end{align}
The term $V_{\phi\phi}$ comes from expanding the derivative of the potential $V_\phi$ around the large wavelength contribution
\begin{equation}
\dv(\phi)\simeq \left. \dv\bareval+\left.\ddv\bareval(\phi(\bx,N)-\bar{\phi}(N))=\left. \dv\bareval+\left.\ddv\bareval \phi_q(\bx,N)  \,,
\label{eq:expansionpotential}
\end{equation}
with $\left. \dv\bareval$ having negligible Fourier transform for large $k$ (due to homogeneity). By using the decomposition of Eq.\,\eqref{ec:a3.5} in the system of Eq.\,\eqref{ec:a2} and Eq.\,\eqref{ec:a3}, a large set of terms vanishes by virtue of Eqs.\,\eqref{ec:a7.1}, \eqref{ec:a8.1}, \eqref{ec:a7.2} and \eqref{ec:a8.2}. All terms whose integrand is proportional to $W(\sigmak-k)$ can be absorbed into $\bphi$, $\bpi$ except one:
\begin{equation}
 \xi^>(x,N)\equiv\int  \dfrac{\diff^3 \bk}{(2 \pi)^{3/2}} \, W(\sigmak - k) e^{i\bk\cdot \bx} \xi_\bk(N)\,.
\end{equation}
{This term corresponds to the effect of the noise on long wavelengths and can be neglected by realizing that the thermal noise settles the value of $\phi^{(i)}_{\bm{k}}$ on superhorizon scales already at horizon crossing $k=a\,H$ (or even before, $k=a\sqrt{\Gamma\, H}$, if $\Gamma \gg H$ \cite{Hall:2003zp}), i.e. superhorizon modes of $\phi^{(i)}_{\bm{k}}$ are not affected by the thermal noise.}\footnote{
{\begin{align} \nonumber
\langle \xi^>(x,N) \xi^>(x,N')\rangle 
& = \int \frac{\diff^3\bk}{(2\pi)^{3/2}} \frac{\diff^3\bk'}{(2\pi)^{3/2}} W(k_\sigma(N)-k) W(k_\sigma(N')-k')\frac{2\Gamma T}{(aH)^3}\langle \xi_{\bk}(N)\xi_{\bk'}(N')\rangle e^{i\bx\cdot(\bk+\bk')}\\
&=\frac{2\Gamma T}{(aH)^3} \delta(N-N')\int \frac{\diff^3\bk}{(2\pi)^{3}}W(k_\sigma(N)-k)  \to 0 \quad \text{for} \quad \sigma\to 0\,. \end{align}}
}
We obtain a system of equations for the large wavelength contributions
\begin{align}
\bar{\phi}'&=\bar{\pi}+\noise,\label{ec:a10}\\
\bar{\pi}' &= -(3+3Q-\epsilon)\bar{\pi}-\frac{\left.\dv\bareval}{H^2}+\pnoise,\label{ec:a11}
\end{align}
where we define
\begin{align}\label{ec:a12}
\noise &= -\int \dfrac{\diff^3 \bk}{(2 \pi)^{3/2}} \, W'(k-\sigmak) \left(\phih\ann + \phih^*\cre + \phii \hat \idd\right)e^{i\bk\cdot\bx}, \\
\pnoise &= -\int \dfrac{\diff^3 \bk}{(2 \pi)^{3/2}} \, W'(k-\sigmak) \left(\pi^{(h)}_k\ann + {\pi^{(h)}_k}^*\cre + \pi^{(i)}_\bk  \hat \idd\right)e^{i\bk\cdot\bx}\,.\label{ec:a12b}
\end{align}
In these expressions $ W'$ is the derivative of the window function with respect to $N$. At this stage, the set of Eqs.\,\eqref{ec:a10} and \eqref{ec:a11} is formally identical to those in cold inflation, with the natural addition of the dimensionless dissipative term $Q$ and the extra inhomogeneous (stochastic) term in Eqs.\,\eqref{ec:a12} and \eqref{ec:a12b}. In cold inflation Eqs.\,\eqref{ec:a10} and \eqref{ec:a11} are treated as a system of Langevin equations where $\bar{\pi},\bar{\phi}$ and the noise $\noise, \pnoise$ are stochastic variables. To prove that this is also the case in warm inflation, we must show that the equal-time commutator of $\noise$ and $\pnoise$ vanishes, i.e.\ that they behave as classical variables. Since the identity operator commutes with every other operator, this commutator reduces to
\begin{equation}
\left[\noise(\bx,N),\pnoise(\bx^\prime,N)\right] = \iint \dfrac{\diff^3 \bk  \diff^3 \bk^\prime}{(2\pi)^3}  W'(k-\sigmak)W'(k'-\sigmak) \left[\phih\,\ann + \phih^*\,\cre,\pi^{(h)}_{k'}\,\annp + {\pi^{(h)}_{k'}}^*\,\crep\right] \,.
\end{equation}
From here, the calculation is identical to the one performed in cold inflation (see e.g. \cite{Ballesteros:2020sre}), which gives $\left[\noise(\bx,N),\pnoise(\bx^\prime,N)\right]\rightarrow 0$ in the limit $\sigma\to 0$. This allows to treat Eqs.\,\eqref{ec:a10} and \eqref{ec:a11} as a system of Langevin equations for the stochastic variables $\bar{\phi}$, $\bar{\pi}$, where the noises  $\noise$, $\pnoise$ should not be confused with the thermal noise $\xi_\bk$ (whose influence resides in the $\phii$ and $\pi^{(i)}_\bk$ in \eqref{ec:a12}, \eqref{ec:a12b}).\par
\medskip
\noindent
\textbf{The Fokker-Planck equation.} By treating Eqs.\,\eqref{ec:a10} and \eqref{ec:a11} as a system of Langevin equations, the power spectrum of large wavelength perturbations can be related to statistical moments of $\bar{\phi}$ and $\bar{\pi}$. In order to write down the Fokker-Planck equation for the probability distribution function $P(\bar \phi,\bar \pi,N)$ (which will be introduced in \eqref{eq:FPstochastic}) and compute such moments, we need the average over realizations of the thermal noise of the two-point correlator of $\noisef$ and $\noiseg$ with the pair $\{f,g\}$ picking values in the set $\{\phi,\pi\}$:
\begin{equation}\label{ec:a13a}
\begin{split}
\langle\langle 0|\noisef(\bx,N)\noiseg(\bx^\prime,\tilde N)|0\rangle\rangle=&\iint \dfrac{\diff^3 \bk  \diff^3 \bk^\prime}{(2\pi)^3} W'
(k-\sigmak(N))W'
(k'-\sigmak(\tilde N)) e^{i(\bk\cdot\bx+\bk'\cdot\bx')}\\
\times \langle & \langle 0| \left({f^{(h)}_k}\,\ann + {f^{(h)}_k}^*\,\cre + {f^{(i)}_\bk}\, \hat \idd\right)\left({g^{(h)}_{k'}}\,\annp + {g^{(h)}_{k'}}^*\,\crep + {g^{(i)}_{\bk'}} \,\hat \idd\right)|0\rangle\rangle \,.
\end{split}
\end{equation}
By using commutation relations for the creation and annihilation operators, the second line in the previous equations can be written as
\begin{equation}
\langle\langle 0| \left({f^{(h)}_k}\ann + {f^{(h)}_k}^*\cre + {f^{(i)}_\bk} \hat \idd\right)\left({g^{(h)}_{k'}}\annp + {g^{(h)}_{k'}}^*\crep + {g^{(i)}_{\bk'}} \hat \idd\right)|0\rangle\rangle  = \delta(\bk+\bk'){f^{(h)}_k}{g^{(h)}_{k'}}^* + \langle {f^{(i)}_\bk} {g^{(i)}_{\bk'}}\rangle.
\end{equation}
We recall that $\langle\langle 0|\ldots|0\rangle\rangle$ means that the operator inside is evaluated on  the vacuum state and averaged over thermal noise realizations. See the comment below Eq.\,\eq{ec:q3}. Defining\footnote{ Notice that this definition is consistent with the one in \eqref{ec:qu7}.}
\begin{equation}
\mathcal{P}^{(h)}_{fg}(k,N) =\frac{k^3}{2\pi^2} {f^{(h)}_k}{g^{(h)}_{k}}^*,\quad \delta(\bk+\bk')\left\langle\mathcal{P}^{(i)}_{fg}(k,N)\right\rangle=\frac{k^3}{2\pi^2}\langle {f^{(i)}_\bk} {g^{(i)}_{\bk'}}\rangle \, ,
\end{equation}
we can express the power spectrum as a sum of two contributions
\begin{equation}\label{ec:2corr}
\left\langle \mathcal{P}_{fg}(k,N)\right\rangle=\mathcal{P}^{(h)}_{fg}(k,N)+\left\langle \mathcal{P}^{(i)}_{fg}(k,N)\right\rangle \,.
\end{equation}
This allows to express Eq.\,\eqref{ec:a13a} as
\begin{equation}\label{ec:a13}
\begin{split}
\langle\langle 0|\noisef(\bx,N)\noiseg(\bx^\prime,\tilde N)|0\rangle\rangle=\delta(N-\tilde N)\frac{\diff \sigmak}{\diff N}\int\frac{\diff^3 \bm{k}}{(2\pi)^3} \delta(k-\sigmak(N)) \frac{2\pi^2}{k^3}\langle\mathcal{P}_{fg}(k,N)\rangle e^{i\bk\cdot(\bx-\bx')}\\
\end{split}\,,
\end{equation}
where the integration over $\diff^3 \bm{k}'$ has been performed using the following relation
\begin{equation}\label{ec:a14}
W'(k-\sigmak(N))W'(k'-\sigmak(\tilde N))=\delta(k-\sigmak(N))\delta(N-\tilde N)\frac{\diff \sigmak}{\diff N} \, .
\end{equation}
The integration over $\diff^3 \bm{k}$ in Eq.\,\eqref{ec:a13} can be performed straightforwardly, giving
\begin{equation}
\langle\langle 0|\noisef(\bx,N)\noiseg(\bx^\prime,\tilde N)|0\rangle\rangle=\frac{\diff\log\sigmak}{\diff N}\langle\mathcal{P}_{fg}(\sigmak,N)\rangle\sinc(\sigmak|\bm{x}-\bm{x}'|) \delta(N-\tilde N)\,,
\end{equation}
which in the limit $\sigma\to 0$ yields
\begin{equation}\label{ec:difussion}
\langle\langle 0|\noisef(\bx,N)\noiseg(\bx^\prime,\tilde N)|0\rangle\rangle=(1-\epsilon)\langle\mathcal{P}_{fg}(\sigmak,N)\rangle  \delta(N-\tilde N)\equiv D_{fg}\delta(N-\tilde N)\,,
\end{equation}
where we used $ \diff \log \sigmak / \diff N = 1- \epsilon $. The reasoning above (from Equation \eqref{ec:a13a} onwards) can be analogously performed if the inflaton is in an excited state instead of the vacuum, as discussed in section \ref{quantumn}. Equation \eqref{ec:2corr} would become
\begin{equation}
\left\langle \mathcal{P}_{fg}(k,N)\right\rangle=\Theta\,\mathcal{P}^{(h)}_{fg}(k,N)+\left\langle \mathcal{P}^{(i)}_{fg}(k,N)\right\rangle \,
\end{equation}
therefore modifying the numerical value of the $D_{fg}$ defined in \eqref{ec:difussion}. The $D_{fg}$ are the entries of the \emph{diffusion matrix} of the system of Langevin equations. They appear in the Fokker-Planck equation for the probability distribution $P\left(\bphi,\bpi,N\right)$ as
\begin{equation}
\frac{\partial P\left(\bphi,\bpi,N\right)}{\partial N} = -\left[\frac{\partial}{\partial \bphi}\left(\drphi\,P\right)
+\frac{\partial}{\partial \bpi}\left(\drpi\,P\right)\right] + \frac{1}{2}\left[D_{\phi\phi}\frac{\partial^2P}{\partial\bphi^2}
+2D_{\phi\pi}\frac{\partial^2P}{\partial\bpi\partial\bphi}
+D_{\pi\pi}\frac{\partial^2P}{\partial\bpi^2}\right]\,,
\label{eq:FPstochastic}
\end{equation}
where $\drphi$ and $\drpi$ are the components of the \emph{drift vector} following
\begin{equation}\label{ec:drift}
\drphi = \bpi\,,\qquad
\drpi =-(3+3Q-\epsilon)\bpi + \frac{\left.\dv\bareval}{H^2}\,.
\end{equation}
Notice that, since Eqs. \eqref{ec:a7.1}, \eqref{ec:a8.1}, \eqref{ec:a7.2}, \eqref{ec:a8.2} are formally the same equations that $\delta\phi$, $\delta\phi'$ satisfy in linear perturbation theory (once metric and radiation perturbations and slow-roll suppressed terms are neglected), $\langle\mathcal{P}_{\phi\phi}\rangle$ (as defined above) coincides with the average over realizations of the thermal noise of the power spectrum of inflaton perturbations as computed using standard cosmological (linear) perturbation theory.\par
\medskip
\noindent
\textbf{The power spectrum.} Let us define the quantities
\begin{equation}
\phcl\equiv \text{E}({\bar{\phi}}), \qquad \pcl \equiv \text{E}(\bar{\pi}), \qquad \dphis \equiv \bphi-\phcl, \qquad \dpis \equiv \bpi-\pcl,
\label{eq:splitclVSst}
\end{equation}
where $\dphis, \dpis$ are stochastic variables and $\text{E}\left(\cdot\right)$ denotes the expectation value with respect to the probability density $P$. $\phcl, \pcl$ are deterministic quantities satisfying
\begin{align}
\frac{\diff \phcl}{\diff N}&=\pcl,\label{ec:cl1}\\
\frac{\diff \pcl}{\diff N}&=-(3+3Q-\epsilon)\pcl - \frac{\left.\dv\bareval}{H^2}\,.\label{ec:cl2}
\end{align}
Fluctuations around $\phcl$, $\pcl$ can be expressed in terms of the variance of $\dphis$. This variance can be related to the power spectrum for the large wavelength perturbations  $\Delta_{\dphis}^2(k)$ via
\begin{equation}
\text{E}(\dphis^2)(N)=\int_0^{\sigmak(N)} \frac{\diff k}{k}\,\Delta_{\dphis}^2(k)\,,
\end{equation}
which implies
\begin{equation}\label{ec:a30}
\Delta^2_{\dphis}(\sigmak)=\frac{1}{1-\epsilon}\frac{\diff \text{E}(\dphis^2)}{\diff N}\,.
\end{equation}
Neglecting both metric and radiation perturbations, the quantity $\Delta_{\dphis}^2(k)$ can be directly related to the power spectrum of curvature perturbations. In order to do this we have to compute the time derivative of the moment in the right-hand side of Eq.\,(\ref{ec:a30}), which we proceed to do next.

\subsection{Differential equations for the statistical moments}

As seen in \eqref{ec:a30}, in order to compute the power spectrum in the context of the stochastic formalism, we must compute time derivatives of moments of the kind
\begin{equation}\label{ec:a15}
\text{E}\left(\dphis^b\dpis^c\right)=\iint \diff \bar{\phi}\diff \bar{\pi} \left(\bphi-\phcl\right)^b\left(\bpi-\pcl\right)^c P(\bphi,\bpi,N)\,,
\end{equation}
where $b,c$ are generic integers. This can be straightforwardly expressed as a sum of three terms
\begin{equation}\label{ec:b1}
\begin{split}
\frac{\diff}{\diff N}\text{E}\left(\dphis^b\dpis^c\right) =& \iint \diff \bphi \diff \bpi \left( \partial_N\left[\bphi-\phcl(N)\right]^b \right)\left[\bpi-\pcl(N)\right]^c P(\bphi,\bpi,N) \\
+& \iint \diff \bphi \diff \bpi  \left[\bphi-\phcl(N)\right]^b \left(\partial_N\left[\bpi-\pcl(N)\right]^c \right) P(\bphi,\bpi,N)\\
+& \iint \diff \bphi \diff \bpi  \left[\bphi-\phcl(N)\right]^b\left[\bpi-\pcl(N)\right]^c \left( \partial_N P(\bphi,\bpi,N) \right).
\end{split}
\end{equation}
The first two terms can be computed using the chain rule, while the third can be expressed in terms of partial derivatives of $P$ with respect to $\bphi$, $\bpi$ using the Fokker-Planck Eq.\,(\ref{eq:FPstochastic}). Analogously to Eq.\,(\ref{eq:splitclVSst}), one can define components of the drift vector evaluated on the trajectories
\begin{equation}
 \drphicl \equiv \text{E}(\drphi) = \phcl', \qquad \drpicl \equiv \text{E}(\drpi) = \pcl'\,,
\end{equation}
which correspond to right-hand sides of Eqs.\,\eqref{ec:cl1} and \eqref{ec:cl2}. Substituting these quantitites in the first two lines of \eqref{ec:b1} and integrating by parts the third line (and making use of the Fokker-Planck equation) yields
\begin{multline}\label{ec:b3}
\frac{\diff}{\diff N}\text{E}\left(\dphis^b\dpis^c\right) = 
b\,\text{E}\left(\dphis^{b-1}\dpis^c\drdphi\right)+
c\,\text{E}\left(\dphis^{b}\dpis^{c-1}\drdpi\right)+\frac{b(b-1)}{2}D_{\phi\phi}\,\text{E}\left(\dphis^{b-2}\dpis^c\right)
\\
+\frac{c(c-1)}{2}D_{\pi\pi}\,\text{E}\left(\dphis^{b}\dpis^{c-2}\right) + b\,c\,D_{\phi\pi}\,\text{E}\left(\dphis^{b-1}\dpis^{c-1}\right)\,,
\end{multline}
where we introduced
\begin{equation}
\drdphi \equiv \drphi - \drphicl, \quad \drdpi \equiv \drpi - \drpicl \,.
\end{equation}
For $b=2$ and $c=0$, we obtain the following complete system of first order differential equations
\begin{align}
\frac{\diff}{\diff N}\text{E}(\dphis^2) &= 2\text{E}(\dphis \drdphi)+D_{\phi\phi},\label{ec:stoch:1}\\
\frac{\diff}{\diff N}\text{E}(\dphis\dpis) &= \text{E}(\dpis\drdphi)+\text{E}(\dphis\drdpi)+D_{\phi\pi},\label{ec:stoch:2}\\
\frac{\diff}{\diff N}\text{E}(\dpis^2) &= 2\text{E}(\dpis \drdpi)+D_{\pi\pi},\label{ec:stoch:3}
\end{align}
whose solution provides the desired quantity appearing in the right-hand side of Eq.\,(\ref{ec:a30}). In order to be solved, the system of Eqs.\,\eqref{ec:stoch:1}, \eqref{ec:stoch:2} and \eqref{ec:stoch:3} requires further simplifications.

\subsection{Expansion of the drift vector}

Up to now, the only property of $\dphis$, $\dpis$ that we have used is that these variables have zero mean. However, in practice they are small deviations with respect to the expected values $\phcl$, $\pcl$, i.e. their standard deviations should be much smaller than $\phcl$, $\pcl$. By expanding the drift vector around $\phcl$, $\pcl$ to first order in $\dphis, \dpis$, a straightforward computation of $\drdphi$ at first order gives
\begin{equation}
 \drdphi=\dpis\,.
\end{equation}
Expanding $\drdpi$ is more involved, since
\begin{equation}
\drpi=-\left.(3+3Q-\epsilon)\right|_{\bphi,\bpi}\bpi+\left.\frac{\dv}{H^2}\right|_{\bphi,\bpi}\,.
\end{equation}
Therefore we have to expand $\epsilon$ and $Q$ in $\dphis$ and $\dpis$.\footnote{We recall that we are assuming $\rho_r$ that lies on the background attractor.} The first Friedmann equation allows to rewrite
\begin{equation}\label{ec:c1}
\drpi = -\left.(3+3Q-\epsilon)\right|_{\bphi,\bpi}\bpi+\left.\left[3\mm^2-\frac{1}{2}\pi^2\left(1+\frac{3Q}{2}\right)\right]\right|_{\bphi,\bpi}\left.(\log V)_{\phi}\bareval.
\end{equation}
Thus we have to expand every quantity in \eqref{ec:c1} to first order in $\dphis$, $\dpis$. First, we have
\begin{equation}\label{ec:c3}
\left.(\log V)_{\phi}\bareval \simeq \left.(\log V)_{\phi}\cleval + \left.(\log V)_{\phi\phi}\cleval \dphis\,.
\end{equation}
To expand $Q=\Gamma/(3H)$, we remind the reader that we are assuming a constant $\Gamma$, therefore the only dependence on background quantities of $Q$ is through $H$. We can express $H$ as a function of $\phi$, $\pi$ by using the first Friedmann equation and leaving the constant $\Gamma$ explicit
\begin{equation}
H=\frac{\pi^2\Gamma +\sqrt{192\mm^2 V-32\pi^2 V + \pi^4\Gamma^2}}{4(6\mm^2-\pi^2)}\,.
\end{equation}
This allows to expand $Q$ around $\phcl$, $\pcl$ as
\begin{multline}\label{ec:d1}
\left.Q\bareval = \qcl + 	\frac{\Gamma}{24\vcl}\left(-2\pcl\Gamma +\frac{2\pcl^3\Gamma^2-32\pcl \vcl}{\sqrt{\pcl^4\Gamma^2-32(\pcl^2-6\mm^2)\vcl}}\right)\dpis\\+\frac{1}{3}\left(\frac{-\pcl^4\Gamma^3\dvcl-96\mm^2\Gamma\vcl\dvcl+16\pcl^2\Gamma\vcl\dvcl}{8\vcl^2\sqrt{\pcl^4\Gamma^2-32(\pcl^2-6\mm^2)\vcl}}+\frac{\pcl^2\Gamma^2\dvcl}{8\vcl^2}\right)\dphis \,,
\end{multline}
where the subscript ``cl" for each quantity indicates that it is evaluated on $\phcl$, $\pcl$. This provides the last term in Eq.\,(\ref{ec:c1}). In order to simplify the expression of $\drpi$ in Eq.\,(\ref{ec:c1}), we identify the dominant terms in both strong ($Q \gg 1$) and weak ($Q \ll 1$) dissipative regimes. We will use the expression of $\epsilon$ in terms of $Q$:
\begin{equation}\label{ec:c0}
\epsilon = -\frac{H'}{H}=\frac{1}{\mm^2}\left[\frac{1}{2}\pi^2+\frac{2}{3}\frac{\rho_r}{H^2}\right] = \frac{1}{2}\frac{\pi^2}{\mm^2}\left(1+Q\right).
\end{equation}

\subsubsection*{Strong dissipative regime $Q \gg 1$} 
In this regime $\epsilon \sim (\pi/\mn)^2Q $ and $\epsilon_V \sim \epsilon Q$. Therefore, Eq.\,\eqref{ec:d1} becomes
\begin{align}\label{ec:c-1} 
\left.Q\bareval & = \qcl\left[1-\frac{\qcl}{4}\frac{\pcl}{\mm}\frac{\dpis}{\mm}-  \left(\frac{\left.\epsilon_V\cleval}{2}\right)^{1/2} \frac{\dphis}{\mm}\right]\\ & \sim \qcl\left[1+\mathcal{O}\left((\epsilon_\cl \qcl)^{1/2}\right)\frac{\dpis}{\mm}+\mathcal{O}\left((\epsilon_\cl \qcl)^{1/2}\right)\frac{\dphis}{\mm}\right]
\end{align}
Every other term is of lower order in $\qcl\gg 1$ or higher in $\epsilon_\cl\ll 1$, i.e. more suppressed. Expanding $\epsilon$ in $\dphis$ and $\dpis$  and substituting Eq.\,\eqref{ec:c-1} in Eq.\,\eqref{ec:c0} yields
\begin{equation}\label{ec:c-2}
\left.\epsilon\bareval = \ecl - \frac{1}{4}\left(\frac{\pcl}{\mm}\right)^2\left(2\left.\epsilon_V\cleval\right)^{1/2}  \qcl \,\frac{\dphis}{\mm} + \left(1+\qcl-\frac{\pcl^2\,\qcl^2}{8\mm^2}\right)\frac{\pcl}{\mm}\,\frac{\dpis}{\mm}\,.
\end{equation}
Using Eqs.\,\eqref{ec:c3}, \eqref{ec:c-1} and \eqref{ec:c-2}, we can expand \eqref{ec:c1} to first order in $\dpis$, $\dphis$ and compute $\drdpi$. The highest power of $\qcl\gg 1$ in the terms appearing in the expansion is 1. Keeping only terms of order zero in $\epsilon \ll 1$ gives
\begin{equation}
\drdpi = -3(1+\qcl)\dpis\,.
\end{equation} 
This expression, in the limit $\qcl \to 0$, yields the usual cold inflation result \cite{Ballesteros:2020sre}.

\subsubsection*{Weak dissipative limit $Q \ll 1$}

For $Q \ll 1$, $\epsilon \sim \epsilon_V \sim (\pi/\mm)^2$ and Eq.\,\eqref{ec:d1} reduces to
\begin{equation}
\left.Q\bareval = \qcl\left[1- \frac{\pcl}{6\mm}\frac{\dpis}{\mm} -  \left(\frac{\left.\epsilon_V\cleval}{2}\right)^{1/2} \frac{\dphis}{\mm} \right]\sim \qcl\left[1+\mathcal{O}\left(\epsilon_{\rm cl}^{1/2}\right)\frac{\dpis}{\mm}+\mathcal{O}\left(\epsilon_{\rm cl}^{1/2}\right)\frac{\dphis}{\mm}\right]\,.
\end{equation}
Every other contribution is of higher order in $\epsilon_{\rm cl} \ll 1$ or ${Q_\text{cl}}\ll 1$, and is therefore neglected. Substituting the expanded $Q$ in \eqref{ec:c0} gives
\begin{equation}
\left.\epsilon\bareval = \ecl - \frac{1}{4}\left(\frac{\pcl}{\mm}\right)^2 \left(2\left.\epsilon_V\cleval\right)^{1/2} \qcl \frac{\dphis}{\mm} + \left(1+\qcl-\frac{\pcl^2\,\qcl}{12\mm^2}\right)\frac{\pcl}{\mm}\frac{\dpis}{\mm}\,.
\end{equation}
Keeping terms of order at most one in $\qcl\ll 1$ and of order zero in $\epsilon_\cl$ in $\drdpi$, we obtain the same result as in the strong dissipative limit,
\begin{equation}
\drdpi = -3(1+\qcl)\dpis.
\end{equation} 
In other words, this expression for $\drdpi$ is valid both in the weak and strong dissipative regimes to lowest order in slow-roll parameters.

\subsection{The power spectrum}

Substituting $\drdphi$ and $\drdpi$ in  \eqref{ec:stoch:1}-\eqref{ec:stoch:3}, we obtain the following system of differential equations for the moments
\begin{align}
\frac{\diff}{\diff N }\text{E}(\dphis^2)&=2\text{E}(\dphis\dpis)+D_{\phi\phi},\label{ec:tps1}\\
\frac{\diff}{\diff N}\text{E}(\dphis\dpis)&=\text{E}(\dpis^2)-3(1+\qcl)\text{E}(\dphis\dpis)+D_{\phi\pi},\label{ec:tps2}\\
\frac{\diff}{\diff N}\text{E}(\dpis^2)&=-6(1+\qcl)\text{E}(\dpis^2)+D_{\pi\pi}.\label{ec:tps3}
\end{align}
This system can be solved analytically to obtain $\text{E}(\dphis^2)$. Taking the large $N$ limit of $\text{E}(\dphis^2)$ and substituting its derivative in \eqref{ec:a30} yields\footnote{This limit of $\diff\text{E}(\dphis^2)/\diff N$ can also be obtained by neglecting the derivative in the left-hand sides of \eqref{ec:tps2}, \eqref{ec:tps3} and substituting the corresponding algebraic equations into the right-hand side of \eqref{ec:tps1}.} 
\begin{equation}
\Delta^2_{\dphis}=\frac{1}{1-\epsilon_\cl}\frac{\diff}{\diff N}\text{E}(\dphis^2) \approx \frac{1}{1-\epsilon_\cl}\left(D_{\phi\phi}+\frac{6 (1 +Q_\cl) D_{\pi\phi}+D_{\pi\pi}}{9(1+Q_\cl)^2}\right).
\end{equation}
In slow-roll inflation, when taking the $\sigma \to 0$ limit, the only non-vanishing diffusion coefficient is $D_{\phi\phi}$, while $D_{\pi\pi},\,D_{\pi\phi}\to 0$. In the case of warm inflation, we see that the latter are further suppressed by the dissipative coefficient $Q$. We thus conclude that, in the $\sigma\to 0$ limit,
\begin{equation}
\Delta^2_{\dphis} =  \frac{1}{1-\epsilon_\cl}D_{\phi\phi}= \langle \mathcal{P}_{\phi\phi} \rangle.
\end{equation}
This expression implies that the stochastic inflation formalism result coincides with standard perturbation theory at linear order in small fluctuations.

\par \medskip
\noindent\textbf{Comparison with previous results.} To the best of our knowledge, the stochastic-inflation formalism has been employed once in the context of warm inflation, in Ref.~\cite{Ramos:2013nsa}. Our analysis yields different results from those of Ref.~\cite{Ramos:2013nsa}. First, we emphasize that considering a Bose-Einstein distribution for the inflaton fluctuation occupation number, which may stem from a population thermalized with the radiation plasma (see e.g.\ \cite{Hall:2003zp}, \cite{Ramos:2013nsa}), is not a requirement {of the stochastic inflation formalism}. As we have discussed, the correction that such a population of inflaton fluctuations induces on the spectrum of curvature fluctuations {can be equally obtained through the quantization of the homogeneous solution of linear perturbations, as in Sec.\,\ref{quantumn}}. Our treatment also differs from Ref.~\cite{Ramos:2013nsa} in other respects. In Ref.\,\cite{Ramos:2013nsa}, quantum diffusion {is described to be} a second source of (effectively classical) noise, independent of the thermal noise. This is a consequence of assuming that the thermal noise does not appear in the classical equation of motion for the Fourier modes of $\delta \phi$. In our work, we show that starting from a consistent quantization procedure for the inflaton implies that quantum effects and the classical thermal noise jointly contribute to the dynamics of large scales. In particular, we show that the thermal noise is implicit in the quantum noise through the short-wavelength Fourier modes of the inhomogeneous solution of the inflaton perturbations. Moreover, our work shows the agreement between the stochastic inflation approach and linear perturbation theory when the quantization scheme we introduced is consistently applied. As a final consistency check, we find that linear perturbation theory itself naturally recovers the cold inflation limit if the homogeneous contribution to the power spectrum is not discarded. This conclusion is supported by the results of App.\,\ref{app:stochasticv2}, which follows the approach of \cite{Ramos:2013nsa} and agrees with the main results of this section.

\section{Summary and conclusions} \label{summconc}

We have explored the possibility for the inflaton to release a fraction of its energy into a thermal radiation bath during inflation in the context of a scenario known as {\it{warm inflation}}. Such a dissipative effect manifests as an additional friction term in the background equation for the inflaton field and as a stochastic source for its perturbations. We have considered several monomial inflaton potentials and power-law dissipation coefficients, which are summarized in Tab.\,\ref{tab:my_table2}.
\begin{table}[h!] \vspace{0.3cm}
    \centering
    \begin{tabular}{c||c|c|c}
   &  $\phi^6$  &  $\phi^4$  &    $\phi^2$  
\\
        \hline
        \hline
   $T$    & \faCheck  & \faCheck & \faTimes  \\ \hline
      $T^3$  & \faTimes & \faCheck & \faTimes  \\  \hline
       $T^3/\phi^2$ & \faTimes & \faTimes & \faTimes  \\ 
    \end{tabular}
    \caption{\small \it Summary of the compatibility with CMB data of the dissipation rates and inflaton potentials considered in this work. \faCheck$\,$ means compatible. \faTimes$\,$   means incompatible. These results assume no  thermal correction due to the occupation number of inflaton perturbations.}
    \label{tab:my_table2}
\end{table}

\noindent
Our main results and methods are the following:\vspace{-0.2cm} \newline \newline 
\noindent
\textbf{Analytical estimates.} We have described a new analytical approximation that solves the system of stochastic differential equations for the perturbations in the slow-roll attractor. It accounts for the backreaction of radiation fluctuations that acts as a source in the equation for inflaton fluctuations, which was not included in previous purely analytical analyses (i.e. Eq.\,\eqref{ec:frompapers} \emph{without} the numerically-fitted correction $G(Q)$). The inclusion of this backreaction term makes our purely analytical approximation improve over the previous one, but does not prevent it from eventually failing at large $Q$. We have derived expressions for the curvature power spectrum and scalar spectral index, $n_s$, that require only a numerical evaluation of background quantities. We detail various approximations and simplifications which allow to achieve a good accuracy for dissipative coefficients $Q_*<\mathcal{O}(10^{-1})$, relevant for current CMB limits. Beyond this point, as mentioned above, purely analytical results fail to capture the power spectrum, which therefore has to be computed numerically. We have also described a quantization scheme for the perturbations sourced by the classical noise giving rise to the usual classical equations for the mode functions encountered in warm inflation.  We find (analytically) that the Gaussian nature of the noise implies a universal (scale-independent) distribution for the power spectrum that was noted in \cite{Ballesteros:2022hjk}. \par \medskip

\noindent
\textbf{Numerical results and comparison with CMB limits.} We have employed a method introduced in Ref.\,\cite{Ballesteros:2022hjk} in order to solve the system of stochastic differential equations for the perturbations numerically. It consists of a Fokker-Planck approach to obtain a system of deterministic linear differential equations for the two-point correlation function of the perturbations. This approach, to which we refer in the text as the \textit{matrix formalism}, allows to compute the power spectrum of scalar perturbations. Armed with this efficient tool, we have confronted the nine models we have considered with current constraints on the amplitude and spectral index of curvature fluctuations, $A_s$ and $n_s$, and the tensor-to-scalar ratio, $r$. Our results are summarized in Tab.\,\ref{tab:my_table2}. The models that are excluded by the data are so, not because they feature a too large tensor-to-scalar ratio, but because their scalar spectral index is too large. Models such as $\lambda\phi^4$ with a dissipation coefficient proportional to the temperature are close to being excluded for this reason, making conservative assumptions about reheating (see below). Although we have not checked it in detail, we conjecture that monomial potentials with even $n>6$ are ruled out for the dissipation coefficients considered in this work. \par \medskip

\noindent
\textbf{Reheating.} For each case, we have studied the post-inflationary evolution of the background inflaton and radiation energy densities depending on the value of the dissipative coefficient at the end of inflation. We have deduced the conditions required for a transition to a radiation dominated universe and used these conditions to account for the uncertainties in the post-inflationary cosmology when constraining $r$ and $n_s$. Some models allow a smooth transition into radiation domination after inflation, whereas in some cases extra species are definitely required for successful reheating. For the sake of generality, we consider the possibility that the Universe is either matter or kination dominated after inflation when we compare specific models with CMB data. \par \medskip

\noindent
\textbf{Stochastic inflation formalism.} We have applied the stochastic inflation formalism to warm inflation, in order to quantify the effects of small wavelength perturbations on large scales. We show that starting with a consistent quantization procedure for the inflaton implies that quantum effects and the classical thermal noise jointly contribute to affect large scales. With this approach, in the slow-roll limit, we recover the power spectrum of curvature fluctuations on large scales from linear perturbation theory. Consistently, we recover the cold-inflation result in the limit where the dissipation coefficient is negligible. We show that a (Bose-Einstein) correction for the primordial spectrum that has been frequently used in the context of warm inflation does not follow directly from the stochastic inflation formalism. The correction may arise when the fluctuations of the inflaton are quantized, assuming that they have a large occupation number. \par \medskip

\noindent
\textbf{The occupation number of inflation fluctuations and predictiveness.} Whether this Bose-Einstein correction is actually required is a model dependent issue that needs to be analyzed for each Lagrangian implementation of warm inflation using the methods of finite temperature quantum field theory. In fact the term coming from the occupation number of inflaton fluctuations, if needed in concrete examples, may take a different form from the commonly assumed Bose-Einstein form \cite{Bastero-Gil:2017yzb}. The effective description of warm inflation we have used in this work does not provide an answer to these questions and we have not included this correction in our main results. Ultimately, this means that a proper assessment of the predictions of inflationary models featuring a thermal bath (or in general large particle production) needs to be done at the particle physics level.
\par \medskip

\noindent
\textbf{Accuracy of our results.} We have solved the Langevin equations that describe the system of metric-inflation-radiation fluctuations for multiple realizations of the stochastic source term and averaged over the resulting spectra. Our results with this technique are found to agree at the percent level with results from the matrix formalism. Limited by finite sampling effects, solving for $n$ realizations of the Langevin equations can only allow to achieve accurate results at large $n$. We argue that percent level accuracy can only be achieved for $n\sim \mathcal{O}(10^4)$ realizations. We compare our results with semi-analytical, numerically-fitted expressions for the amplitude of the power spectrum and scalar spectral index commonly encountered in the literature. We find that such expressions (at least with the functional form for the numerically-fitted correction $G(Q)$ commonly assumed) fail to capture the properties of the power spectrum with the $\gtrsim\mathcal{O}(1\%)$ accuracy required to confront model of inflation with CMB data. We conclude that only the matrix formalism approach (introduced in~\cite{Ballesteros:2022hjk}) allows for a fast and reliable determination of the scalar power spectrum and we advocate its use in future studies of warm inflation.  \par \medskip

We expect that this work contributes to set warm inflation on firmer ground at a phenomenological level and help to compare this scenario of inflation with CMB bounds.

\appendix

\begin{acknowledgments}
The authors thank Pasquale Serpico for useful discussions and Rudnei Ramos for comments. APR thanks Arturo de Giorgi for discussions about the quantization of stochastic scalar fields.  The work of GB has been funded by a Contrato de Atracci\'on de Talento (Modalidad 1) de la Comunidad de Madrid (Spain), 2017-T1/TIC-5520 and 2021-5A/TIC-20957. This work is partially supported by PID2021-124704NB-I00 funded by MCIN/AEI/10.13039/501100011033 and by ERDF A way of making Europe. This work is partially supported as well  by the Spanish Research Agency (Agencia
Estatal de Investigación) through the Grant IFT Centro de Excelencia Severo
Ochoa No CEX2020-001007-S, funded by MCIN/AEI/10.13039/501100011033. APR has been supported by Universidad Aut\'onoma de Madrid with a PhD contract {\it contrato predoctoral para formaci\'on de personal investigador (FPI)}, call of 2021. MP acknowledges support by the Deutsche Forschungsgemeinschaft (DFG, German Research Foundation) under Germany's Excellence Strategy – EXC 2121 “Quantum Universe” – 390833306. MP thanks the IFT UAM-CSIC for hospitality during part of the realization of this work.
\end{acknowledgments}

\section{Matrices of the matrix formalism} \label{matrices}

In this appendix we provide the matrices needed to compute the primordial spectrum of curvature fluctuations following the method discussed in Sec.\,\ref{sec:computationspectrum}. These expressions were first presented in Ref.\,\cite{Ballesteros:2022hjk}, where this matrix method was introduced. The $4\times 4$ matrix ${\bm A}$ is given by
\begin{equation}\label{ec:matrixA}
{\bm A}=
\begin{pmatrix}
f_\psi & f_\rho & f_{\d\phi} & f_\phi \\

g_\psi+4\rho_rf_\psi & g_\rho+4\rho_rf_\rho & g_{\d\phi}+4\rho_rf_{\d\phi} & g_\phi+4\rho_rf_\phi \\

h_\psi+4(\d\phi/\d N)f_\psi & h_\rho+4(\d\phi/\d N)f_\rho & h_{\d\phi}+4(\d\phi/\d N)f_{\d\phi} & h_\phi+4(\d\phi/\d N)f_\phi \\

0 & 0 & -1 & 0 

\end{pmatrix},
\end{equation}
where 
\begin{align}
f_\psi&=1+\frac{k^2}{3a^2H^2}-\frac{1}{6M_P^2}\bigg(\frac{\d\phi}{\d N}\bigg)^2, & g_\psi&=\Gamma H\bigg(\frac{\d\phi}{\d N}\bigg)^2-\frac{k^2}{3a^2} \bigg[2M_P^2\frac{k^2}{a^2H^2}-\bigg(\frac{\d\phi}{\d N}\bigg)^2\bigg],\nonumber
\\
f_\rho&=\frac{1}{6M_P^2H^2}, & g_\rho&=4-\Gamma_T\frac{HT}{4\rho_r}\bigg(\frac{\d\phi}{\d N}\bigg)^2-\frac{k^2}{3a^2H^2},\nonumber
\\
f_{\d\phi}&=\frac{1}{6M_P^2}\frac{\d\phi}{\d N}, & g_{\d\phi}&=-\left(\frac{k^2}{3a^2}	+2\Gamma H\right)\frac{\d\phi}{\d N}, \nonumber
\\
f_\phi&=\frac{V_\phi}{6M_P^2H^2}, & g_\phi&=-\frac{k^2}{3a^2H^2}\left(3H^2\frac{\d\phi}{\d N}+V_\phi\right)-H\Gamma_\phi\left(\frac{\d\phi}{\d N}\right)^2,\nonumber
\\
h_\psi&=2\frac{V_\phi}{H^2}+\frac{\Gamma}{H}\frac{\d\phi}{\d N}, & h_\rho&=\frac{T\,\Gamma_T}{4H\rho_r}\frac{\d\phi}{\d N},\nonumber
\\
h_{\d\phi}&=3+\frac{\Gamma}{H}+\frac{1}{H}\frac{\d H}{\d N}, & h_\phi&=\frac{k^2}{a^2H^2}+\frac{V_{\phi\phi}}{H^2}+\frac{\Gamma_\phi}{H}\frac{\d\phi}{\d N}.
\end{align}

The column vectors ${\bm B}$ and ${\bm C}$ are
\begin{align} \label{eq:BandC}
{\bm B}&=\begin{pmatrix}
0 \\
-\sqrt{2\Gamma T H/a^3}\left(\d\phi/\d N\right) \\
\sqrt{2\Gamma T/(aH)^3} \\
0 \\
\end{pmatrix},
&
{\bm C}&=\frac{1}{3H^2\left(\frac{\d\phi}{\d N}\right)^2+4\rho_r}
\begin{pmatrix}
2M_P^2k^2/a^2-4H^2\left(\frac{\d\phi}{\d N}\right)^2-4\rho_r \\
1 \\
H^2\frac{\d\phi}{\d N} \\
V_{\phi}\\
\end{pmatrix}.
\end{align}
The initial conditions are given by 
\begin{equation}
\label{eq:initialmatrix}
{\bm Q}_i=
\frac{1}{2ka^2(N_\text{ini})}\begin{pmatrix}
 0 & 0 & 0 & 0\\
 0 & 0 & 0 & 0\\
 0 & 0 & 1+(k/k_i)^2 & -1-i(k/k_i)\\
 0 & 0 & -1+i(k/k_i) & 1
\end{pmatrix},
\end{equation}
where, as discussed in Sec.\,\ref{seclav}, $k_i$ is the scale that crosses the horizon at the time at which the initial conditions are set. In practice, we can start integrating at some time $N_{\rm ini}$ such that $k/k_i\simeq100$. As discussed in Ref.\,\cite{Ballesteros:2022hjk}, the choice of initial conditions does not affect strongly the result provided that they are set early enough.

\paragraph*{A more efficient implementation of the matrix formalism}

The numerical integration of \eqref{matrix_eq} for the $\bm{A}$ matrix in \eqref{ec:matrixA} with the initial conditions in \eqref{eq:initialmatrix} can be quite expensive from the point of view of computation time. The reason behind this is that, due to the Bunch-Davies initial conditions for $\ddp$, $\ddp'$, the system oscillates towards the thermal attractor (if dissipation is strong enough) or towards standard cold-inflation freeze-out (if dissipation is weak). Since all perturbations are coupled, not only inflaton perturbations oscillate, but also $\ddr$. These oscillations of $\ddr$ are, however, futile and computationaly costly. Since $\langle|\ddr|^2\rangle$ is ultimately driven to a thermal attractor by the noise, its evolution deep inside the horizon does not affect its final value. Nevertheless, numerically evolving oscillating systems is very time consuming, hence the slowness of the code.\par
The following trick allows to circumvent this issue and integrate numerically \eqref{matrix_eq} faster. If we restore the dependence on $\delta q_r$ in \eqref{ec:b01}-\eqref{ec:b03} through \eqref{ec:b04}, we get a system of five stochastic differential equations for the variables \(\ddr,\delta q_{r,\bm{k}},\psi_{\bm{k}},\ddp,\ddp'\) whose $\bm{A}$ matrix reads
\begin{equation}\label{ec:5t5}
{\bm A}=
\begin{pmatrix}
G_\rho +4\rho_r f_\rho& -H k^2/(a^2H^2) & G_\psi +4\rho_r f_\psi & G_\phi +4\rho_r f_\phi& G_{\d\phi} +4\rho_r f_{\d\phi}\\

1/(3H) & 3 & 4\rho_r/(3H) & \Gamma (\d\phi/\d N) & 0 \\

f_\rho & 0 & f_\psi & f_\phi & f_{\d\phi} \\

0 & 0 & 0 & 0 & -1 \\

h_\rho +4(\d\phi/\d N)f_\rho& 0 & h_\psi +4(\d\phi/\d N)f_\psi& h_\phi +4(\d\phi/\d N)f_\phi& h_{\d\phi} +4(\d\phi/\d N)f_{\d\phi}
\end{pmatrix},
\end{equation}
where
\begin{align}
G_\rho&=g_\rho+\frac{k^2}{3a^2H^2},\\
G_\psi&=g_\psi+\frac{k^2}{3a^2} \bigg[2M_P^2\frac{k^2}{a^2H^2}-\bigg(\frac{\d\phi}{\d N}\bigg)^2\bigg],\\
G_\phi&=g_\phi+\frac{k^2}{3a^2H^2}\bigg[3H^2 \frac{\d\phi}{\d N}+V_\phi\bigg],\\
G_{\d\phi}&=g_{\d\phi}+\frac{k^2}{3a^2}\frac{\d\phi}{\d N},
\end{align}
and the functions $f$, $g$ and $h$ are the same as above. After some algebra, the new $\bm{B}$ and $\bm{C}$ vectors are 
\begin{align}
{\bm B}&=\begin{pmatrix}
-(\d\phi/\d N)\sqrt{2\Gamma T H/a^3} \\
0 \\
0 \\
0 \\
\sqrt{2\Gamma T/(aH)^3} \\
\end{pmatrix},
&
{\bm C}&=\frac{1}{3H^2(\d\phi/\d N)^2+4\rho_r}
\begin{pmatrix}
0 \\
3H \\
-3H^2(\d\phi/\d N)^2-4\rho_r \\
-3H^2(\d\phi/\d N)\\
0
\end{pmatrix}.
\end{align}
Appropriately setting the initial conditions is what actually makes this approach significantly faster. Choosing
\begin{equation}
{\bm Q}_i=
\frac{1}{2ka^2(N_\text{ini})}\begin{pmatrix}
 0 & 0 & 0 & 0 & 0\\
 0 & 0 & 0 & 0 & 0\\
 0 & 0 & 0 & 0 & 0\\
 0 & 0 & 0 & 1 & -1+i(k/k_i)\\
 0 & 0 & 0 & -1-i(k/k_i) & 1+(k/k_i)^2
\end{pmatrix},
\end{equation} 
imposes $\delta q_{r,\bm{k}}=0$ initially. The superhorizon backreaction of the (naturally oscillating) $\ddp$, $\ddp'$ into $\ddr$ is removed. Therefore, $\ddr$ reaches its thermal attractor (the same one as in the $4\times 4$ system), but it does so without oscillating, which saves significant computation time.

\section{Statistics and the determination of the power spectrum}
\label{app:statistics}

In this appendix, we briefly review some elements about statistics of stochastic variables in order to ease the connection between the Langevin approach employed to compute the power spectrum and the matrix formalism. We start by general considerations about the precision of the Langevin approach. In a second part, we discuss the connection between the curvature perturbation and the power spectrum.

\subsection{Intrinsic limitation of the Langevin approach}
\label{app:limitationLangevin}Let us consider $n$ copies of a stochastic variable $X$ and label them $X_i$, with $i=1,\ldots,n$. If $\mu$ and $\sigma^2$ are the average and variance of $X$, then the average and variance of $\mathcal{S}=\sum_{i=1}^n X_i$ are $\bar{\mathcal S} = n\,\mu$ and $\sigma_{\mathcal S}^2= n\,\sigma^2$, respectively. If we were to determine $\mu$ by drawing a value from each stochastic variable $X_i$ and averaging over, the expected error would be given by the ratio
\begin{align} \label{eq:stdevoverE}
\frac{\sqrt{\sigma_{\mathcal S}^2}}{\bar{\mathcal S}}=\dfrac{\sigma}{ \sqrt{n}\mu}\,.
\end{align}
This shows that the determination of $\mu$ is intrinsically limited by the variance $\sigma^2$, but the precision on its evaluation increases with $n$.

 We can apply this consideration to the determination of the power spectrum of curvature fluctuations from the Langevin approach, considering that each time we solve the system of stochastic differential equations to compute $\mathcal{P}_\mathcal{R}$ we are drawing values for the real and imaginary parts of $\mathcal{R}$ from a stochastic variable. Since the distributions of the real and imaginary parts of $\mathcal{R}$ are Gaussians centered around zero, the standard deviation and average for the stochastic variable $\mathcal{P}_\mathcal{R} \propto |\mathcal{R}|^2$ are of the same order, i.e. $\sigma_{\mathcal{P}_\mathcal{R}} \sim \mu_{\mathcal{P}_\mathcal{R}}$. The relative error on the expected value of $\mathcal{P}_\mathcal{R}$ achieved by averaging over $n$ realizations can be estimated by setting $\sigma \simeq \mu$ in Eq.\,(\ref{eq:stdevoverE}) as
\begin{equation}
\frac{1}{\sqrt{n}}\simeq \, 3\,\% \,\left(\dfrac{n}{10^3} \right)^{-1/2} \, \simeq \, 1\,\% \,\left(\dfrac{n}{10^4} \right)^{-1/2} \, \simeq \, 0.3\,\% \,\left(\dfrac{n}{10^5} \right)^{-1/2} \,.
\end{equation}
Therefore, to achieve an $\mathcal{O}(1\%)$ or better precision on the determination of $\langle \mathcal{P}_\mathcal{R} \rangle$, one would need at least $\mathcal{O}(10^4)$ stochastic realizations.
\subsection{Statistics for the curvature perturbation and power spectrum}

\label{app:distributionforR}

In the following we consider a fixed value of $k$ in Fourier space. The real ($x\equiv$ Re($\mathcal{R}_k$)) and  imaginary ($y\equiv$ Im($\mathcal{R}_k$)) parts of the stochastic variable $\mathcal{R} \equiv \mathcal{R}_k$ are themselves stochastic variables. Their joint probability distribution function $f_{xy}(x,y)$ satisfies the normalization condition
\begin{equation}
    \int_{- \infty}^{ \infty} f_{xy}(x,y) \diff x \diff y \, = \, 1 \,.
\end{equation}
Assuming that both variables, $x$ and $y$, are independent Gaussians with zero average ($\mu=0$) and standard deviation $\sigma$, the probability distribution function $f_{xy}(x,y)$ can be expressed as a product of Gaussian distributions satisfying
\begin{equation}
    \int_{- \infty}^{ \infty} \dfrac{1}{\sigma \sqrt{2\pi}}e^{-x^2/(2\sigma^2)}\diff x    \int_{- \infty}^{ \infty} \dfrac{1}{\sigma \sqrt{2\pi}}e^{-y^2/(2\sigma^2)} \diff y \, = \, 1 \,.
    \label{eq:jointdistribution}
\end{equation}
We can define the new variable
\begin{equation}
  \Delta \, \equiv \, \log_{10} \mathcal{P}_\mathcal{R} - \log_{10} (\langle \mathcal{P}_\mathcal{R} \rangle)  \qquad \text{with} \qquad \mathcal{P}_\mathcal{R} \, \equiv \, \dfrac{k^3}{2\pi^2} |\mathcal{R}|^2 \, = \, \dfrac{k^3}{2\pi^2}\big( x^2 + y^2 \big)  \, .
\end{equation}
The brackets $\langle ... \rangle$ denote the ensemble average, i.e. the average over stochastic realizations. Eq.\,(\ref{eq:jointdistribution}) can be expressed in terms of $\Delta$ as
\begin{equation}
\int_{-\infty}^\infty \log(10) \, 10^\Delta \exp(-10^\Delta) \diff \Delta \, \equiv \,  \int_{-\infty}^\infty \mathcal{F}(\Delta)\diff \Delta \, = \,1 \,,
\end{equation}
where $\mathcal{F}(\Delta)$ is the probability distribution function for the stochastic variable $\Delta$. The probability  distribution for $\Delta$ does not depend on the variance $\sigma^2$ of the distribution for the real and imaginary parts of $\mathcal{R}$ and it does not depend on the scale $k$. Indeed, $\mathcal{F}$ does not depend on any free parameter. Therefore we expect this distribution to be universal (i.e. $k$-independent) provided that $\mathcal{R}$ is Gaussian distributed for all $k$ and that both real and imaginary parts are independent variables.

Following the same assumptions, the variance of the stochastic variable  $\mathcal{R} \equiv \mathcal{R}_k$  is given by
\begin{equation}
    \langle |\mathcal{R}|^2 \rangle \, =  \, \langle x^2 + y^2 \rangle \, =  \, \int \dfrac{1}{\sigma \sqrt{2\pi}} \, e^{-x^2/(2 \sigma^2)} \diff x\int  \dfrac{1}{\sigma \sqrt{2\pi}} \, e^{-y^2/(2 \sigma^2)} \diff y  \, \Big(x^2+y^2 \Big) \, = \,2  \sigma^2 \, .
\end{equation}
From the definition of the power spectrum $\mathcal{P}_\mathcal{R} \, \equiv \, k^3/(2\pi^2) |\mathcal{R}|^2 $ it follows that
\begin{equation}
  2 \sigma^2 \, = \,   \langle |\mathcal{R}|^2 \rangle \, = \, \dfrac{2\pi^2}{ k^3} \langle \mathcal{P}_\mathcal{R} \rangle \, ,
\end{equation}
which implies
\begin{equation}
     \langle \mathcal{P}_\mathcal{R} \rangle \, = \, k^3\dfrac{\sigma^2}{\pi^2 } \,.
\end{equation}
This quantity can be estimated by averaging over several stochastic realizations, which is the same quantity that is determined with the matrix formalism.

\section{Reheating}
\label{app:reheating}

In this section, we study the asymptotic post-inflationary behaviour of the radiation and inflaton energy densities for a generic power law $\Gamma(T)$ and potential $V(\phi)$,
\begin{equation}\label{ec:q19.5}
V(\phi)=\frac{\lambda}{n}\left(\frac{\phi}{\mm}\right)^n \mm^{4}\,, \qquad \qquad \Gamma(T) = C\left(\frac{T}{\mm}\right)^\beta\mm \,.
\end{equation}
Our goal is to determine whether a radiation dominated universe can be achieved after inflation without requiring extra couplings or fields. We seek to find asymptotic solutions to the continuity equations 
\begin{align}
\rp ' &= -3(1+Q)\,H^2\,\phi'^2\,, \label{ec:q20}\\
\rr ' &= -4\rr + 3Q\,H^2\,\phi'^2\,,\label{ec:q21}
\end{align}
outside of the inflationary attractor. We approach the problem in the weak ($Q\ll 1$) and strong ($Q\gg 1$) dissipation regimes. To lighten the notation, we set $N=0$ to be the end of inflation (defined by the condition $\epsilon=1)$.  

\subsection{Weak dissipation limit ($Q\ll 1$)}\label{ss:wd}

If $Q\ll 1$ at the end of inflation, $\rr(0) \ll \rp(0)$. In this limit, we can neglect the $Q$ term in \eqref{ec:q20}. We then have
\begin{equation}\label{ec:q22}
\rp ' = -3H^2\,\phi'^2\,.
\end{equation}
We assume inflation ends close to a minimum of the potential, around which the inflaton oscillates.\footnote{In cold inflation, the inflaton is subject to Hubble friction. When it reaches the minimum of its potential, it oscillates with a decreasing amplitude. This means that the damping due to Hubble friction is undercritical. If $Q\ll 1$, the damping $3H(1+Q)$ is still undercritical, hence there are also oscillations of the inflaton around its minimum.} In this scenario, Eq.\,\eqref{ec:q22} can be solved (see e.g.\ \cite{turner1983coherent} for a detailed calculation). For a potential as in \eqref{ec:q19.5}, one has
\begin{equation}\label{ec:q26}
\rp(N) = \rp(0) \,e^{-\frac{6n}{n+2} N} \qquad \text{and} \qquad   H^2\langle\phi'^2\rangle=\frac{2n}{n+2} \rp\,,
\end{equation}
where $\langle\cdot\rangle$ represents the average over oscillations of the inflation around the minimum of its potential.\footnote{Not to be confused with an average over realizations of the thermal noise. In this appendix, there is no thermal noise since we work only at background level.} The dissipation rate can be expressed as
\begin{equation}\label{ec:q28}
3Q = \frac{\Gamma}{H} = \frac{\sqrt{3} C}{\kappa^{\beta/4}} \left(\frac{\rr}{\mm^4}\right)^{\beta/4}\left(\frac{\rp}{\mm^4}\right)^{-1/2}\,,
\end{equation}
with $\kappa \equiv g_\star \pi^2 /30$.  Substituting \eqref{ec:q26} and \eqref{ec:q28} into \eqref{ec:q21} yields
\begin{equation}\label{ec:q29}
\rr ' = -4\rr + A_1\left(\frac{\rr}{\mm^4}\right)^{\beta/4}e^{-\frac{3n}{n+2}  N}\mm^4\,,
\end{equation}
where $A_1$ is a constant, see Eq.\,(\ref{eq:defA1A2}). Eq.\,(\ref{ec:q29}) can be solved analytically. \par \medskip

\noindent
\textbf{For $\beta=4$.} In this case the radiation energy density can be expressed as
\begin{equation}
\rr(N) = c_1 \exp{ \left(-4N - \frac{(2+n)A_1}{3n}e^{-\frac{3n}{2+n}N}\right)} \mm^4\simeq c_1 e^{-4N}\mm^4\,,
\end{equation}
where $c_1$ is a dimensionless integration constant. \par \medskip

\noindent
\textbf{For $\beta\neq 4$.} The solution to (\ref{ec:q29}) is given by
\begin{equation}\label{ec:q29.5}
\rr(N) = A_2\, e^{-\frac{12n}{(4-\beta)(n+2)}N} \mm^4+ c_2\, e^{-4N}\mm^4 \sim 
\begin{cases}
A_2\,\exp\left({-\frac{12n}{(4-\beta)(n+2)}N} \right)\mm^4   \quad &\text{if} \quad \frac{12n}{(4-\beta)(n+2)}<4\,,\\
c_2 \,e^{-4N}\mm^4 \quad &\text{if}\quad  \frac{12n}{(4-\beta)(n+2)}>4\,,
\end{cases}
\end{equation}
where $c_2$ is a dimensionless integration constant and
\begin{equation}
A_1\equiv \frac{\sqrt{3\rp(0)\,\mm^{-4}}\,C }{\kappa^{\beta/4}}\frac{2n}{n+2}\,, \qquad \qquad A_2 \equiv \left[\frac{(4-\beta)\,(n+2)\,A_1}{4\,[(n+2)\,\beta-(n+8)]}\right]^{\frac{4}{4-\beta}}\,.
\label{eq:defA1A2}
\end{equation}
In order to complete this analysis, we need to determine the time evolution of $Q$. We do so by substituting \eqref{ec:q26} and \eqref{ec:q29.5} into \eqref{ec:q28}. If $Q$ decreases with $N$, the weak dissipation regime persists and the analysis above holds indefinitely. However, if $Q$ increases with $N$, at some point $Q$ is no longer small and we enter the strong dissipative regime (which we study next).

\subsection{Strong dissipative regime ($Q\gg 1$)}\label{ss:sd}

If $Q\gg 1$ at the end of inflation, $\rr(0) \simeq \rp(0)$. After the end of inflation, large values of $Q$ favour an efficient energy transfer from $\rp$ to $\rr$. Hence, we can make the rough approximation that the total energy is rapidly dominated by radiation. As in the previous subsection, we need to decouple one of the continuity equations. Unlike in the $Q\ll 1$ case, it is impossible to do so neglecting $Q$ in \eqref{ec:q20}. Under the reasonable assumption that $\rr$ redshifts away after the end of inflation ($\rr'<0$), from \eqref{ec:q21} we conclude that such condition can be achieved only if $\rr > QH^2\phi'^2$ shortly after the end of inflation.\footnote{Another way of understanding this is: since the inflaton component is subdominant at this stage, it cannot efficiently source $\rho_r$ via $Q$ in \eqref{ec:q21} whose right-hand side is dominated by the redshift term.} Hence, we assume that the $Q$ term in \eqref{ec:q21} can be neglected. The equation for $\rr$ then yields $\rr ' = -4\rr $. This implies $ \rr = \rr(0) e^{-4N}$, which expresses the usual redshift of radiation. The equation for $\rp$ reads
\begin{equation}\label{ec:q40.5}
\rp' =  -3H^2\,(1+Q)\,\phi'^2\approx -3H^2Q\,\phi'^2\,.
\end{equation}
Assuming $\rr>\rp$, as mentioned above, the Hubble rate can be expressed only in terms of $\rr$
\begin{equation}\label{ec:q41}
3H^2 \simeq \frac{\rr}{\mm^2} = \frac{\rr(0)e^{-4N}}{\mm^2} \,.
\end{equation}
The dissipation rate $Q$ can therefore be expressed as
\begin{equation}\label{ec:q42}
Q = \frac{\Gamma}{3H} = \frac{C}{\sqrt{3}\kappa^{\beta/4}} \left(\frac{\rr}{\mm}\right)^{\beta/4-1/2} = \frac{C}{\sqrt{3}\kappa^{\beta/4}}\,\left(\frac{\rr(0)}{\mm^4}\right)^{\beta/4-1/2}\,e^{-(\beta-2)N}\,. 
\end{equation}
Due to the strong dissipation, the inflaton does not oscillate around its minimum but instead asymptotically relaxes towards it.\footnote{In the weak dissipative regime, the damping due to Hubble friction is undercritical. However, in the strong dissipative regime, $Q$ becomes large enough for the damping due to dissipative friction to be (over)critical.} Hence, the dominant contribution to the inflaton energy density is the potential, which allows to write
\begin{equation}\label{ec:q43}
\rp' \simeq  \dv \,\phi' = \lambda \left(\frac{\phi}{\mm}\right)^{n-1}\frac{\phi'}{\mm}\, \mm^{4}\,.
\end{equation}
Substituting \eqref{ec:q41}, \eqref{ec:q42} and \eqref{ec:q43} into \eqref{ec:q40.5}, provides a differential equation satisfied by $\phi$
\begin{equation}\label{ec:q45}
\left(\frac{\phi}{\mm}\right)^{n-1} = -A_3\,  e^{-(2+\beta)N} \,\left(\frac{\phi'}{\mm}\right),\qquad \text{with} \qquad A_3\equiv \frac{C}{\sqrt{3}\,\lambda\,\kappa^{\beta/4}}\,\left(\frac{\rr(0)}{\mm^4}\right)^{1/2+\beta/4}>0\,.
\end{equation}
The solution for $n\neq 2$ is
\begin{equation}\label{ec:q46}
 \phi = \left(\frac{1}{A_3}\,\frac{n-2}{\beta+2}\right)^{-\frac{1}{n-2}}e^{-\frac{\beta+2}{n-2}N} \mm\,,
\end{equation}
where we have neglected an irrelevant integration constant. Hence, we can express the inflaton energy density as
\begin{equation}
\rho_\phi(N) \simeq  \frac{\lambda}{n}\left(\frac{1}{A_3}\,\frac{n-2}{\beta+2}\right)^{-\frac{n}{n-2}}e^{-\frac{(\beta+2)n}{n-2}N} \mm^4\,.
\end{equation}
For the $n=2$ case, we have
\begin{equation}
\phi(N)= c_3\, \exp{\left(-\frac{e^{(\beta+2)N}}{(\beta+2)A_3}\right)} \mm\,,
\end{equation}
which implies
\begin{equation}
\rp(N) \simeq \frac{ \lambda \,c_3^2}{2} \exp{\left(-\frac{2e^{(\beta+2)N}}{(\beta+2)\,A_3}\right)}\mm^4\,,
\end{equation}
where $c_3$ is a dimensionless integration constant. This completes our analysis of the strong dissipation regime.

\subsection*{Transition between regimes}
Notice that the dominant species after the end of inflation is not fully determined by the value of $Q_{\text{end}}$. If $Q_{\text{end}}\ll 1$, then $\rp(0)\gg \rr(0)$. On the contrary, if  $Q_{\text{end}}\gg 1$, $\rp(0)\simeq \rr(0)$ (this can be seen from the slow-roll attractor equations). However, as already advanced, the later tendency of $\rp$ and $\rr$ might imply a change of the dominant species.\par
The dissipative regime that holds at the end of inflation (weak or strong) does not necessarily last indefinitely. If dissipation is weak at the end of inflation but $Q$ grows in time, the strong dissipative regime is eventually reached, and the behaviour of $\rp$, $\rr$ changes from the one described in Sec.\,\ref{ss:wd} to the one described in \ref{ss:sd}. Reciprocally, starting in strong dissipation and having a decreasing $Q$ eventually leads to the weak dissipation regime. \par

An important remark is that a change in the dissipative regime \emph{might} be accompanied by a change in the dominant species, but it does not necessarily have to. This is particularly relevant for the following reason. The behaviour of $\rp$ and $\rr$ is determined by the regime ($Q \gg 1$ or $Q \ll 1$). However, the behaviour of $Q$ depends also on the dominant species (which determines the main contribution to $H$). We will summarize the possible scenarios and provide detailed examples in the next couple of sections.

\subsection{Summary and cases of interest}
\label{sec:summary}

Tab.\,\ref{tab:1} summarizes the asymptotic time dependence of $\rho_\phi$, $\rho_r$ and $Q$ in the strong dissipation (SD) and weak dissipation (WD) regimes. The Tabs.\,\ref{tab:2}, \ref{tab:3} and \ref{tab:4} show the same quantities for quadratic, quartic and sextic potential respectively, for linear and cubic dissipation rates. These tables indicate the dominant species in the energy budget of the universe: inflaton domination (ID) or radiation domination (RD). In addition, transitions to a different regime or dominant species are indicated with an arrow.

\begin{table}[H]
\centering
\caption{Asymptotic behaviour of $\rp$, $\rr$ and $Q$ after the end of inflation for generic exponents $n$ and $\beta$ defined in Eq.\,(\ref{ec:q19.5}).}
\vspace{0.2in}
\begin{tabular}{c||c|c}\label{tab:1}
		&	$Q\ll 1$	& $Q\gg 1$ \\ \hline \hline &&\\[-1em]
$\rp$	& $\sim e^{-\frac{6n}{n+2} N}$	& \( \begin{array}{c} 
\sim e^{-\frac{2e^{(\beta+2)N}}{(\beta+2)A_3}} \quad \text{if} \quad n=2 \\[5pt] 
\sim e^{-\frac{(\beta+2)n}{n-2}N} \quad \text{otherwise}   \end{array} \) \\ \hline &&\\[-1em]
$\rr$ & \( \begin{array}{l} 
 \sim e^{-\frac{12n}{(4-\beta)(n+2)}N}   \quad \text{if} \quad \frac{12n}{(4-\beta)(n+2)}<4 \text{ and } \beta\neq4\\[5pt] 
 \sim e^{-4N} \quad \text{otherwise}  \end{array} \) & $\sim e^{-4N}$ \\ \hline &&\\[-1em]
$Q$ & \( \begin{array}{l} 
 \sim e^{\frac{6n(2-\beta)}{(4-\beta)(n+2)}N}   \quad \text{if} \quad \frac{12n}{(4-\beta)(n+2)}<4 \text{ and } \beta\neq4\\[5pt]
 \sim e^{\left(\frac{3n}{n+2}-\beta\right)N} \quad \text{otherwise}   \end{array} \) & $\sim e^{(2-\beta)N}$
\end{tabular}
\end{table}

\begin{table}[H]
\centering
\caption{Results for $V(\phi)\propto \phi^2$ with $\Gamma \propto T$ (left) and $\Gamma \propto T^3$ (right)  : asymptotic behaviour of $\rp$, $\rr$ and $Q$ after the end of inflation, dominant species for strong and weak dissipation regimes. See Sec.\,\ref{sec:summary} for details.}
\vspace{0.2in}
\begin{tabular}{c||c|c}\label{tab:2}
$\Gamma \propto T$		&	$Q_{\text{end}}\ll 1$	& $Q_{\text{end}}\gg 1$ \\ \hline \hline &&\\[-1em]
$\rp$	& $ e^{-3 N}$	& $ e^{-\frac{2e^{3N}}{3A_3}}$ \\ \hline &&\\[-1em]
$\rr$ & $ e^{-2N}$ & $ e^{-4N}$ \\ \hline &&\\[-1em]
$Q$ & $ e^{N}$ & $ e^{N}$\\ \hline &&\\[-1em]
Regime & WD $\to$ SD &  SD \\ \hline &&\\[-1em]
Dom. sp. & ID $\to$ RD &  RD
\end{tabular}\qquad  \qquad 
\begin{tabular}{c||c|c}
$\Gamma \propto T^3$		&	$Q_{\text{end}}\ll 1$	& $Q_{\text{end}}\gg 1$ \\ \hline \hline &&\\[-1em]
$\rp$	& $\sim e^{-3N}$	& $e^{-\frac{2e^{5N}}{5A_3}}$ \\ \hline &&\\[-1em]
$\rr$ & $\sim e^{-4 N}$ & $\sim e^{-4N}$ \\ \hline &&\\[-1em]
$Q$ & $\sim e^{-3 N/2}$ & $\sim e^{-N}$\\ \hline &&\\[-1em]
Regime & WD &  SD $\to$ WD \\ \hline &&\\[-1em]
Dom. sp. & ID &  RD $\to$ ID (late)
\end{tabular}
\end{table}

\begin{table}[H]
\centering
\caption{Results for $V(\phi)\propto \phi^4$ with $\Gamma \propto T$ (left) and $\Gamma \propto T^3$ (right): asymptotic behaviour of $\rp$, $\rr$ and $Q$ after the end of inflation, dominant species for strong and weak dissipation regimes. See Sec.\,\ref{sec:summary} for details.}
\vspace{0.2in}
\begin{tabular}{c||c|c}\label{tab:3}
$\Gamma \propto T$		&	$Q_{\text{end}}\ll 1$	& $Q_{\text{end}}\gg 1$ \\ \hline \hline &&\\[-1em]
$\rp$	& $ e^{-4 N}$	& $ e^{-6N}$ \\ \hline &&\\[-1em]
$\rr$ & $ e^{-8N/3}$ & $ e^{-4N}$ \\ \hline &&\\[-1em]
$Q$ & $ e^{4N/3}$ & $ e^{N}$\\ \hline &&\\[-1em]
Regime & WD $\to$ SD &  SD \\ \hline &&\\[-1em]
Dom. sp. & ID $\to$ RD &  RD
\end{tabular}\qquad  \qquad
\begin{tabular}{c||c|c}
$\Gamma \propto T^3$		&	$Q_{\text{end}}\ll 1$	& $Q_{\text{end}}\gg 1$ \\ \hline \hline &&\\[-1em]
$\rp$	& $\sim e^{-4N}$	& $e^{-10N}$ \\ \hline &&\\[-1em]
$\rr$ & $\sim e^{-4 N}$ & $\sim e^{-4N}$ \\ \hline &&\\[-1em]
$Q$ & $\sim e^{- N}$ & $\sim e^{-N}$\\ \hline &&\\[-1em]
Regime & WD &  SD $\to$ WD \\ \hline &&\\[-1em]
Dom. sp. & ID &  RD
\end{tabular}
\end{table}

\begin{table}[H]
\centering
\caption{Results for $V(\phi)\propto \phi^6$ with $\Gamma \propto T$ (left) and $\Gamma \propto T^3$ (right): asymptotic behaviour of $\rp$, $\rr$ and $Q$ after the end of inflation in addition to dominant species for strong and weak dissipation regimes. See Sec.\,\ref{sec:summary} for details.}
\vspace{0.2in}
\begin{tabular}{c||c|c}\label{tab:4}
$\Gamma \propto T$		&	$Q_{\text{end}}\ll 1$	& $Q_{\text{end}}\gg 1$ \\ \hline \hline &&\\[-1em]
$\rp$	& $ e^{-9 N/2}$	& $ e^{-9 N/2}$ \\ \hline &&\\[-1em]
$\rr$ & $ e^{-3N}$ & $ e^{-4N}$ \\ \hline &&\\[-1em]
$Q$ & $ e^{3N/2}$ & $ e^{N}$\\ \hline &&\\[-1em]
Regime & WD $\to$ SD &  SD \\ \hline &&\\[-1em]
Dom. sp. & ID $\to$ RD &  RD
\end{tabular}\qquad   \qquad
\begin{tabular}{c||c|c}
$\Gamma \propto T^3$		&	$Q_{\text{end}}\ll 1$	& $Q_{\text{end}}\gg 1$ \\ \hline \hline &&\\[-1em]
$\rp$	& $\sim e^{-9N/2}$	& $e^{-15 N/2}$ \\ \hline &&\\[-1em]
$\rr$ & $\sim e^{-4 N}$ & $\sim e^{-4N}$ \\ \hline &&\\[-1em]
$Q$ & $\sim e^{- 3N/4}$ & $\sim e^{-N}$\\ \hline &&\\[-1em]
Regime & WD &  SD $\to$ WD \\ \hline &&\\[-1em]
Dom. sp. & ID $\to$ RD &  RD
\end{tabular}
\end{table}

\subsection{Two detailed examples}

\begin{figure}[t]
\begin{center}
\includegraphics[width=.95\textwidth]{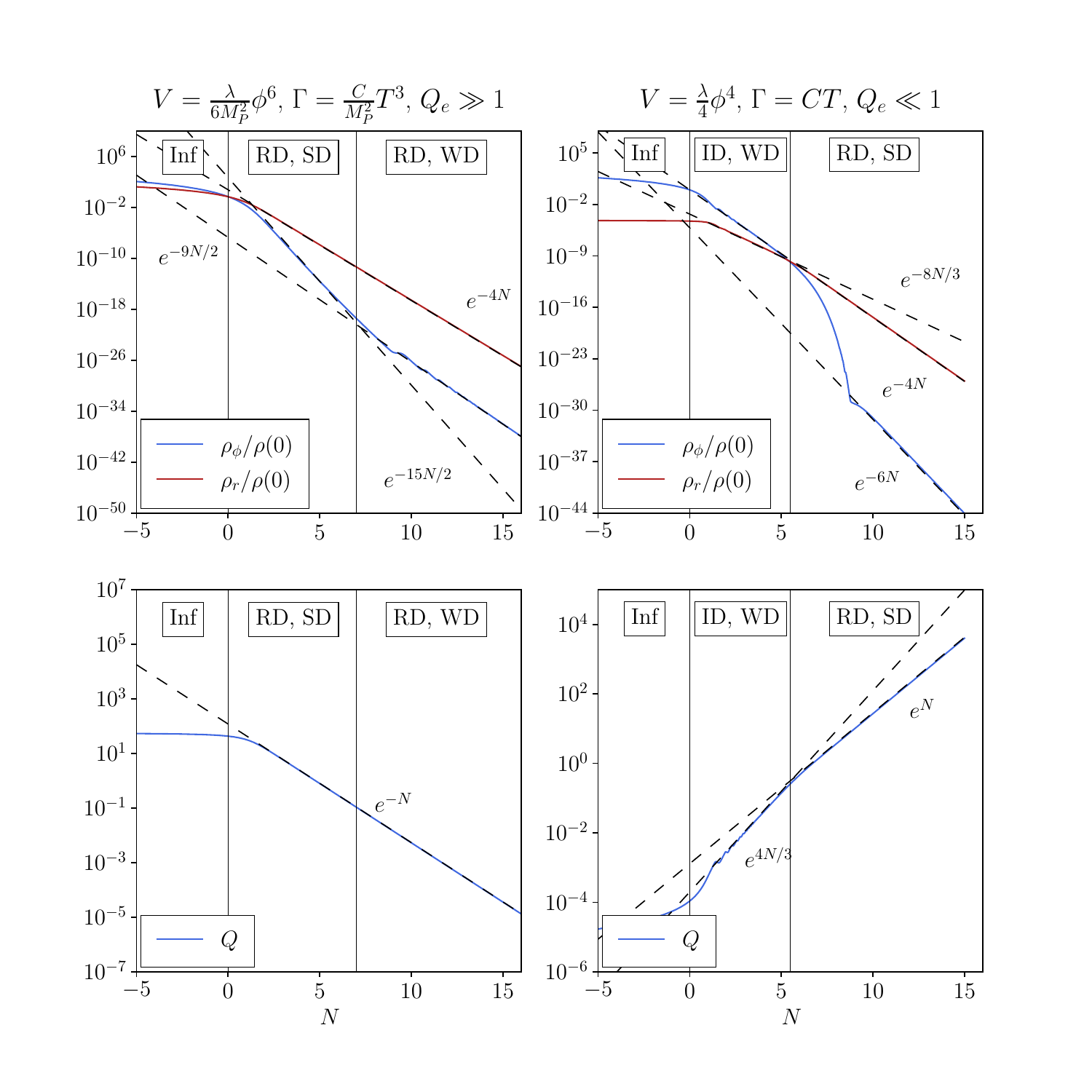}
\caption{Evolution of $\rr$, $\rp$ normalized by the total energy at the end of inflation $\rho(0)$ (top panels) and $Q$ (bottom panels) for two different models (left and right). Solid lines are numerical solutions of the background equations. Dashed lines represent different asymptotic exponential behaviours (indicated by the labels inside the plot). Recall that inflation ends at $N=0$. Black vertical lines  separate different phases, labeled in the upper part and indicating the dominating species (``Inf'' for quasi-de Sitter inflation, ``RD'' for radiation domination, ``ID" for inflaton domination) and the post-inflationary dissipative regime (``SD" for strong dissipation, ``WD" for weak dissipation).}
\label{fig:reheating}
\end{center}
\end{figure}

In this section, we discuss in detail the post-inflationary dynamics of two models: For $V\propto \phi^6$, $\Gamma \propto T^3$ and $Q_\text{end}\gg 1$ and for $V\propto \phi^4$, $\Gamma \propto T$ and $Q_\text{end}\ll 1$. The results, illustrated in Fig.\,\ref{fig:reheating}, are discussed in the following.

\paragraph{$V\propto \phi^6$, $\Gamma \propto T^3$ and $Q_{\text{end}} \gg 1$. } In this case we have $\rr(0)\simeq \rp(0)$ at the end of inflation. To illustrate the behavior of the various quantitites, we focus on the right panel of Tab.\,\ref{tab:4}. Initially, $\rr$, $\rp$ and $Q$ evolve as shown in the right column: $\rr \sim e^{-4N}$, $\rp \sim e^{-15N/2}$ and $Q \sim e^{-N}$. The universe is dominated by radiation and the dissipation rate is large. Due to the latter, the inflaton is overdamped  and does not reach the minimum of its potential, hence there are no oscillations in $\rp$. At $N\simeq 7$, $Q$ becomes smaller than one, hence we enter weak dissipation. As a consequence, $\rp$ and $\rr$ are now described by the left column  of Tab.\,\ref{tab:4}: $\rp \sim e^{-9N/2}$ and, still, $\rr \sim e^{-4N}$. Also, wiggles appear in $\rp$, indicating newly-appeared underdamped oscillations of the inflaton. Regarding $Q$, recall that its evolution depends on the dominant species. Since the dominant species is still radiation, and the evolution of radiation has not changed in the transition, still $Q\sim e^{-N}$ (in other words, since this is a radiation domination phase, a change in the behaviour of $\rp$ does not affect $Q$). From this point onwards, weak dissipation and radiation domination will continue as long as \eqref{ec:q20} and \eqref{ec:q21} hold (indeed, both $Q$ and $\rp/\rr$ keep decreasing).

\paragraph{$V\propto \phi^4$, $\Gamma \propto T$ and $Q_{\text{end}} \ll 1$. } In this case we have $\rp(0)\gg \rr(0)$. To illustrate the behavior of the various quantitites, we focus on the left panel of Tab.\,\ref{tab:3}. Initially, $\rr$, $\rp$ and $Q$ evolve as shown in the left column: $\rp \sim e^{-4N}$, $\rr \sim e^{-8N/3}$ and $Q \sim e^{4N/3}$. Since dissipation is weak, the inflaton undergoes underdamped oscillations, which manifest as wiggles in $\rp$. At $N \simeq 5.5$, $Q$ becomes larger than one \emph{and} $\rp \simeq \rr$. The transition to strong dissipation (which ends at $N\simeq 8$) implies that $\rr$ and $\rp$ are now described by the right column ($\rr\sim e^{-4N}$, $\rp \sim e^{-6N}$) and wiggles in $\rp$ no longer appear (the inflaton is overdamped). The change in the dominant species means that $Q$ is now \emph{also} described by the right column, $Q\sim e^N$ (unlike in the previous case in which the behaviour of $Q$ did not change despite the change of regime). From this point onwards, strong dissipation and radiation domination will persist as long as \eqref{ec:q20} and \eqref{ec:q21} hold (indeed, both $Q$ and $\rr/\rp$ keep increasing).

\section{Stochastic warm inflation: a second approach}
\label{app:stochasticv2}

In this appendix, we follow the same approach to stochastic inflation as in Ref.\,\cite{Ramos:2013nsa}. This procedure differs from the one in Sec.\,\ref{sec:stoch} and other recent works~\cite{Ezquiaga:2018gbw,Ballesteros:2020sre,Figueroa:2020jkf,Tomberg:2022mkt} in specific points that we discuss below. The results that follow reach the same conclusion as in Sec.\,\ref{sec:stoch}: the power spectrum obtained by means of stochastic inflation is (to first order in perturbation theory) equivalent to the one that can be computed from linear perturbation theory. \par \medskip

\noindent
We start by considering the equation of motion for the full (i.e.\ space-time dependent)  inflaton field $\phi(\bx,t)$: \begin{equation}\label{ec:ar1}
\ddot{\phi}(\bx,t) + 3H(1+Q)\dot{\phi}(\bx,t) - \frac{\nabla^2}{a^2}\phi(\bx,t) + V_\phi = \sqrt{\frac{2\Gamma T}{a^3}}\, \xi(\bx,t) \,.
\end{equation}
For simplicity, background quantities (such as $H$) are taken to be constant. We decompose the field into three parts as\footnote{This kind of decomposition differs from others in the literature, where the gradient (in other words, the $\bm{x}$ dependence) of long-wavelength modes is neglected (see e.g.\ \cite{Starobinsky:1994bd}, and more recently e.g.\ \cite{Figueroa:2021zah,Tomberg:2023kli}).}
\begin{equation}\label{ec:ar1.01}
\phi(\bx,t)=\bar{\phi}(t)+\delta\varphi(\bx,t)+\phi_q(\bx,t)\,.
\end{equation}
The term $\bar{\phi}(t)$ is the average (background) field, i.e. the solution to $\ddot{\bar{\phi}}+3H(1+Q)\dot{\bar{\phi}} + V_\phi=0$. The term $\delta\varphi(\bx,t)$  corresponds to the long-wavelength perturbations, while $\phi_q(\bx,t)$ corresponds to the short-wavelength perturbations (considered to be unsourced by thermal noise, see \eqref{ec:ar1.1} below). Injecting this decomposition into \eqref{ec:ar1}, the background terms cancel by virtue of the background equation of motion, and $\delta\varphi(\bx,t)$ is sourced by two terms: the thermal noise $\sqrt{\frac{2\Gamma T}{a^{3}}}\, \xi(\bx,t)$ and the terms depending on $\phi_q$ and its derivatives. Implicitly, this approach assumes that the thermal noise only acts on long-wavelength modes, and therefore it does not source $\phi_q$. Therefore, the equation of motion for $\phi_q$ is
\begin{equation}\label{ec:ar1.1}
\ddot{\phi}_q+3H(1+Q)\,\dot{\phi}_q-\frac{\nabla^2}{a^2}\phi_q + V_{\phi\phi}\,\phi_q + \Gamma_\phi\,\dot{\phi}\,\phi_q=0\,,
\end{equation}
with
\begin{equation}\label{ec:ar4}
\phi_q = \int\frac{\diff^3\bk}{(2\pi)^{3/2}}\, e^{i\bk\cdot \bx}\,W(k-k_\sigma)[\phi_k\, \hat{a}_{\bk}+\text{h.c.}]\,,
\end{equation}
where the modes $\phi_k$ are classical solutions of \eqref{ec:ar1.1} in Fourier space (cf.\ Sec.\,\ref{sec:analyticalestimates}). Notice the absence of the thermal contribution because of restricting the thermal noise to long wavelengths. In \eqref{ec:ar4}, $W$ is some window function selecting short-wavelength modes with respect to a reference scale $k_\sigma = \sigma \,a\, H$, $\sigma\ll 1$ (later, we will take this function to be a Heaviside step function). Also, $\hat{a}_{\bk}$, $\hat{a}^\dagger_{\bk}$ are the standard annihilation and creation	 operators. Ref.\,\cite{Ramos:2013nsa} further assumes that dissipation does not affect short-wavelength modes, i.e. that $Q=\Gamma_{\phi}=0$ in \eqref{ec:ar1.1}, therefore obtaining that the quantum component of the power spectrum corresponds to what we have called the \emph{cold-like} power spectrum (up to the correction due to the occupation number of inflaton perturbations). In this calculation, we do not make this assumption, in order to check that the procedure described in this appendix once again recovers the linear perturbation theory result obtained in Sec.\,5.  Eq.\,\eqref{ec:ar1} becomes\footnote{The form of the short-wavelength term (in brackets on the right-hand side) differs from the more common one in the literature, where only one time derivative appears. Usually, the second-order differential equation for the inflaton is recast into a system of two coupled, first-order differential equations\,\cite{Ezquiaga:2018gbw,Ballesteros:2020sre}, with independent noise contributions as done in Sec.\,\ref{sec:stoch}. Both noises can be related though in the slow-roll limit.}
\begin{multline}\label{ec:ar2}
\delta\ddot{\varphi}+3H(1+Q)\,\delta\dot{\varphi}-\frac{\nabla^2}{a^2}\delta\varphi + V_{\phi\phi}\,\delta\varphi + \Gamma_\phi\,\dot{\phi}\,\delta\varphi \\= -[\ddot{\phi}_q+3H(1+Q)\,\dot{\phi}_q-\frac{\nabla^2}{a^2}\phi_q + V_{\phi\phi}\,\phi_q + \Gamma_\phi\,\dot{\phi}\,\phi_q] + \sqrt{\frac{2\,\Gamma\, T}{a^3}}\, \xi(\bx,t)\,.
\end{multline}
From here on, the calculations are very similar to the ones performed in Sec.\,5 and Sec.\,6. Decomposing $\delta \varphi$ in Fourier space modes:
\begin{equation}
\delta \varphi = \int\frac{\diff^3\bk}{(2\pi)^{3/2}}\, e^{i\bk\cdot \bx}\,\delta \varphi_k,
\qquad\delta \varphi_k \, \equiv \, \left[\phi_k(z)\,\hat{a}_\bk+\text{h.c.}\right] W(k_\sigma-k)\,,
\end{equation}
terms proportional to $W$ on the right-hand side of Eq.\,(\ref{ec:ar2}) vanish by virtue of $\phi_{k}$ being a solution of \eqref{ec:ar1.1}. Changing the time variable to $z\equiv k/(aH)$ (assuming constant $H$), this equation yields
\begin{equation}\label{ec:ar5}
\frac{\diff^2\delta\varphi_k}{\diff z^2}+\frac{1-2\nu}{z}\frac{\diff\delta\varphi_k}{\diff z}+\left[1+\frac{1}{z^2H^2}(V_{\phi\phi}+\Gamma_\phi\,\dot{\phi})\right]\delta\varphi_k = \tilde{\xi}_\bk + \sqrt{\frac{2\,\Gamma\, T}{k^3}}\xi_{\bk}(z)\,,
\end{equation} 
where $\nu=\frac{3}{2}(1+Q)$ and the effective noise $\tilde{\xi}_{\bk}$ is defined as
\begin{equation}\label{ec:ar5.5}
\tilde{\xi}_{\bk}\, \equiv \, -\left[f_k(z)\,\hat{a}_\bk+\text{h.c.}\right]\,,
\end{equation}
with
\begin{equation}
f_k(z)=\frac{\diff^2W(k-k_\sigma)}{\diff z^2}\,\phi_k+2\frac{\diff W(k-k_\sigma)}{\diff z}\,\frac{\diff \phi_k}{\diff z}+\frac{1-2\nu}{z}\frac{\diff W(k-k_\sigma)}{\diff z}\,\phi_k\,.
\end{equation}
A formal solution to \eqref{ec:ar5} is 
\begin{equation}
\delta\varphi_k(z)=\int_z^\infty \, \diff z' \, G(z,z' )\, \tilde{\xi}_{\bm{k}}(z')+\sqrt{\frac{2\,\Gamma\, T}{k^3}}\int_z^\infty \, \diff z' \, G(z,z' )\,\xi_{\bk}(z') \,,
\end{equation}
where $G(z,z^\prime)$ is the Green's function of the differential operator appearing on the right-hand-side of Eq.\,(\ref{ec:ar2}). The corresponding power spectrum has therefore two distinct components\footnote{It should be noted that the average in the first integral is a quantum average, while the one in the second integral is an average over realizations of the thermal noise. Even though the quantum character of $\tilde{\xi}$ is manifest from \eqref{ec:ar5.5}, it is easy to show that $[\tilde{\xi}(\bx,z),\tilde{\xi}(\bx',z')]=0$, which allows to treat $\tilde{\xi}$ as a stochastic classical source for long-wavelength modes. }
\begin{align}
\label{ec:ar6}
\mathcal{P}_{\delta\varphi}\,\delta(\bk+\bk')=\frac{k^3}{2\pi^2}\langle \delta\varphi_k(z) \delta\varphi_{k'}(z)\rangle =	\left(\mathcal{P}_{\delta\varphi}^{(q)}+\mathcal{P}_{\delta\varphi}^{(th)}\right)\delta(\bk+\bk')\,,
\end{align}
where
\begin{align}
{\mathcal{P}_{\delta\varphi}^{(q)}\delta(\bk+\bk')}={\frac{k^3}{2\pi^2}\int_z\, \diff z_1\,\diff z_2 \,G(z,z_1)\,G(z,z_2)\langle \tilde{\xi}_{\bm{k}}(z_1)\,\tilde{\xi}_{\bm{k}'}(z_2)\rangle}\,, \\
\mathcal{P}_{\delta\varphi}^{(th)}\delta(\bk+\bk') = {\frac{\Gamma T}{\pi^2}\int_z\, \diff z_1\,\diff z_2 \,G(z,z_1)\,G(z,z_2)\langle {\xi}_{\bm{k}}(z_1)\,{\xi}_{\bm{k}'}(z_2)\rangle}\,.
\end{align}
The term $\mathcal{P}_{\delta\varphi}^{(th)}$ is called {\it thermal component} in Ref.\,\cite{Ramos:2013nsa} (as it arises from the thermal noise $\xi$ sourcing the inflaton perturbations). It corresponds exactly to what we called \emph{inhomogeneous component} in the context of linear perturbation theory, see Eq.\,(\ref{ec:b46}).\footnote{In Eq.\,(\ref{ec:b46}) there is an extra term (last term in brackets) which accounts for the backreaction of $\delta\rho_r$ into $\delta\phi$. Such backreaction term is discarded in this calculation for simplicity.} The term $\mathcal{P}_{\delta\varphi}^{(q)}$ is called {\it quantum component} in Ref.\,\cite{Ramos:2013nsa}, as it arises from the evolution of quantum perturbations unsourced by thermal noise (re-expressed as an effective source itself, $\tilde{\xi}$). Next, we will prove that this quantum component  corresponds exactly to what we called \emph{homogeneous component} in the context of perturbation theory in Sec.\,\ref{sec:analyticalestimates}. This component can be expressed as
\begin{equation} \label{ec:ar7}
\mathcal{P}_{\delta\varphi}^{(q)}\,\delta(\bk+\bk')=\frac{k^3}{2\pi^2}\,\int_z^\infty\int_z^\infty \, \diff z_1\,\diff z_2 \, G(z,z_1)\, G(z,z_2)\,[f_k(z_1)\,f^*_{k'}(z_2)\langle\hat{a}_\bk\,\hat{a}^\dagger_{\bk'}\rangle+\text{h.c.}]\,.
\end{equation}
As per the same discussion as in Sec.\,\ref{quantumn}, the quantum expectation value of the number operator and its complex conjugate are\footnote{Notice how the appearance of a possible Bose-Einstein thermal correction is strictly due to the expectation value of the number operator and its hermitian conjugate, exactly as in standard perturbation theory, see Eq.~\,(\ref{ec:thermal_correction}).}
\begin{equation}
\langle\hat{a}_\bk\,\hat{a}^\dagger_{\bk'}\rangle = (n_{BE}+1)\,\delta(\bk+\bk'), \qquad \langle\hat{a}^\dagger_\bk\,\hat{a}_{\bk'}\rangle = n_{BE}\,\delta(\bk+\bk')\,,
\end{equation}
if inflaton perturbations are in equilibrium with the thermal bath, and
\begin{equation}
\langle\hat{a}_\bk\,\hat{a}^\dagger_{\bk'}\rangle = \delta(\bk+\bk'), \qquad \langle\hat{a}^\dagger_\bk\,\hat{a}_{\bk'}\rangle = 0\,,
\end{equation}
if inflaton perturbations are in vacuum. Following the notation introduced in Sec.\,\ref{sec:analyticalestimates}, we denote
\begin{equation}
\langle\hat{a}_\bk\hat{a}^\dagger_{\bk'}\rangle+\text{h.c.} = \Theta \, \delta(\bk+\bk')\,.
\end{equation}
Eq.\,\eqref{ec:ar7} thus leads to
\begin{equation}\label{ec:ar7.5}
\mathcal{P}_{\delta\varphi}^{(q)} = \frac{k^3}{2\pi^2}  \, |I_k(z)|^2 \, \Theta\,,
\end{equation}
with
\begin{equation}
I_k(z)\, \equiv \, \int_z^\infty \, \diff z'\, G(z,z')\, f_k(z')\,.
\end{equation}
One can evaluate $I_k(z)$, by integrating by parts the term involving $\diff^2W/\diff z^2$, yielding
\begin{equation}\label{ec:ar9}
I_k(z)=\int_z^\infty \diff z'\, \delta(z'-\sigma) \left[G(z,z')\frac{\diff \phi_k(z)}{\diff z'}+\left(\frac{1-2\nu}{z'}G(z,z')-\frac{\partial G(z,z')}{\partial z'}\right)\phi_k(z')\right]\,,
\end{equation}
where we chose the window function to be a Heaviside step function. At this point, we have to introduce a functional form for $G(z,z')$ and $\phi_k(z')$. Crucially, the differential operator which defines Eq.\,\eqref{ec:ar1.1} (satisfied by $\phi_k$) is the same one defining the left-hand side of \eqref{ec:ar2}.\footnote{This is not the case in Ref.\,\cite{Ramos:2013nsa}, where dissipative terms are neglected in \eqref{ec:ar1.1} but not in \eqref{ec:ar2}.} This means that, given two independent solutions for \eqref{ec:ar1.1}, which we denote $\phi^{(1)}_k$, $\phi^{(2)}_k$, the Fourier mode $\phi_k$ is given by
\begin{equation}\label{ec:ar10}
\phi_k = A\,\phi^{(1)}_k + B\,\phi^{(2)}_k\,,
\end{equation} 
with constants $A$ and $B$ to be fixed by initial conditions. The Green's function can be expressed in terms of these independent solutions as
\begin{equation}\label{ec:ar11}
G(z,z')=-\frac{\phi^{(1)}_k(z)\,\phi^{(2)}_k(z')-\phi^{(2)}_k(z)\,\phi^{(1)}_k(z')}{\phi^{(1)}_k(z')\,\dfrac{\diff\phi^{(2)}_k(z')}{\diff z'}-\phi^{(2)}_k(z')\,\dfrac{\diff \phi^{(1)}_k(z')}{\diff z'}}\,\theta(z'-z)\,.
\end{equation}
One can chose the following set of two real independent solutions of Eq.\,\eqref{ec:ar1.1}:
\begin{equation}
\phi^{(1)}_k = z^\nu J_\mu(z), \qquad \phi^{(2)}_k = z^\nu Y_\mu(z)\,,
\end{equation}
with\footnote{As a consequence of Ref.\,\cite{Ramos:2013nsa} neglecting dissipative terms in \eqref{ec:ar1.1} but not in \eqref{ec:ar2}, the Bessel functions in \eqref{ec:ar10} have $\nu=3/2$, $m=3\eta_V$, while the Bessel functions in \eqref{ec:ar11} have $\nu=\frac{3}{2}(1+Q)$, $m=3\eta_V-\frac{3Q}{1+Q}\beta_V$.}
\begin{equation}
\nu\equiv \frac{3}{2}(1+Q), \qquad \mu \equiv \sqrt{\nu^2-m}, \qquad m \equiv  3\,\eta_V-\frac{3Q}{1+Q}\beta_V\,, \qquad \beta_V \equiv M_P^2 \dfrac{\Gamma_\phi V_\phi}{\Gamma V}.
\end{equation}
By substituting these quantities into \eqref{ec:ar9}, all the dependence in $z'$ inside the brackets vanishes, and we are left with
\begin{equation}
I_k(z)= \left(A\,\phi^{(1)}_k(z) + B\,\phi^{(2)}_k(z)\right) \int_z^\infty \diff z' \, \delta(z'-\sigma).
\end{equation}
For $z<\sigma$ (i.e. sufficiently outside the horizon), the integral is equal to 1 and we have\begin{equation}
I_k(z)= A\,\phi^{(1)}_k(z) + B\,\phi^{(2)}_k(z), \qquad z<\sigma\,.
\end{equation}
Going back to \eqref{ec:ar7.5}, we find
\begin{equation}
\mathcal{P}^{(q)}_{\delta\varphi} = \frac{k^3}{2\pi^2}\left|A\,\phi^{(1)}_k(z) + B\,\phi^{(2)}_k(z)\right|^2\,\Theta, \qquad z<\sigma\,.
\end{equation}
This is exactly the same result obtained in linear perturbation theory, i.e. the \emph{homogeneous component} of the power spectrum --See Eq.\,(\ref{ec:b42}) and Eq.\,(\ref{eq:homoegenoust}) for the functional form and Eq.\,(\ref{ec:correctedbythermal}) for the thermal correction-- in agreement with the main conclusion of Sec.\,\ref{sec:stoch}. \par \medskip

This result coincides, to first order in $\sigma$, with the one in Eq.\,(4.14) of Ref.\,\cite{Ramos:2013nsa} by substituting $\nu=3/2(1+Q)\to\nu= 3/2$ and $m=3\eta_V-\frac{3Q}{1+Q}\beta_V\to m=3\eta_V $. The discrepancy can be traced back to the neglect of dissipation terms in \eqref{ec:ar1.1} in Ref.\,\cite{Ramos:2013nsa}. This difference seems to be at the origin of the $\Theta$-term multiplying the ``cold-like" component in Eq.\,(\ref{ec:frompapers}) instead of the \textit{homogeneous} contribution. As illustrated in Fig.\,\ref{fig:comparison_nBE_older} and Fig.\,\ref{fig:comparison} on the left panel, this discrepancy causes significant differences on the parameter space giving rise to dissipation coefficients in the intermediate regime $Q\sim [10^{-3}-1]$.

\newpage

\addcontentsline{toc}{section}{References}
\bibliographystyle{utphys}
\bibliography{references}

\end{document}